\documentclass[12pt]{article}
\usepackage{amsmath,amssymb,epsfig}
\usepackage{ulem}

\usepackage{color}
\input{colordvi.tex}
\def\unit{{\relax{\rm 1\kern-.26em I}}}

\def\6#1{{\underline{#1}}}
\def\m6#1{{\underline{#1}\,}}

\newdimen\Tdim
\def\ispan{{\setbox0=\hbox{i}%
\Tdim\ht0\advance\Tdim\dp0\rule[-\dp0]{0pt}{\Tdim}}}
\def\jspan{{\setbox0=\hbox{j}%
\Tdim\ht0\advance\Tdim\dp0\rule[-\dp0]{0pt}{\Tdim}}}
\def\Tspan#1{{\setbox0=\hbox{#1}%
\Tdim\ht0\advance\Tdim\dp0\advance\Tdim.55ex\rule[-\dp0]{0pt}{\Tdim}\box0}}

\def\be{\begin{eqnarray}}
\def\ben{\begin{eqnarray*}}
\def\ee{\end{eqnarray}}
\def\een{\end{eqnarray*}}
\def\Tr{{\rm Tr}}

\def\p{\partial}

\def\sech{\mathop{\mathrm{sech}}\nolimits}

\def\=:{=\hspace{-.7em}\raisebox{1.1ex}{.}\hspace{.1em}\raisebox{-0.2ex}{.} }

\newcommand {\beq}{\begin{eqnarray}}
\newcommand {\eeq}{\end{eqnarray}}
\newcommand {\non}{\nonumber\\}

\newcommand{\capfont}{\small}

\makeatletter

\renewcommand\section{\@startsection {section}{1}{\z@}%
                                   {-3.5ex \@plus -1ex \@minus -.2ex}%
                                   {2.3ex \@plus.2ex}%
                                   {\normalfont\large\bfseries}}

\renewcommand\subsection{\@startsection{subsection}{2}{\z@}%
                                     {-3.25ex\@plus -1ex \@minus -.2ex}%
                                     {1.5ex \@plus .2ex}%
                                     {\normalfont\normalsize\bfseries}}

\newcount\hour \newcount\minute
\hour=\time \divide \hour by 60
\minute=\time
\def\now{%
\ifnum \hour<13
  \ifnum \hour=0 \advance \hour by 12 \number\hour:\else \number\hour:\fi%
     \ifnum \minute<10 0\fi%
     \number\minute%
\ A.M.%
\else \advance \hour by -12 \number\hour:%
  \ifnum \minute<10 0\fi%
  \number\minute%
  \ P.M.%
\fi%
}

\makeatother

\begin{document}

\baselineskip=18pt  
\numberwithin{equation}{section}  
\allowdisplaybreaks  



%
%


\thispagestyle{empty}

\vspace*{-2cm}
\begin{flushright}
{\tt YGHP-14-01}
\end{flushright}

\begin{flushright}
\end{flushright}

\begin{center}

\vspace{-.5cm}

\vspace{0.5cm}
{\bf\Large Dynamics of slender monopoles and anti-monopoles\\
 in non-Abelian superconductor}
\vspace*{1.5cm}

{\bf
$^{a,b}$Masato Arai, $^{b,c}$Filip Blaschke, $^{d}$Minoru Eto, $^{e}$Norisuke Sakai}
\vspace*{0.5cm}
 
\vspace*{0.5cm}
 
{\it 
$^a$Fukushima National College of Technology, 30 Nagao, Kamiarakawa, Taira, Iwaki, Fukushima 970-8034, Japan\\
$^b$Institute of Experimental and Applied Physics, \\
Czech Technical University in Prague, Horsk\'a 3a/22, 128 00 Prague 2, Czech Republic\\
$^c$Institute of Physics, Silesian University in Opava, Bezru\v{c}ovo n\'am\v{e}st\'i 1150/13, 746 01 Opava, Czech Republic\\
$^d$Department of Physics, Yamagata University, Kojirakawa-machi 1-4-12, Yamagata, Yamagata 990-8560, Japan\\
$^e$Department of Physics, and Research and Education Center for Natural Sciences, Keio University,
Hiyoshi 4-1-1, Yokohama, Kanagawa 223-8521, Japan
}

\end{center}

\vspace{1cm} \centerline{\bf Abstract} \vspace*{0.5cm}

Low energy dynamics of 
magnetic monopoles and anti-monopoles in the $U(2)_{\rm C}$ gauge 
theory is studied in the Higgs (non-Abelian superconducting) phase. 
The monopoles in this superconducting phase are not 
spherical but are of slender ellipsoid which are pierced by a vortex string. 
We investigate scattering of the slender monopole and anti-monopole, 
and find that they do not always decay into radiation, contrary 
to our naive intuition. 
They can repel, make bound states (magnetic mesons) 
or resonances. 
Analytical solutions including any number of monopoles and anti-monopoles 
are obtained in the first non-trivial order of rigid-body 
approximation.
We point out that some part of solutions of slender monopole 
system in $1+3$ dimensions can be mapped exactly onto the sine-Gordon 
system in $1+1$ dimensions. 
This observation allows us to visualize dynamics of monopole 
and anti-monopole scattering easily.

\newpage
\setcounter{page}{1} 





\section{Introduction}

Topological solitons in classical and quantum field theories 
are widely appreciated as important objects in diverse areas of 
modern physics \cite{Manton:2004tk,Tong:2005un,Eto:2006pg,Shifman:2007ce}. 
Among various topological solitons, magnetic monopoles are 
one of the most 
fascinating solitons in high energy physics. 
If a monopole exists, electric charges in the universe are 
quantized by the Dirac's quantization condition \cite{Dirac:1931kp}: 
product of electric charge $Q_{\rm e}e$ and magnetic charge $Q_{\rm m}g$ must be 
proportional to integers: $Q_{\rm e}eQ_{\rm m}g \propto n$. 
This implies that a weak electric 
coupling $e$ corresponds to a strong magnetic coupling $g$. 
This strong-weak coupling duality is a powerful tool to understand 
strong coupling physics. 
Furthermore, monopoles are expected to play an important role to explain 
the confinement in QCD. 
It has been proposed that the QCD vacuum is in dual color 
superconductor where magnetic monopoles condense \cite{Nambu:1974zg,Mandelstam:1974vf,'tHooft:1981ht}. 
Then the color-electric fluxes from quarks and anti-quarks 
are squeezed to form stringy flux tubes, resulting in a linear 
confining potential. 
It is, however, very difficult to verify this idea, since QCD is 
strongly coupled at low energies because of the asymptotic freedom. 
It is only in the case of supersymmetric gauge theories \cite{Seiberg:1994rs} that
this confinement mechanism has been demonstrated explicitly.

In contrast, QCD becomes weakly coupled at high baryon density, and
enters into the color-superconducting 
phase where di-quarks condense \cite{Alford:1997zt,Rapp:1997zu}. 
Then magnetic monopoles should be confined and color magnetic 
fields form flux tubes. 
The flux tube in the color-flavor locked (CFL) phase at 
asymptotically high density is a non-Abelian vortex string (called semi-superfluid vortex string)
\cite{Balachandran:2005ev,Nakano:2007dr}. 
A distinctive feature of non-Abelian vortex compared to the 
usual Nielsen-Olesen Abelian vortex is that the non-Abelian 
vortex breaks a non-Abelian global symmetry of the vacuum 
state and non-Abelian orientational moduli \cite{Hanany:2003hp,Auzzi:2003fs,Shifman:2004dr} emerge. 
This orientational moduli has been found to give
a confining state of monopole and anti-monopole, 
namely magnetic meson was predicted 
in high density QCD \cite{Gorsky:2011hd,Eto:2011mk,Eto:2013hoa}.

If we tune couplings to critical values which enable us 
to embed the theory into a supersymmetric one, we can have 
Bogomolnyi-Prasad-Sommerfield (BPS) solitons \cite{Bogomolny:1975de}, which preserve 
a part of supersymmetry \cite{Witten:1978mh}. 
An analytic solution \cite{Bogomolny:1975de} can be  obtained for the
't Hooft-Polyakov monopole \cite{'tHooft:1974qc,Polyakov:1974ek} in the Coulomb phase
where $SU(2)$ gauge symmetry is broken to a $U(1)$ subgroup.
Since there is no static forces between BPS monopoles, positions 
of BPS monopoles become parameters of the solution, which are 
generically called moduli. 
The moduli space of monopoles must be a hyper-K\"ahler manifold, 
reflecting the underlying 
supersymmetry with eight supercharges.
The dynamics of slowly moving BPS solitons are well-approximated 
by a geodesic motion on the moduli space \cite{Manton:1981mp}. 
This is called the moduli approximation. 
Asymptotic metrics on moduli space have been known for 
well-separated monopoles \cite{Gibbons:1995yw}. 
For the particular case of scattering of two monopoles, the 
exact metric is known, and it is explicitly shown that
slowly-moving BPS monopoles scatter with the right 
angle when they collide head-on \cite{Atiyah:1988jp}. 
Although the moduli approximation is useful,
one should note that it can be applied neither for  scattering of BPS
solitons with high momentum nor for non-BPS systems.
Among non-BPS solutions, an interesting non-BPS ``bound state'' 
of a monopole and an anti-monopole in the Coulomb phase has 
been rigorously established \cite{Taubes:1982ie}. 
Subsequent numerical studies also revealed a vast array of 
monopole and anti-monopole composite states \cite{Oleszczuk:1987dx,Ioannidou:1999xq,Kleihaus:1999sx}.  
Although these ``bound states'' eventually decay due to 
unstable modes, they can play a significant and 
remarkable role in understanding the dynamics of monopole and 
anti-monopole system in the Coulomb phase. 
It has been noted that D-brane and anti-D-brane configurations 
play an important role in D-brane dynamics even though they are 
unstable \cite{Sen:1998rg,Sen:1998ii,Sen:1998tt}.

If the non-Abelian gauge symmetry is completely broken, 
we are in the Higgs phase, namely the non-Abelian 
superconducting phase. In contrast to the Coulomb phase, 
monopoles in the Higgs phase have several distinctive features. 
Firstly, they are pierced by a vortex string. 
In other words, the magnetic fields coming out of the monopole 
are squeezed into flux tubes attached to both ends of the monopole. 
In the non-Abelian superconductor, there can be several 
species of magnetic flux tubes and a magnetic monopole 
can be interpreted as a junction \cite{Hindmarsh:1985xc} on which two different 
species of magnetic fluxes meet. 
In fact, the static BPS monopole in the Higgs phase has been found as a 
static BPS kink solution in the $1+1$-dimensional low-energy effective field 
theory on the vortex \cite{Tong:2003pz}. 
Secondly, shape of a monopole is not spherical. This can be seen as follows.
In the non-Abelian superconductor, there are two types of 
topological solitons, vortex string and monopole. 
Thus we have two fundamental length scales:
One is a transverse size $L_{\rm T}$ of the flux tube 
and the other is a length $L_{\rm L}$ of the monopole. 
Since a monopole resides on a vortex, its shape is generally 
not spherical depending on the ratio of the two scales. 
The $1+1$ dimensional effective theory description was used
in Refs.~\cite{Auzzi:2003fs,Tong:2003pz,Shifman:2004dr}  assuming
\begin{equation}
L_{\rm T} \ll L_{\rm L}, 
\label{eq:slender_cond}
\end{equation}
where monopoles are of a slender ellipsoidal shape. 
This approximation has been used previously \cite{Cipriani:2012pa} 
to obtain an effective action of 1/4 BPS non-Abelian monopole-vortex 
complex. Another work to obtain effective action of 
monopole-vortex complex appeared recently \cite{Evslin:2013wka}.

In this work, we will consider a straight vortex string where both monopoles 
and anti-monopoles reside. 
Since certain 
interaction 
exists between a monopole and anti-monopole, 
such solutions can no longer be static.
The monopoles and anti-monopoles move along the vortex string.
Therefore, 
scatterings are all head-on collisions. 
In contrast to lots of studies on scattering of the BPS monopoles,
there is very few studies about the dynamics of BPS monopoles 
and anti-monopoles, especially in the Higgs phase. 
This is mainly due to the inapplicability of the moduli 
approximation to the non-BPS monopoles and anti-monopoles 
system, which necessitates other methods such as numerical 
analysis. 
Instead of numerical methods, we consider a systematic 
expansion in powers of the ratio $L_{\rm T}/L_{\rm L}$ 
of length scales of the model, which allows us to obtain 
analytic solutions. 
At the first order of the expansion, we obtain the rigid-body 
approximation, where distortion of vortex profile can be 
neglected during the collision. 
These approximate analytic solutions are very 
useful to understand the dynamics of monopole and 
anti-monopole system. 
As explained above, the monopole and anti-monopole dynamics 
here is essentially $1+1$-dimensional.
This observation is crucial in this work. 
We will observe that a part of the dynamics can be mapped 
on to the integrable sine-Gordon model. 
In the rigid-body approximation, we can compute magnetic and 
electric fields and other induced fields with the aid of an 
exact mapping between $1+3$-dimensional field configurations 
of gauge theory and $1+1$-dimensional field configurations 
of the sine-Gordon model. 
This allows us to visualize scattering of monopoles and 
anti-monopoles easily.
As a result, we obtain that monopoles and anti-monopoles can 
repel each other, make bound states (magnetic mesons), or resonances, 
as in the case of dense QCD \cite{Gorsky:2011hd,Eto:2011mk,Eto:2013hoa}. 
Another reason why there are only few studies of 
monopole and anti-monopole system may be possible 
instabilities of the system. 
Since the total topological charge for the monopole and 
anti-monopole pair vanishes, one may be tempted to conclude 
that they eventually annihilate into radiation. 
In spite of this intuition, we find an intriguing result 
that they can repel, make bound states or resonances. 
Thus, our study may open a new direction in understanding 
the monopole dynamics.

This paper is organized as follows. 
The model is introduced in Sec.~2, where we also discuss the 
geometry of the vacuum manifold and of the associated moduli 
space. 
In Sec.~3 we derive BPS equations and describe two types of 
static solutions
the exact 1/2 BPS vortex string 
solution and 
the approximate 1/4 BPS solution, 
describing a static monopole attached to the host vortex string. 
In Sec.~4, we rederive these solutions using 
a systematic expansion in powers of $L_{\rm T}/L_{\rm L}$. 
At the first order of the expansion we obtain the {\it rigid-body} 
approximation. 
We provide a mapping between $1+3$-dimensional dynamics 
onto a $1+1$-dimensional effective dynamics. 
In Sec.~5 we fully utilize this mapping in order to discuss 
scattering of monopoles and anti-monopoles on the vortex 
string and also their bound states. 
In Sec.~6 we study dyonic extension of monopoles by using the rigid-body approximation.
We conclude and discuss
future directions of our work
in Sec. 7.

\section{Model}

Let us consider a $U(2)_{\rm C}$ Yang-Mills-Higgs system 
\beq
{\cal L} &=& \Tr\left[
- \frac{1}{2g^2}F_{\mu\nu}F^{\mu\nu} + D_\mu H (D^\mu H)^\dagger + \frac{1}{g^2}D_\mu \Sigma D^\mu \Sigma
\right] - V, \label{eq:lag}\\
V &=& \Tr\left[
\frac{g^2}{4}\left(HH^\dagger - v^2{\bf 1}_2\right)^2 + \left(\Sigma H - HM\right)\left(\Sigma H - HM\right)^\dagger
\right],
\eeq
where the field strength and the covariant derivatives are defined by
\beq
F_{\mu\nu} &=& \p_\mu A_\nu - \p_\nu A_\mu + i \left[A_\mu,A_\nu\right],
\label{eq:coder1}\\
D_\mu H &=& \p_\mu H + i A_\mu H,
\label{eq:coder2}\\
D_\mu \Sigma &=& \p_\mu \Sigma + i \left[A_\mu,\Sigma\right].
\label{eq:coder3}
\eeq
The $N_F$ species of Higgs fields in the fundamental 
representation of the $U(2)_{\rm C}$ gauge group is denoted by a 
$2 \times N_{\rm F}$ matrix $H$. 
We concentrate on $N_{\rm F} = 2$ case in the following. 
Another Higgs field $\Sigma$ in the adjoint representation 
of the $U(2)_{\rm C}$ gauge group is denoted by a real $2 \times 2$ matrix. 
The quartic scalar coupling is given in terms of the gauge 
coupling constant $g$, which allows our model to be embeded in 
a supersymmetric theory. 
The parameter $v$ giving the vacuum expectation value of the 
Higgs field $H$ comes from the so-called Fayet-Illiopoulos 
(FI) parameter in the supersymmetric context. 
We assume $v > 0$ in what follows.
We take the mass matrix $M$ for $H$ in the following form 
\beq
M = \frac{m}{2} \sigma_3.
\eeq
Global symmetry of the model depends on the mass parameter $m$. 
If $m=0$, the flavor symmetry is $SU(2)_{\rm F}$. 
If $m\neq0$, the flavor symmetry reduces to 
$U(1)_{\rm F} \subset SU(2)_{\rm F}$ generated by the third 
component of $SU(2)_{\rm F}$. 

The 
vacuum of the model is determined by the condition $V=0$: 
\beq
H H^\dagger = v^2 {\bf 1}_2,\quad \Sigma = M. 
\eeq
Fixing $\Sigma=M$, the most general solution for $H$ is given 
for $m\not=0$ by 
\beq \label{eq:vach}
H=v e^{i\alpha} e^{i\beta \sigma_3/2}\; ,  
\eeq
and for $m=0$ in terms of $U \in U(2)$ as
\beq
H=v U.
\eeq
Therefore the vacuum manifold in the $m\not =0$ case is 
\beq
\dfrac{\dfrac{U(1)_{\rm C0}\times U(1)_{\rm C3}}{{\mathbb Z}_2} 
\times U(1)_{\rm F}}{U(1)_{\rm C3+F}} 
\simeq \dfrac{U(1)\times U(1)}{\mathbb{Z}_2} \simeq T^2,
\eeq
and in the $m=0$ case is given as
\beq
\frac{U(2)_{\rm C}\times SU(2)_{\rm F}}{SU(2)_{\rm C+F}}\simeq U(2).\label{v-U}
\eeq 
In the massive case $U(1)_{\rm C0}$ stands for the overall $U(1)$ 
subgroup of $U(2)_{\rm C}$, while $U(1)_{\rm C3}$ is a $U(1)$ subgroup of 
$SU(2)_{\rm C}$ generated by $\sigma_3/2$. 
The $U(1)_{\rm C3+F}$ is a diagonal group of simultaneous 
$U(1)_{\rm C3}$ and $U(1)_{\rm F}$ rotations which is unbroken 
by the vacuum. 
The $\mathbb{Z}_2$ center of $U(2)_{\rm C}$ here is needed to 
prevent over-counting, since in Eq.~(\ref{eq:vach}) both 
$H = e^{i\alpha}e^{i\beta\sigma_3/2}$ and 
$H=e^{i(\alpha+\pi)}e^{i(\beta+\pi)\sigma_3/2}$ represent 
the same point on the vacuum manifold.
In either massive or massless case, however, we can make 
$U(2)_{\rm C}$ gauge transformations to bring 
the fields to the following representative value 
\beq
H = v {\bf 1}_2,\quad \Sigma = M.
\eeq
Therefore all the points in the vacuum manifold are physically 
equivalent, and the vacuum moduli space consists of only one point. 
We call this vacuum 
the color-flavor locking (CFL) vacuum. 
The vacuum is in the Higgs phase, where the gauge symmetry is 
completely broken. 

In this work, we consider $m\not=0$ case unless stated otherwise. 

\section{Monopole in the Higgs phase}

\subsection{The BPS equations}
\label{sec:bps_eq}

The model (\ref{eq:lag}) admits rich topological excitations; 
vortex strings and magnetic monopoles. 
In the Higgs vacuum, magnetic field can only exist by having an 
unbroken normal vacuum in a small neighborhood of the zero of the 
Higgs field. Hence magnetic field is squeezed into a vortex, 
which we call a vortex string. 
The vortex string is topologically stable due to a non-trivial 
fundamental homotopy group in the massive case 
\beq
\pi_1(T^2) = \mathbb{Z} \times \mathbb{Z}.
\label{eq:winding_number}
\eeq
There are two kinds of vortex quantum numbers, 
corresponding to two kinds of vortex strings, 
which we call the N-vortex and S-vortex. 
A magnetic monopole is a source of the conserved magnetic 
fluxes which are squeezed into vortex strings in this Higgs vacuum. 
Therefore a stable magnetic monopole is possible only as a 
composite soliton in the middle of a vortex string, 
but cannot exist as an isolated soliton, which can also be 
understood from the trivial homotopy
\beq
\pi_2(T^2) = 0.
\eeq

In the $U(2)_{\rm C}$ Yang-Mills Higgs theory, 
solutions for the magnetic monopole pierced by vortex strings
have been found \cite{Tong:2003pz,Isozumi:2004vg}, which preserve a quarter of supersymmetry 
charges when embedded into the supersymmetric theory. 
To see these Bogomol'nyi-Prasad-Sommerfield (BPS) solutions, 
we rewrite the total energy $E$ of static fields (and $A_0 = 0$) 
as a sum of perfect squares (plus boundary terms) as follows
\beq
E\!\! &=&\!\! \int d^3x ~
\Tr\left[
\frac{1}{g^2}\left\{
\left(F_{12}-D_3\Sigma - \frac{g^2}{2}\left(HH^\dagger 
- v^2 {\bf 1}_2\right)\right)^2
+ \left(F_{23} - D_1\Sigma\right)^2  + \left(F_{31} - D_2\Sigma\right)^2
\right\}\right.\non
&+&~4\bar D H(\bar DH)^\dagger + (D_3 H + \Sigma H - H M)
(D_3 H + \Sigma H - H M)^\dagger \non
&+& \left. \frac{1}{g^2}\epsilon_{ijk}\p_i(\Sigma F_{jk}) 
- v^2 F_{12} 
+i\left\{\partial_{1} (HD_{2} H^\dagger)- \partial_{2} (HD_{1} H^\dagger)\right\}
\right].
\label{eq:BPS_comp}
\eeq
We define
\beq
& \displaystyle z = x^1+ix^2,\quad \bar z = x^1-ix^2, \quad 
\partial = \frac{\p_1-i\p_2}{2}, \quad
\bar \partial = \frac{\p_1+i\p_2}{2},& \\
& \displaystyle A = \frac{A_1 - i A_2}{2},\quad
 \bar A = \frac{A_1 + i A_2}{2},\quad 
 D = \frac{D_1 - i D_2}{2},\quad 
 \bar D = \frac{D_1 + i D_2}{2}.&
\eeq
The last term in Eq. (\ref{eq:BPS_comp}) is the total derivative term 
which does not contribute to the total energy.
In deriving Eq. (\ref{eq:BPS_comp}), we have used the following identities 
\beq
D_iH(D_iH)^\dagger &=& 2 \left(DH(DH)^\dagger 
+ \bar D H (\bar D H)^\dagger\right),\label{id1} \\
DH (DH)^\dagger &=& \bar D H (\bar D H)^\dagger 
- \frac{1}{2}HH^\dagger F_{12}+
\frac{i}{2}\left\{\p_1(H(D_2H)^\dagger) - \p_2(H(D_1H)^\dagger)\right\}. \label{id2}
\eeq
Other useful formulas are 
\beq
\left[D_1,D_2\right] &=& -iF_{12},\label{id3} \\
\p(H(DH)^\dagger)-\bar\p(H(\bar D H)^\dagger) &=&
\frac{i}{2}\left\{\p_1(H(D_2H)^\dagger) - \p_2(H(D_1H)^\dagger)\right\}. \label{id4}
\eeq

The total energy $E$ is bounded by the sum of two topological 
charges representing the monopole energy $M_{\rm mono}$ and 
the vortex energy $M_{\rm vor}$ 
\beq
E &\ge& M_{\rm mono} + M_{\rm vor},\label{eq:bogomolnyi_bound} \\
M_{\rm mono} &=& \frac{1}{g^2}\int d^3x\ 
\Tr[\epsilon_{ijk}\p_i\left(\Sigma F_{jk}\right)],
\label{eq:mass_mono} \\
M_{\rm vor} &=& - v^2 \int d^3x ~ \Tr[F_{12}].
\label{eq:mass_vortex}
\eeq
This bound is saturated when the 
following BPS equations are satisfied 
\beq
F_{12}-D_3\Sigma - \frac{g^2}{2}\left(HH^\dagger 
- v {\bf 1}_2\right) = 0,\label{eq:BPS1}\\
F_{23} - D_1\Sigma=0,\label{eq:BPS2}\\
F_{31} - D_2\Sigma=0,\label{eq:BPS3}\\
\bar DH = 0,\label{eq:BPS4}\\
D_3 H + \Sigma H - H M = 0.\label{eq:BPS5}
\eeq

If we define a $2\times 2$ matrix field $S$ taking values in 
$GL(2,\mathbb{C})$ whose elements are functions 
of $x^{1,2,3}$ as 
\beq
\bar A = -iS^{-1}\bar\p S,\quad A_3 - i\Sigma = - i S^{-1}\p_3 S,
\label{eq:def_s}
\eeq
we can solve the equations (\ref{eq:BPS2}) -- (\ref{eq:BPS5}) in 
terms of a holomorphic matrix $H_0(z)$ 
\beq
 H = v S^{-1}(x^{1},x^{2},x^{3})H_0(z) e^{Mx^3}. 
\label{eq:mm_sol}
\eeq
This method to solve the BPS equation
is called the moduli matrix formalism \cite{Isozumi:2004vg, Eto:2006pg}. 
The following $V$-transformations leave the physical fields 
$H$ in Eq.~(\ref{eq:mm_sol}) and  $A_i$ and 
$\Sigma$ in Eq.~(\ref{eq:def_s}) unchanged. 
\beq
S(x^1,x^2,x^3) \to V(z) S(x^1,x^2,x^3),\quad H_0(z) \to V(z)H_0(z),
\qquad V(z) \in GL(2,\mathbb{C}),
\eeq
where elements of the $GL(2,\mathbb{C})$ matrix $V(z)$
are holomorphic functions in $z$. 
Therefore the moduli space of the monopole vortex complex becomes 
the moduli matrices divided by the $V$-equivalence relation.

The $U(2)_{\rm C}$ gauge 
transformations act on $S^{-1}$ from 
the left as 
\beq
S^{-1} \to U_{\rm C} S^{-1}. 
\eeq
By defining $U(2)_{\rm C}$ gauge invariant matrices $\Omega$ and 
$\Omega_0$, 
\beq
\Omega = SS^\dagger,\quad
\Omega_0 = H_0e^{2Mx^3}H_0^\dagger, 
\eeq
we can cast the remaining BPS equation (\ref{eq:BPS1}) into the 
following master equation 
\beq
\frac{1}{g^2v^2}\left[4\bar\p\left(\p\Omega \Omega^{-1}\right) + \p_3\left(\p_3\Omega\Omega^{-1}\right)\right] = {\bf 1}_2 -  \Omega_0\Omega^{-1}.
\label{eq:master}
\eeq
This master equation should be solved with the boundary condition 
\beq
\Omega \to \Omega_0\qquad {\rm as}\qquad \left|\vec x\right| \to \infty.
\eeq
The $U(2)_{\rm C}$ gauge invariants $\Omega$ and $\Omega_0$ are covariant 
under the $V$-transformations.

\subsection{Vortex strings}

Before describing the monopole-vortex complex, let us first 
explain a simpler configuration of vortex strings without monopoles. 
Let us consider a vortex with unit vorticity corresponding to the 
Higgs field $H$ with a single zero. 
Since $S^{-1}$ is defined to have no singularities nor zeros, 
the zero should be placed either in upper-left or lower-right 
corner\footnote{The zero can be placed only on the diagonal, 
otherwise one cannot solve the master equation.} of the moduli 
matrix $H_0(z)$. 
Hence, in the $m\not = 0$ case, we have two different types 
of Abelian vortices 
(see also the argument around Eq.~(\ref{eq:winding_number})). 
When moduli matrix has a zero in the upper-left corner, 
we obtain 
\beq
H_0 = \left(
\begin{array}{cc}
z & 0\\
0 & 1
\end{array}
\right),\quad S = 
\left(
\begin{array}{cc}
e^{\frac{\psi(x^1,x^2)}{2}} & 0\\
0& 1
\end{array}
\right)e^{Mx^3},
\label{eq:N_vor}
\eeq
which is called an N-vortex.\footnote{
This terminology comes from the fact that the vortex 
is sitting at the north pole of the moduli space of the 
non-Abelian vortex for $m=0$.
} 
Hereafter, we choose $\psi$ to be real by a gauge choice. 
\begin{figure}[ht]
\begin{center}
\begin{tabular}{ccc}
\includegraphics[width=7.5cm]{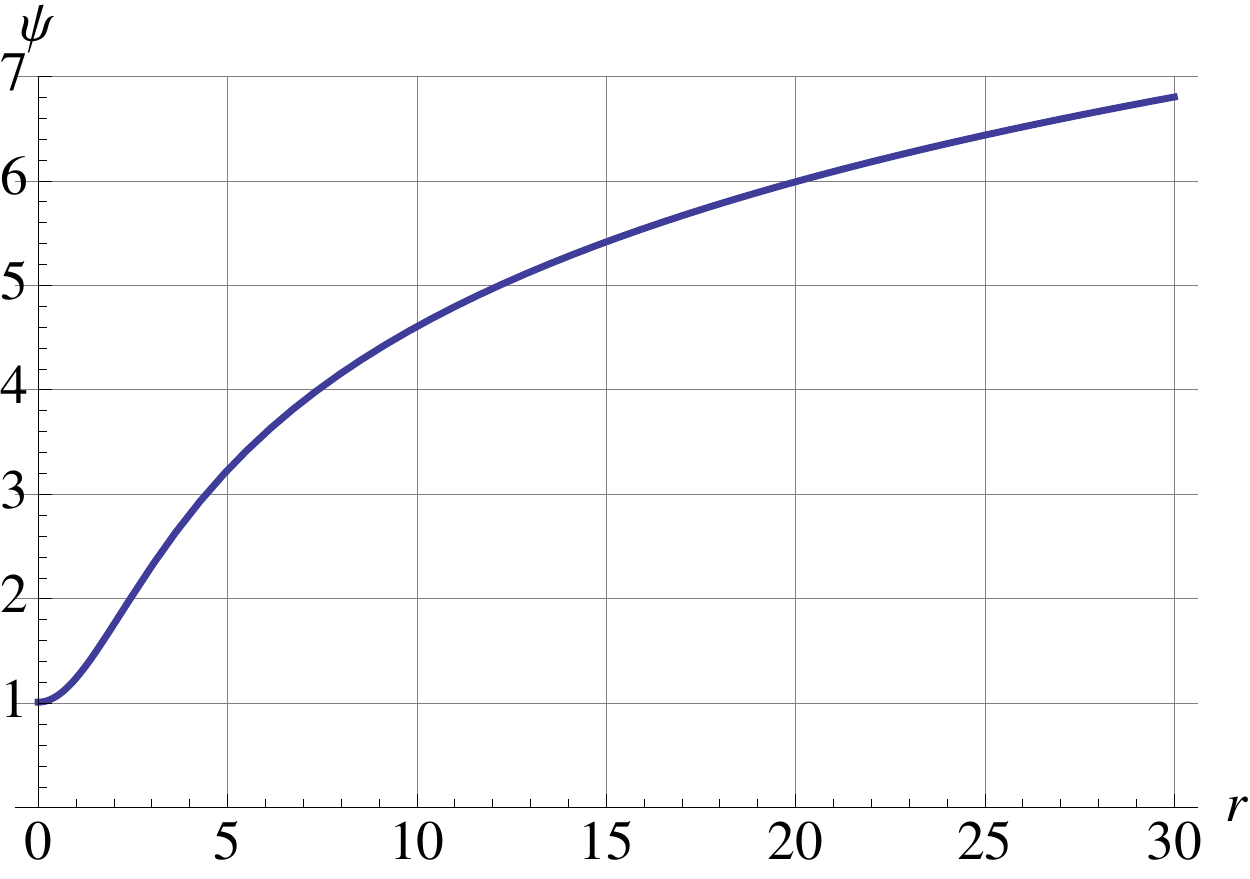} & &
\includegraphics[width=7.5cm]{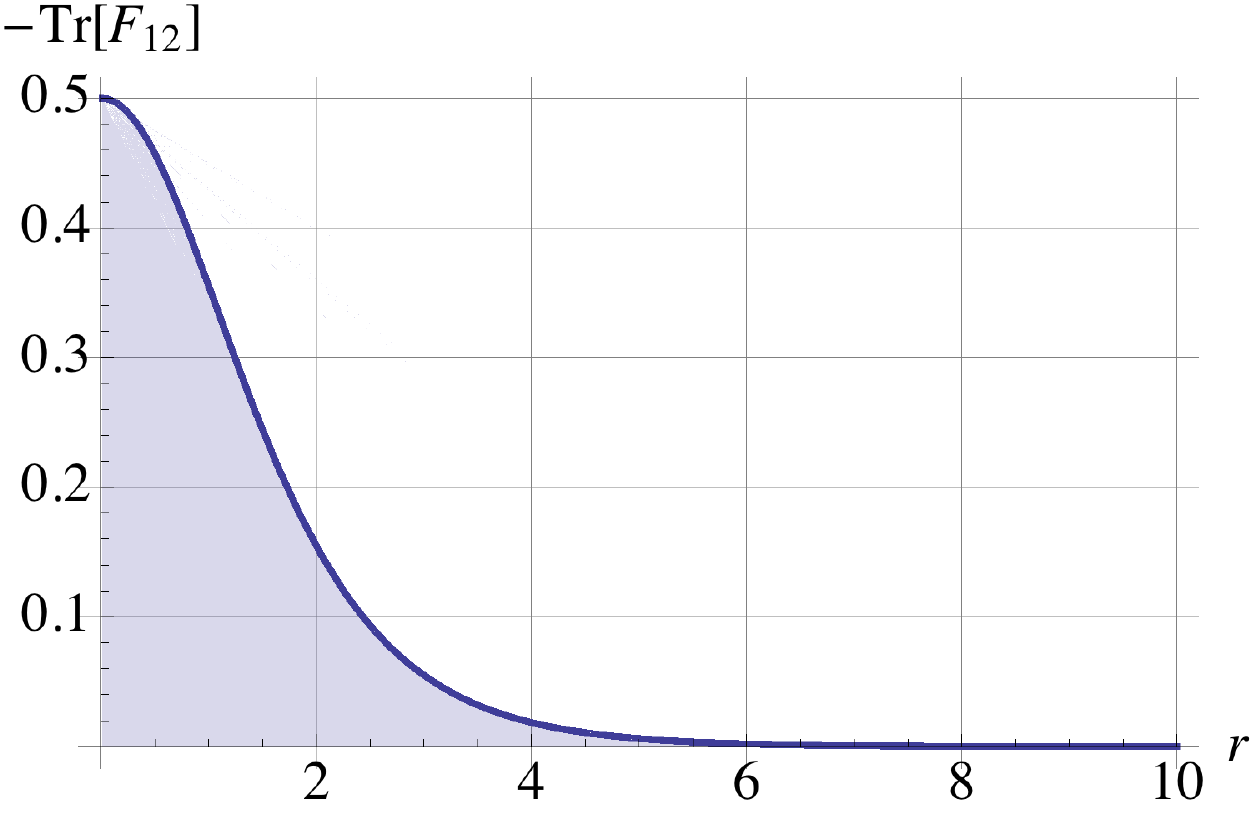}
\end{tabular}
\caption{\capfont The left figure shows numerical solution $\psi(r)$ 
for Eq.~(\ref{eq:master_ANO}) and the right figure shows the 
magnetic field $\Tr[F_{12}] = -2 \p\bar\p\psi$ for 
the single vortex string (a cross-section of the flux tube at any $z$).
We set the parameters by $gv = 1$.} \label{fig:ANO}

\end{center}
\end{figure}
One can easily see that the above matrices give us the following 
physical fields 
\beq
H_{\text{N-vor}} = v
\left(
\begin{array}{cc}
e^{-\frac{\psi}{2}} z & 0\\
0&1
\end{array}
\right),\quad
\Sigma = 
M,\quad
\bar A_{\text{N-vor}} = \left(
\begin{array}{cc}
-\frac{i}{2}\bar\p \psi & 0 \\
0 & 0 
\end{array}
\right),\quad
A_{0,3} = 0.
\eeq
Then the magnetic field 
and gauge invariant $\Omega$ are given by 
\beq
F_{12}^{\text{N-vor}} = \left(
\begin{array}{cc}
-2\p\bar\p \psi & 0 \\
0 & 0
\end{array}
\right), 
\quad 
\Omega^{\text{N-vor}} = 
\left(
\begin{array}{cc}
e^{\psi+mx^3} & 0\\
0 & e^{-mx^3}
\end{array}
\right).
\eeq
We find that the master equation (\ref{eq:master}) 
for the N-vortex (\ref{eq:N_vor}) reduces to the 
master equation for the Abelian vortex 
\beq
\frac{4}{g^2v^2}\p\bar\p\psi = 1-|z|^2 e^{-\psi} .
\label{eq:master_ANO}
\eeq
This equation has no known analytic solution, but can easily 
be solved numerically. 
The asymptotic behavior for $\psi$ is given by 
\beq
\psi \to \log |z|^2 
+ q K_0(gv|z|),
\qquad {\rm as} \quad |z| \to \infty,
\label{eq:asym}
\eeq
where $K_0$ is the modified 
Bessel function of the second kind 
and a constant $q$ can be obtained numerically. 
The function $\psi$ and magnetic field for the 
axially symmetric solution are obtained numerically and shown
in Fig.~\ref{fig:ANO}.
Now it is easy to obtain the mass of the vortex as 
\beq
M_{\rm vor} = L_3\times v^2 \int dx^1dx^2~ 2\p\bar\p \psi
= L_3 \times 2\pi v^2,
\eeq
where $L_3$ is the length of the vortex string along the $x^3$ axis. 

For later convenience, let us decompose the magnetic field $F_{12}$ 
in the $U(2)_{\rm C}=(U(1)_{\rm C0}\times SU(2)_{\rm C})/\mathbb{Z}_2$ 
gauge group into the field $F_{12}^{0}$ for the overall 
$U(1)_{\rm C0}$ and the field $F_{12}^{\Sigma}$ projected along 
the adjoint field $\Sigma$ (this is identical to the third 
component $U(1)_{\rm C3}$ of $SU(2)_{\rm C}$ in the present case) as 
\beq
F_{ij}^{0} = \Tr\left[F_{ij}\frac{{\bf 1}_2}{2}\right] ,\quad
F_{ij}^{\Sigma} =\Tr\left[F_{ij}\frac{\Sigma}{m}\right] .
\label{eq:abel_nonabel_flux}
\eeq
We will call $F_{12}^{0}$ as Abelian magnetic field and $F_{12}^{\Sigma}$ 
as non-Abelain magnetic field. 
Note that these two magnetic fields are associated with 
$(U(1)_{\rm C0}\times U(1)_{\rm C3})/\mathbb{Z}_2 \subset U(2)_{\rm C}$ (asymptotically) 
which is not broken by the adjoint scalar field 
$\Sigma = (m/2)\sigma_3$. 
A linear combination of these $U(1)$ gauge symmetries is restored 
inside vortices. Therefore they are precisely the appropriate 
magnetic fields to measure the magnetic flux flowing to infinity 
through vortices. 
 For the N-vortex, we obtain Abelian and non-Abelian magnetic 
fields as 
\beq
F_{12}^{0} = -\p\bar\p\psi,\quad
F_{12}^{\Sigma} = -\p\bar\p\psi.
\eeq
We see that only the sum $F_{12}^0+F^{\Sigma}_{12}$ 
(not the difference: $F_{12}^0-F^{\Sigma}_{12} = 0$ does not play any role) 
has nonvanishing magnetic field inside the N-vortex. 
This linear combination precisely corresponds to the restored 
$U(1)$ gauge symmetry inside the N-vortex.

Another possibility to place the zero of the Higgs field is 
at the lower-right corner of the moduli matrix as  
\beq
H_0 = \left(
\begin{array}{cc}
1 & 0\\
0 & z
\end{array}
\right),\quad S = 
\left(
\begin{array}{cc}
1 & 0\\
0 & e^{\psi/2}
\end{array}
\right)e^{Mx^3},
\label{eq:S_vor}
\eeq
where $\psi$ is the same function as the N-vortex. 
We call this the S-vortex. 
Physical fields of the S-vortex are given by  
\beq
H_{\text{S-vor}} = v
\left(
\begin{array}{cc}
1 & 0\\
0 & e^{-\frac{\psi}{2}} z
\end{array}
\right),\quad
\Sigma = 
M,\quad
\bar A_{\text{S-vor}} = \left(
\begin{array}{cc}
0 & 0 \\
0 & -\frac{i}{2}\bar\p \psi 
\end{array}
\right),\quad
A_{0,3} = 0.
\eeq
The Abelian $F_{12}^{0}$ and non-Abelian $F_{12}^{3}$ magnetic 
fields of the S-vortex is given as 
\beq
F_{12}^{0} = -\p\bar\p\psi,\quad
F_{12}^{\Sigma} = +\p\bar\p\psi.
\eeq
We see that only the difference $F_{12}^0-F^{\Sigma}_{12}$ 
has nonvanishing magnetic field inside the S-vortex. 
This linear combination is the restored 
$U(1)$ gauge symmetry inside the S-vortex. 

Note that the N-vortex and the S-vortex have the same Abelian 
magnetic fields but their non-Abelian magnetic fields are oriented 
in the opposite directions along the $x^3$ axis. 
The mass of the S-vortex equals to that of the N-vortex.

\subsection{A monopole in the Higgs phase}
\label{sec:monopole_in_higgs_phase}

Now we are ready to understand the magnetic monopole. 
We connect the N-vortex and S-vortex at 
a point on the $x^3$ axis. 
While the Abelian magnetic field 
$F_{12}^{0}$ can be smoothly connected, 
a non-trivial magnetic source is needed at the point to connect oppositely 
directed non-Abelian magnetic fields $F_{12}^{\Sigma}$. 
This source  
is nothing but a magnetic monopole.
This configuration is simply described by the moduli 
matrix \cite{Isozumi:2004vg} 
with a complex moduli parameter $\phi$ 
\beq
H_0(z) = \left(
\begin{array}{cc}
z & 0 \\
-\phi & 1
\end{array}
\right),\qquad
\phi = -e^{-mX_m-i\eta},
\eeq  
where constant real parameters $X_m$ and $\eta$ correspond to the 
position and the phase of the monopole as we see immediately. 
In order to solve the master equation, a useful Ansatz for 
$\Omega$ was proposed with $\tau, \psi_1, \psi_2$ as functions 
of $x^1, x^2, x^3$ \cite{Cipriani:2012pa} 
\beq
\Omega &=& e^{M(X_m + i\eta/m)}
\left(
\begin{array}{cc}
1 & \tau(x^1, x^2, x^3)\\
0 & 1
\end{array}
\right)
\left(
\begin{array}{cc}
e^{\psi_1(x^1, x^2, x^3)} & 0 \\
0 & e^{\psi_2(x^1, x^2, x^3)}
\end{array}
\right) \nonumber \\
&& \times 
\left(
\begin{array}{cc}
1 & 0\\
\bar\tau(x^1, x^2, x^3) & 1
\end{array}
\right)e^{M(X_m - i\eta/m)}.
\label{eq:CFsolution1}
\eeq
Even for the vortex configuration, the master equation cannot 
be solved analytically. 
The situation is the same for the monopole, and  
is even worse for 
the monopole-vortex system. 

In what follows, we study the monopole configuration in the 
parameter region 
\beq
m \ll gv.
\label{eq:rigid_body_cond}
\eeq
It was found \cite{Cipriani:2012pa} that, to the first order 
in powers of 
\beq
\epsilon \equiv  \frac{m}{gv},
\eeq
the master equation (\ref{eq:master}) is solved analytically 
with the Ansatz (\ref{eq:CFsolution1}) as 
\beq
\psi_1(x^1, x^2, x^3) &=& \psi(z,\bar{z}) - \log \left\{2\cosh m(x^3 - X_m)\right\} + {\cal O}(\epsilon^2),
\label{eq:CFsolution2}
\\
\psi_2(x^1, x^2, x^3) &=& \log \left\{2\cosh m(x^3 - X_m)\right\} + {\cal O}(\epsilon^2),
\label{eq:CFsolution3}
\\
\tau(x^1, x^2, x^3) &=& \frac{z e^{m(x^3-X_m)}}{2\cosh m(x^3-X_m)} + {\cal O}(\epsilon^2) , 
\label{eq:CFsolution4}
\eeq 
where $\psi(z,\bar{z})$ satisfies Eq. (\ref{eq:master_ANO}). 
Using the Ansatz in Eq.~(\ref{eq:CFsolution1}) with the solutions in 
Eqs.~(\ref{eq:CFsolution2})--(\ref{eq:CFsolution4}), 
 the physical fields can be cast into the 
following form after a gauge choice\footnote{
In the paper \cite{Cipriani:2012pa}, the configurations in the 
singular gauge are given.}
\beq
H_{\rm mono} &\simeq& U^\dagger(x^3)\left[v\left(
\begin{array}{cc}
ze^{-\frac{\psi(z,\bar{z})}{2}} & 0\\
0 & 1
\end{array}
\right)\right]U(x^3),
\label{eq:sol_BPS1}
\\
\bar A_{\rm mono} &\simeq& U^\dagger(x^3)
\left(
\begin{array}{cc}
- \frac{i}{2}\bar \p\psi(z,\bar{z}) & 0\\
0 & 0
\end{array}
\right) U(x^3),
\label{eq:sol_BPS2}
\eeq
with 
\beq
U(x^3) &=& \frac{1}{\sqrt{1+|\phi(x^3)|^2}}\left(
\begin{array}{cc}
1 & \bar\phi(x^3)\\
-\phi(x^3) & 1
\end{array}
\right) \in SU(2),\\
\phi(x^3) &=& - \exp\left(m(x^3-X_m)-i\eta\right).
\label{eq:U}
\eeq
The structure of this peculiar solution will be explained in 
Sec.~\ref{sec:rigid_body_approx}.
From now on we set $X_m =0$ and $\eta =0$ for simplicity. 
The third component of the gauge field 
$A_3$ and the adjoint field $\Sigma$ are given by 
\beq
A_3 &\simeq& \frac{im}{2}\sech mx^3\,
U^{\dagger}
\left(\begin{array}{cc}
0 & 1-ze^{-\frac{\psi}{2}} \\
\bar{z}e^{-\frac{\psi}{2}}-1 & 0 \\
\end{array}\right) U,
\label{eq:A_CF}\\
\Sigma &\simeq& \frac{m}{2}\sech mx^3\,
U^{\dagger}
\left(\begin{array}{cc}
-\sinh mx^3 & ze^{-\frac{\psi}{2}}\\
 \bar{z}e^{-\frac{\psi}{2}} & \sinh mx^3 \\
\end{array}\right) U.
\label{eq:sigma_CF}
\eeq 
Here $\simeq$ stands for the equality up to ${\cal O}(\epsilon^2)$ terms.
This approximate solution has no unknown functions. 
Although no analytic solution for $\psi$ is known,  
it is the solution 
of the master equation (\ref{eq:master_ANO})
for the Abelian vortex, which we have already solved numerically. 
 
\begin{figure}[ht]
\begin{center}
\includegraphics[width=16cm]{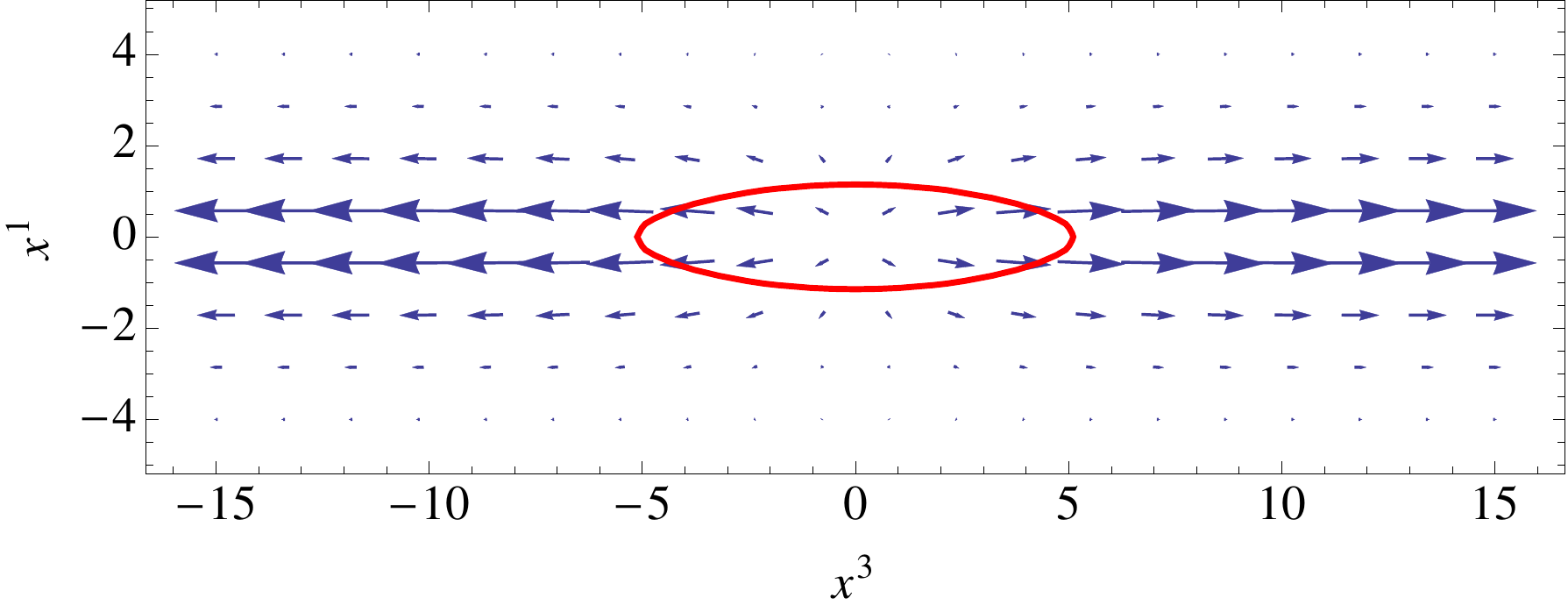}
\caption{\capfont The slender magnetic monopole in the Higgs phase. 
The parameters are set as 
$gv = 1$ and $m=1/5$. The red solid 
curve shows a contour of the 
topological charge density 
${\cal Q}_m = 0.05$. 
The vectors show the magnetic field of 
$(F^\Sigma_{12},F^\Sigma_{23})$. 
Length of the vector is proportional to norm of the magnetic 
field.}
\label{fig:monopole}
\end{center}
\end{figure}

Now we are ready to see the physical meaning of the monopole 
connecting the N-vortex and S-vortex strings. 
The asymptotic behavior of $U(x^3)$ defined in Eq.~(\ref{eq:U}) 
is given by 
\beq
U(x^3) \to \left\{
\begin{array}{ccl}
-i\sigma_2 & & \text{for}\quad x^3 \to \infty\\
{\bf 1}_2 & & \text{for}\quad x^3 \to -\infty
\end{array}
\right..
\eeq
Therefore, the asymptotic behavior of $H$ and $\bar A$ are
\beq
H_{\rm mono} &\to& \left\{
\begin{array}{ccl}
H_{\text{S-vor}} & & \text{for}\quad x^3 \to \infty\\
H_{\text{N-vor}} & & \text{for}\quad x^3 \to -\infty
\end{array}
\right.,\\
\bar A_{\rm mono} &\to& \left\{
\begin{array}{ccl}
\bar A_{\text{S-vor}} & & \text{for}\quad x^3 \to \infty\\
\bar A_{\text{N-vor}} & & \text{for}\quad x^3 \to -\infty
\end{array}
\right..
\eeq 
The $2\times 2$ matrices of the $U(2)$ magnetic field are 
computed as  
\beq
F_{12} &\simeq&
 -2\p\bar\p\psi\ U^{\dagger}
\left(\begin{array}{cc}
1 & 0 \\
0 & 0 \\
\end{array}\right) U
,\\
F_{23} &\simeq& \frac{m}{2} e^{-\frac{1}{2} \psi}  \sech m x^3\, U^{\dagger}
\left(
\begin{array}{cc}
 0 & 
 1-z \partial \psi \\
 1-\bar z \bar\partial \psi & 
 0 \\
\end{array}
\right) U,\\
F_{31} &\simeq&  \frac{i m}{2} e^{-\frac{1}{2} \psi} \sech mx^3\,
U^{\dagger} \left(
\begin{array}{cc}
 0 & 1-z\partial\psi \\
\bar z\bar\partial\psi-1 & 0 \\
\end{array}
\right) U.
\eeq 
Next, we compute the Abelian magnetic field $F_{ij}^{0}$ and 
the non-Abelian magnetic field $F_{ij}^{\Sigma}$ defined in 
Eq.~(\ref{eq:abel_nonabel_flux}) to find  
\beq
F^0_{12} \simeq - \p\bar\p\psi,\quad 
F^0_{23} \simeq 0,\quad F^0_{31} \simeq 0,
\eeq
and
\beq
F^\Sigma_{12} &\simeq& \p\bar\p\psi \tanh m x^3,\\
F^\Sigma_{23} &\simeq& 
\frac{m}{4} \sech^2 mx^3 \partial_1(r^2e^{-\psi}),\\
F^\Sigma_{31} &\simeq&
\frac{m}{4} \sech^2 mx^3 \partial_2(r^2e^{-\psi}),
\eeq
where $r = \sqrt{(x^1)^2 + (x^2)^2}$.
The Abelian (overall $U(1)$) magnetic field 
$F^0_{ij}$ is the same for the N-vortex and the S-vortex, 
whereas the non-Abelian magnetic field $F^\Sigma_{ij}$ is 
non-trivially changed from the N-vortex to the S-vortex. 
The monopole-vortex system has two scales 
\beq
L_{\rm T} = \frac{1}{gv},\quad
L_{\rm L} = \frac{1}{m}.
\eeq
As shown in Eq.~(\ref{eq:asym}), the asymptotic behavior of 
$\psi$ is given by $\psi \sim \log r^2 + q K_0(r/L_{\rm T})$. 
Because of $K_0(r/L_{\rm T}) \sim \exp(-r/L_{\rm T})$ 
for $r/L_{\rm T} \gg 1$, the scale $L_{\rm T}$ represents the 
transverse (orthogonal to the $x^3$ axis) size of the vortex 
string. 
On the other hand, $L_{\rm L}$ is another typical length scale 
within which the N-vortex is changed to the S-vortex. 
Namely, $L_{\rm L}$ is the longitudinal size of the magnetic 
monopole.
Note that our approximation (\ref{eq:rigid_body_cond})  
is valid only for 
\beq
L_{\rm L}  \gg L_{\rm T}.
\eeq
Namely, the monopole in this regime is not spherically 
symmetric but has ellipsoidal shape. 
For this reason we call this solution a 
 {\it slender} monopole. 
In Fig.~\ref{fig:monopole}, we show 
a contour of the topological charge density
\beq
{\cal Q}_m\equiv {1 \over 2g}\epsilon_{ijk}\partial_i
F_{jk}^\Sigma \label{tcd}
\eeq
in order to visualize the slender monopole. 

From Eq.~(\ref{eq:mass_mono}), the total energy of the static monopole is given by 
\beq
M_{\rm mono} 
= \frac{2m}{g^2}\int d^3x\left(\p_1 F_{23}^\Sigma + \p_2 F_{31}^\Sigma + \p_3 F_{12}^\Sigma\right)
= \frac{4\pi m}{g^2}.
\eeq
Note that the monopole contribution to the total energy is of 
order ${\cal O}(\epsilon^2)$, 
while that of the host vortex string is of order ${\cal O}(1)$. 
This can be seen as
\beq
E_{\rm tot} = 2\pi v^2 \times L_{\rm L} + \frac{4\pi m}{g^2} 
= \frac{2\pi v^2}{m}  \left(1 + 2\epsilon^2 \right),
\label{eq:background_energy}
\eeq
where the first term corresponds to the mass of the vortex 
string of length $L_{\rm L}$.
 
\begin{figure}[ht]
\begin{center}
\includegraphics[width=16cm]{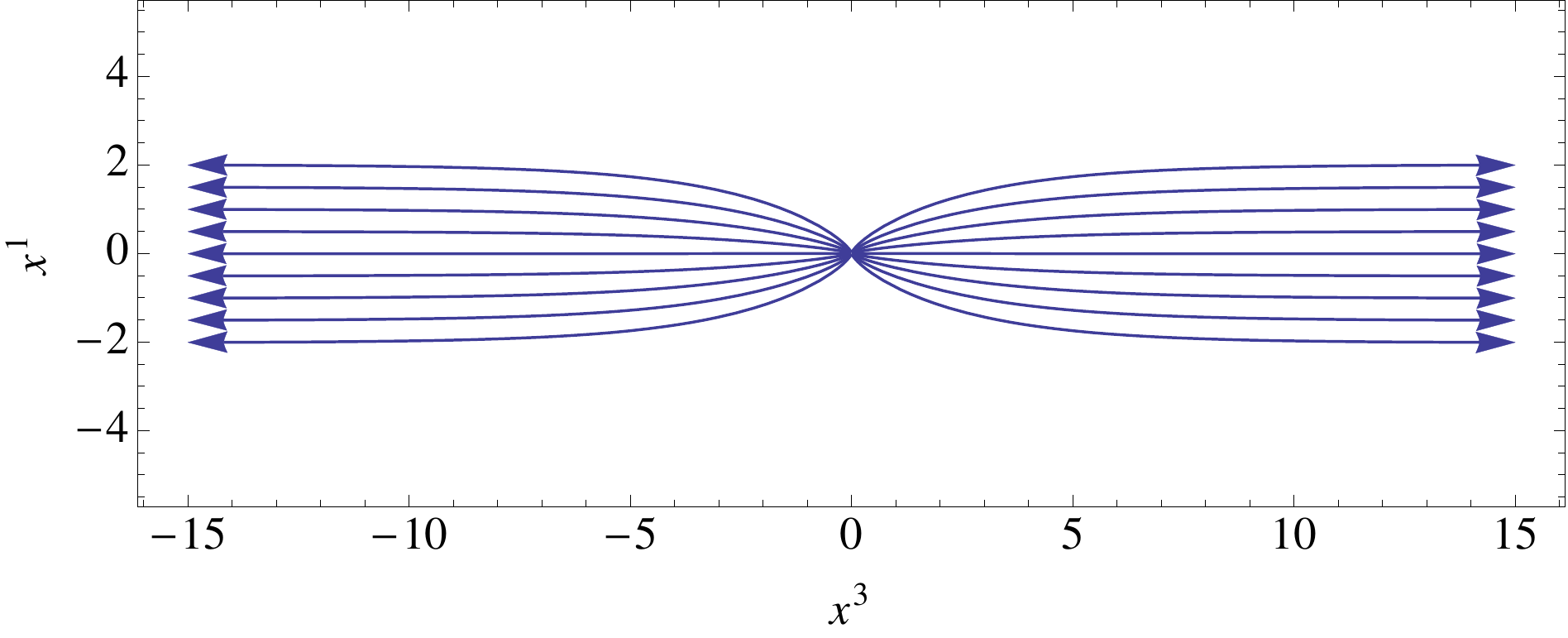}\\
\includegraphics[width=16cm]{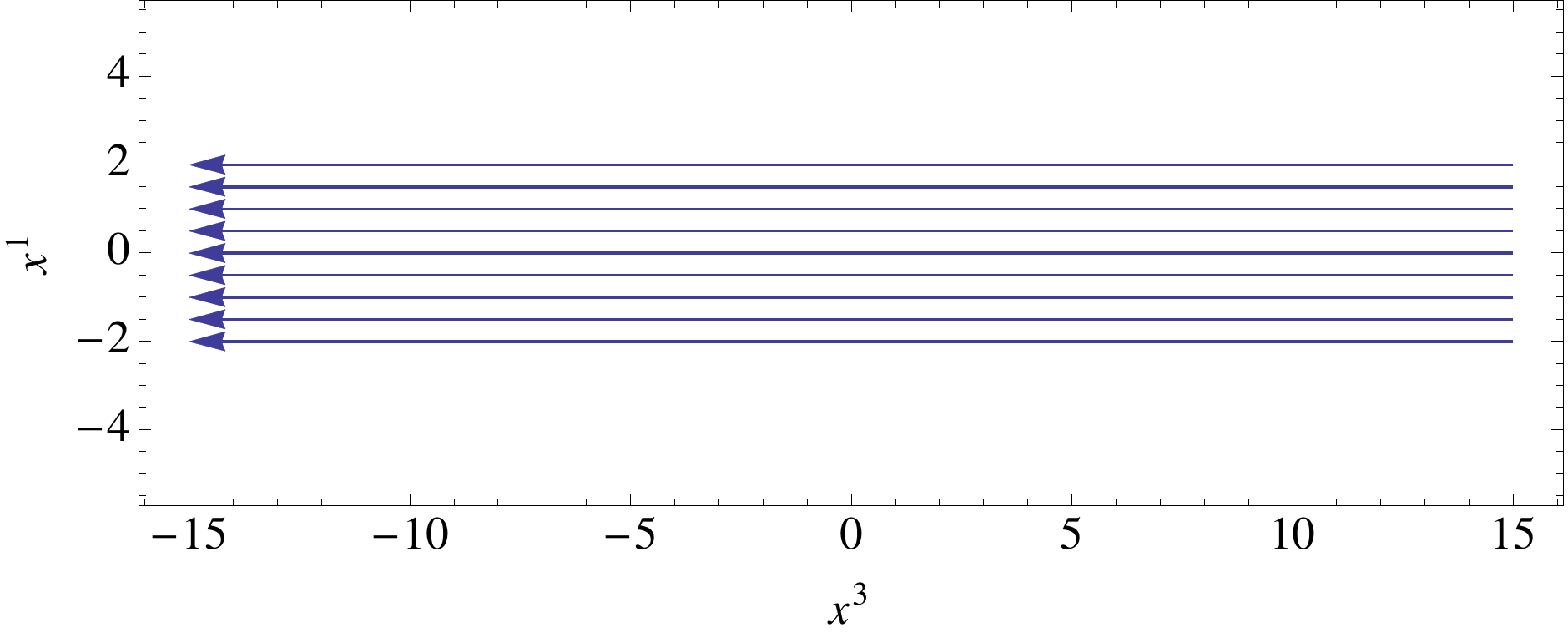}
\caption{\capfont Stream line plots of the Abelian magnetic fields of 
$(F^\Sigma_{12},F^\Sigma_{23})$ (upper) and
$(F^0_{12},F^0_{23})$ (lower). The parameters are the same as those
for Fig.~\ref{fig:monopole}. We pick up the stream lines which pass the points
$x^1= \{\pm 2,\pm1.5,\pm1,\pm0.5,0\}$ for $x^3=\pm 14$.
}
\label{fig:monopole_flux}
\end{center}
\end{figure}

One should note that all of the non-Abelian magnetic field lines of 
$(F^\Sigma_{23},F^\Sigma_{31},F^\Sigma_{12})$ start 
from a point, as illustrated in 
the upper figure of Fig.~\ref{fig:monopole_flux}. 
However, the Abelian magnetic field lines 
$(F^0_{23},F^0_{31},F^0_{12})$ form trivially straight 
lines like a flux tube as depicted  
in the lower figure of Fig.~\ref{fig:monopole_flux}.

\section{Rigid-body approximation}
\label{sec:rigid_body_approx}

\subsection{Formalism}
\label{sec:fomalism}

In this section, we consider the same configuration,  
namely monopoles in the non-Abelian superconducting phase, from 
another perspective. We use a systematic expansion up to the 
next-to-leading order within 
an approximation, which we call 
the rigid-body approximation. As a result, we can understand the 
Ansatz and the analytic solutions of the previous section 
using a more systematic approach. 
Moreover, the results can be extended to consider time-dependent 
solutions as we discuss dynamics of monopoles in the next section. 
The transverse size of the vortex string $L_{\rm T} = 1/(gv)$ 
is associated with a large mass scale $gv$, 
 and the longitudinal monopole size  $L_{\rm L} = 1/m$ is associated 
with a small mass scale. 
Therefore the condition $gv \gg m$ introduces hierarchal 
mass scales in the system: the thin vortex-string is 
generated at the high energy scale $\sim gv$, and the 
slender monopole is generated at the lower energy scale $\sim m$. 

This picture allows us to understand the slender monopole 
as a kink  
in the 1+1 dimensional theory on the vortex 
world-sheet \cite{Tong:2003pz}. Assuming 
\beq
\epsilon = \frac{m}{gv} \sim \frac{\p_\alpha}{\p_i} \ll 1,
\quad (\alpha=0, 3 \mbox{ and } i=1,2), 
\eeq
we expand the fields in power series of $\epsilon$ 
\beq
H &=& H^{(0)} +  H^{(2)} + \cdots,\label{eq:H_exp}\\
A_i &=& A_i^{(0)} +  A_i^{(2)} + \cdots, \qquad (i = 1,2),
\label{eq:A_i_exp}\\
A_{\alpha} &=&  A_\alpha^{(1)} +  A_\alpha^{(3)} + \cdots, 
\qquad (\alpha = 0,3),\label{eq:A_alpha_exp}\\
\Sigma &=&  \Sigma^{(1)} +  \Sigma^{(3)} + \cdots,
\label{eq:Sigma_exp}
\eeq
where the superscript $(n)$ indicates the $n$-th order in 
powers of $\epsilon$. 
Note that $H$ and $A_i$ start from the zeroth order because they 
are nontrivial in the background vortex-string configuration. 
On the other hand, since $A_\alpha$ and $\Sigma$ vanish 
in the background vortex-string configuration, they start from 
the first order.

We would like to solve the full equations of motion:
\begin{align}
D_0^2 H -D_3^2 H -
2(D \bar D + \bar D D )H 
+ \frac{g^2}{2}(HH^\dagger - v^2{\bf 1}_2)H  
=\ & 2\Sigma HM - HM^2-\Sigma^2H,
\label{eq:eom_H} \\
\frac{2}{g^2}D_0F_{0\bar z}-\frac{2}{g^2}D_3F_{3\bar z} 
- \frac{4}{g^2}\bar D F_{z\bar z} 
+ i (H (\bar DH)^\dagger - \bar DHH^\dagger) 
& =- \frac{2i}{g^2}\left[\Sigma,\bar D \Sigma\right], \\
\frac{2}{g^2}D_3F_{03}+\frac{4}{g^2}(D F_{0\bar z}+\bar D F_{0z} )
+ i (H(D_0H)^\dagger-D_0H H^\dagger) 
& =-\frac{2i}{g^2}\left[\Sigma,D_0\Sigma\right], \\
\frac{2}{g^2}D_0F_{03}+\frac{4}{g^2}(D F_{3\bar z}+\bar D F_{3z} )
+ i (HD_3H^\dagger-D_3H H^\dagger) 
& =-\frac{2i}{g^2}\left[\Sigma,D_3\Sigma\right], \\
-\frac{2}{g^2}(D_0^2\Sigma -D_3^2\Sigma
- 2 D\bar D \Sigma - 2 \bar D D \Sigma)
= (\Sigma H - HM)H^\dagger & + H(H^\dagger \Sigma - M H^\dagger).
\label{eq:eom_sigma}
\end{align}

\subsubsection*{Zero-th order: background vortex string for $m=0$}

Retaining only the zero-th order fields in Eqs.~(\ref{eq:H_exp}) -- 
(\ref{eq:Sigma_exp}), we find the following zero-th order 
equations 
\begin{align}
2(D\bar DH^{(0)} + \bar D D H^{(0)}) 
- \frac{g^2}{2}(H^{(0)}H^{(0)\dagger} - v^2{\bf 1}_2)H^{(0)} & =0, 
\label{eq:zeroth1}\\
- \frac{4}{g^2}\bar D F^{(0)}_{z\bar z} 
+ i (H^{(0)} (\bar D H^{(0)})^\dagger - \bar D H^{(0)}H^{(0)\dagger}) & =0.
\label{eq:zeroth2}
\end{align}
The zero-th order solutions can be compactly expressed in the 
moduli matrix formalism as 
\beq
H^{(0)} = v S^{(0)-1} H_0(z),\quad
\bar A^{(0)} = - i S^{(0)-1}\bar\p S^{(0)},
\label{eq:mm_vor}
\eeq
with the master equation for the vortex 
\beq
\frac{4}{g^2v^2}\bar\p\left(\p\Omega^{(0)}\Omega^{(0)-1}\right) 
= {\bf 1}_2 -  \Omega_0^{(0)}\Omega^{(0)-1},\quad
\Omega^{(0)} = S^{(0)}S^{(0)\dagger},\quad
\Omega_0^{(0)} = H_0^{(0)}H_0^{(0)\dagger}.
\label{eq:master_vor}
\eeq
They are identical to the vortex equations (\ref{eq:def_s})
(\ref{eq:mm_sol}), and (\ref{eq:master}) except for the additional 
condition $M=0$. 
When $M=0$, the flavor symmetry is enhanced from $U(1)_{\rm F}$ to 
$SU(2)_{\rm F}$ 
and the symmetry of the vacuum becomes $SU(2)_{\rm C+F}$. 
A single vortex spontaneously breaks this symmetry to $U(1)_{\rm C+F}$. 
Therefore, the Nambu-Goldstone zero modes $\phi$ appear as a moduli 
\beq
\mathbb{C}P^1 = \frac{SU(2)_{\rm C+F}}{U(1)_{\rm C+F}} \simeq S^2.
\eeq
By introducing the moduli parameter $\phi$ as an inhomogeneous 
coordinate of the moduli space $\mathbb{C}P^1 \simeq S^2$, 
we can express the generic moduli matrix $H_0$ with the moduli 
parameter $\phi \in \mathbb{C}$ 
as a color-flavor $SU(2)_{\rm C+F}$ rotation of the N-vortex solution 
together with an accompanying $V$-transformation as 
\beq
H_0^{(0)} &=& 
\left(
\begin{array}{cc}
z & 0\\
-\phi & 1
\end{array}
\right)
= 
V \left(
\begin{array}{cc}
z & 0\\
0 & 1
\end{array}
\right) U 
,
\label{eq:H_0_zeroth}
\\
S^{(0)} &=& 
\left(
\begin{array}{cc}
\frac{e^{\frac{\psi}{2}} + z |\phi|^2}{1+|\phi|^2} 
& \frac{(e^{\frac{\psi}{2}}-z)\bar\phi}{1+|\phi|^2}\\
- \phi & 1
\end{array}
\right)
= 
V \left(
\begin{array}{cc}
e^{\frac{\psi}{2}} & 0 \\
0 & 1
\end{array}
\right) U
,
\label{eq:S_zeroth}
\eeq
with 
\beq
U &=& \frac{1}{\sqrt{1+|\phi|^2}}\left(
\begin{array}{cc}
1 & \bar\phi\\
-\phi & 1
\end{array}
\right) \in SU(2)_{\rm C+F},
\label{eq:flavor_rot}\\
V &=& \frac{1}{\sqrt{1+|\phi|^2}}
\left(
\begin{array}{cc}
1 & - \bar\phi z\\
0 & 1 + |\phi|^2
\end{array}
\right).
\label{eq:V_rot}
\eeq
Here we need the $V$-transformation $V(z)$, in order
for $H_0^{(0)}$ to be a holomorphic function of the moduli 
parameter $\phi$. 
The single vortex solution with the generic moduli $\phi$ can 
be obtained explicitly by inserting the N-vortex solution $\psi$ 
into Eqs.~(\ref{eq:S_zeroth}) and (\ref{eq:mm_vor}).

At the zero-th order in $\epsilon\ll 1$, we obtained a 
moduli parameter $\phi$ in 
Eqs.~(\ref{eq:flavor_rot}) and (\ref{eq:V_rot}) 
as a constant. 
However, our approximation allows the weak dependence 
of $\phi$ on $x^0, x^3$ from the beginning. 
Therefore, we should consider $\phi(x^0, x^3)$ to be a 
slowly varying function of $x^0, x^3$. 
Then, the vortex background configuration depends on $x^0$ 
and $x^3$ only through the moduli field. 
\beq
H^{(0)}(x^1,x^2; \phi(x^0,x^3)),\quad
A_i^{(0)}(x^1,x^2; \phi(x^0,x^3))\qquad (i=1,2).
\label{eq:0thorder_moduli_field}
\eeq
The various fields are then induced by this slowly varying 
$\phi(x^0, x^3)$, 
and are determined by the full equations of motion in 
Eqs.~(\ref{eq:eom_H})--(\ref{eq:eom_sigma}). 
The determination of these is our task in the following. 

Here one important point on the boundary condition is in order. 
In constructing the zero-th order solution, we 
need to use the boundary condition appropriate for $m\not=0$ 
case even though the zero-th order 
equations of motion corresponds to the $m=0$ case. 
Otherwise, our power series expansion cannot work. 
Therefore the vortex moduli $\phi$ should tend to $0$ or $\infty$ 
at asymptotic region $x^3\to \pm \infty$: 
We need to choose the N-vortex or the S-vortex at 
$x^3\to\pm\infty$ as the zero-th order solution. 
As a result, the slowly varying moduli field $\phi(x^0,x^3)$ 
interpolates the N-vortex and/or the S-vortex at asymptotic 
regions $x^3\to\pm\infty$. 

\subsubsection*{First order: Gauss's law constraints}

Let us next solve the first 
order equations:
\beq
\frac{4}{g^2}(D F^{(1)}_{\alpha\bar z}
+\bar D F^{(1)}_{\alpha z} )
+ i (H^{(0)}D_{\alpha}H^{(0)\dagger}
-D_{\alpha}H^{(0)} H^{(0)\dagger}) =0,\quad (\alpha = 0,3),
\hspace*{2cm}\label{eq:eom_NLO_1}\\
\frac{4}{g^2}(D \bar D + \bar D D )\Sigma^{(1)}
-(\Sigma^{(1)} H^{(0)} - H^{(0)}M)H^{(0)\dagger} 
- H^{(0)}(H^{(0)\dagger} \Sigma^{(1)} - M H^{(0)\dagger}) = 0.
\label{eq:eom_NLO_2}
\eeq
We call these the Gauss's law constraints which determine 
$A_\alpha^{(1)}$ ($\alpha=0,3$) and $\Sigma^{(1)}$ 
for a given background vortex configuration 
in Eq.~(\ref{eq:0thorder_moduli_field}) 
with a slowly varying moduli field $\phi(x^0,x^3)$.
Moreover, the deformation of the vortex profile in $x^1, x^2$ 
plane arises from the higher order terms $H^{(2)}, H^{(4)}, \cdots$ 
by including massive modes and higher derivative corrections. 
In this paper, we consider up to the first order in $\epsilon$. 
Therefore the background vortex is given only by the zero-th 
order term $H^{(0)}$ with the slowly varying moduli field 
$\phi(x^0,x^3)$. 
Since the zero-th order fields $H^{(0)}$ and $A_{1,2}^{(0)}$ 
depend on $\phi(x^0,x^3)$
only through the flavor transformation $U \in SU(2)_{\rm F}$, 
energy density of the background vortex string does not depend 
on $x^0,x^3$. 
In other words, the vortex string is treated as a {\it rigid body}.
This is the reason why we call our approximation the rigid-body 
approximation.  

While the $x^0,x^3$ dependence of $H$ and $A_{1,2}$ fields 
comes only through the $U$ flavor rotation, 
the fields $A_{0,3}$ and $\Sigma$ depend in addition on the 
$V$-transformation matrix (\ref{eq:V_rot}) and also on the 
derivatives of both $U$ and $V$. 
This can be seen from the solution \cite{Eto:2006pg,Eto:2006uw,Eto:2012qda} 
of first order equations 
(\ref{eq:eom_NLO_1})--(\ref{eq:eom_NLO_2}) 
for an arbitrary $\phi(x^0,x^3)$ 
\beq
A_\alpha^{(1)} &=& i \left[ (\delta_\alpha S^{(0)\dagger})
S^{(0)\dagger -1} -
S^{(0)-1}\delta_\alpha^\dagger S^{(0)}\right],\quad (\alpha=0,3),
\label{eq:a0_1st}\\
\Sigma^{(1)} &=& M +
 i\left[ (\delta_\phi S^{(0)\dagger})S^{(0)\dagger-1} 
- S^{(0)-1}\delta_\phi^\dagger S^{(0)}\right],
\label{eq:sigma_1st}
\eeq
with 
\beq
\delta_\alpha = \p_\alpha\phi \frac{\delta}{\delta \phi},\quad
\delta_\alpha^\dagger = \p_\alpha\bar\phi \frac{\delta}{\delta \bar\phi},
\quad
\delta_\phi = -im \phi \frac{\delta}{\delta \phi},\quad
\delta^\dagger_\phi = im \bar\phi \frac{\delta}{\delta \bar\phi}.
\eeq

The remaining task is to look for the appropriate configurations of 
 $\phi(x^0,x^3)$ which minimize the energy of the solution. 
To this end, we plug $A^{(1)}_\alpha$ and $\Sigma^{(1)}$ into 
the original Lagrangian (\ref{eq:lag}) and pick up terms up 
to the second order 
in  $\epsilon$. 
After a tedious calculation, one obtains the following 
expression, where 
the $x^{1,2}$ and $x^{0,3}$ dependence are factorized as
\beq
{\cal L} = - v^2 F_{12}(x^1,x^2) + 
\frac{F(x^1,x^2)}{g^2} \times
\frac{|\p_\alpha\phi(x^0,x^3)|^2 - m^2 |\phi(x^0,x^3)|^2}
{(1+|\phi(x^0,x^3)|^2)^2}
+ {\cal O}(\epsilon^4),
\eeq
where we ignore unessential total derivative terms.
The prefactor in the second term depends on only $x^1$ and 
$x^2$ and it is given by
\beq
F(x^1,x^2) = 4\p\bar\p \psi(x^1,x^2).
\label{eq:def_F}
\eeq
Hence, in order to minimize the action to the second order, 
we need to find a stationary point
of 
\beq
{\cal L}^{(2)} = \frac{F(x^1,x^2)}{g^2} \times
\frac{|\p_\alpha\phi(x^0,x^3)|^2 
- m^2 |\phi(x^0,x^3)|^2}{(1+|\phi(x^0,x^3)|^2)^2}.
\label{eq:lag_2nd_order}
\eeq
Since the prefactor $F(x^1,x^2)$ is determined at the zero-th 
order, our task is basically to solve the massive non-linear 
sigma model in two dimensions with the target space $\mathbb{C}P^1$. 
Note that the process here is essentially the same as a well 
known derivation of a low energy effective action in the 
moduli approximation. To obtain the effective action, one just needs to 
integrate the Lagrangian over $x^1$ and $x^2$. 
The resulting overall coefficient is $4\pi =
\int dx^1dx^2\ F(x^1,x^2)$ and thus 
\beq
{\cal L}_{\rm eff} = \frac{4\pi}{g^2} 
\frac{|\p_\alpha\phi(x^0,x^3)|^2 
- m^2 |\phi(x^0,x^3)|^2}{(1+|\phi(x^0,x^3)|^2)^2}.
\eeq

In summary, in order to solve the equations of motion to the 
first order, we just need to solve the equations of motion 
of the effective theory, and to plug the solution $\phi(x^0,x^3)$
into $H(x^1,x^2;\phi(x^0,x^3))$ and $A_{1,2}(x^1,x^2;\phi(x^0,x^3))$.
The remaining fields $A_{0,3}^{(1)}(x^0,x^3)$ and 
$\Sigma^{(1)}(x^0,x^3)$ to the first order 
are obtained through Eqs.~(\ref{eq:a0_1st}) and (\ref{eq:sigma_1st}).

For later convenience, let us 
introduce another parametrization of $\mathbb{C}P^1$ in terms of 
polar angles $0 \le \Theta \le \pi$ and $0\le \Phi \le 2\pi$ as 
\beq
\phi = -e^{i\Phi}\tan\frac{\Theta}{2} .
\label{eq:cp1_angle}
\eeq
The effective Lagrangian is rewritten as
\beq
{\cal L}_{\rm eff} = \frac{\pi}{g^2}\left[
\p_\alpha \Theta\p^\alpha \Theta 
+ \sin^2\Theta\p_\alpha\Phi\p^\alpha \Phi
-  m^2 \sin^2\Theta\right].
\label{eq:eff_lag}
\eeq
The scalar potential $(\pi m^2/g^2) \sin^2\Theta$ is minimized 
at $\Theta = 0$ and $\Theta =\pi$. 
Clearly, these correspond to the N-vortex $(\phi=0)$ and
the S-vortex ($\phi=\infty$).

\subsection{Monopole}

\begin{figure}[ht]
\begin{center}
\includegraphics[height=6cm]{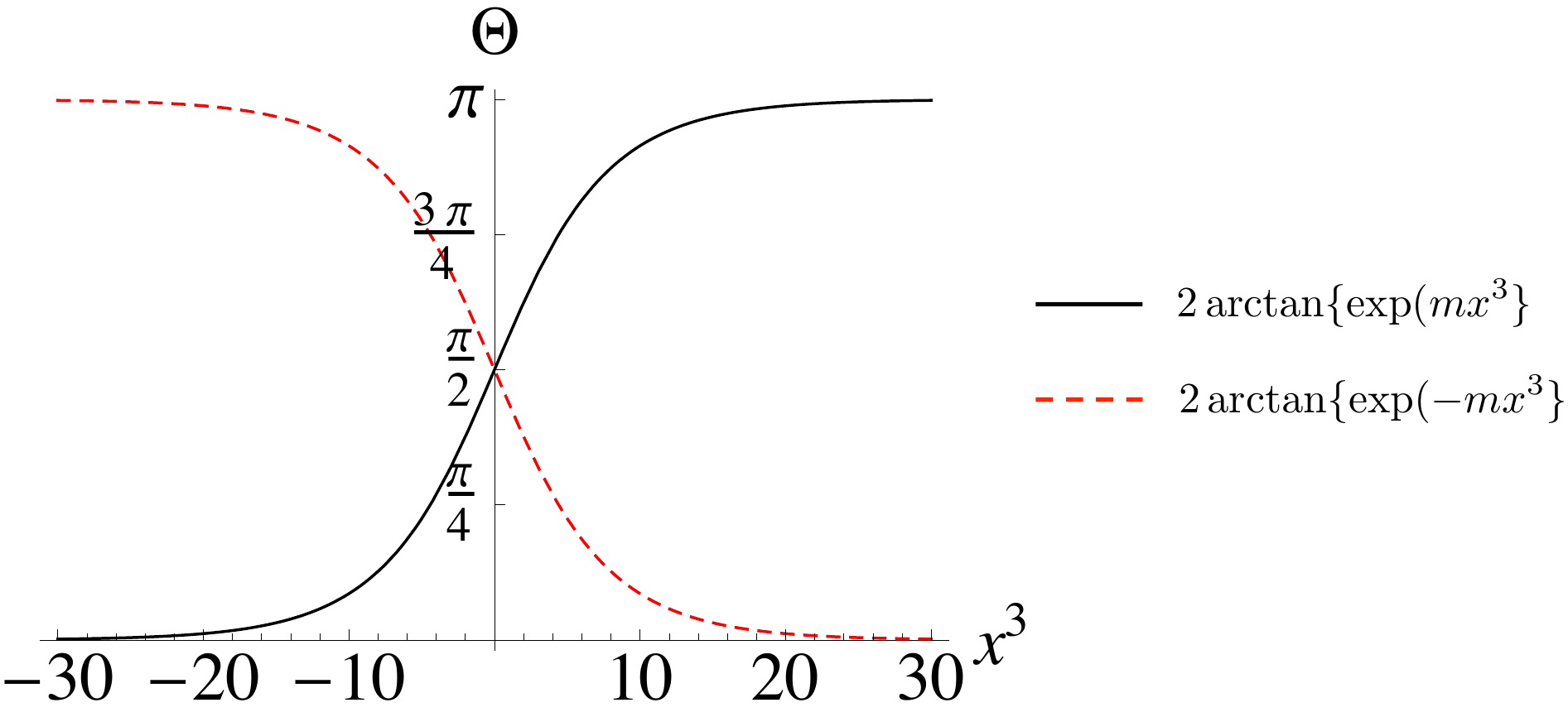}
\caption{\capfont The sine-Gordon kink (black solid line) 
and anti-kink (red dashed line) for $m = 1/5$.
}
\label{fig:sg_kinks}
\end{center}
\end{figure}

We are now ready to reconsider the slender magnetic monopole in the 
Higgs phase in our rigid-body approximation. 
Let us first look for an appropriate moduli configuration 
which minimizes  
the action in 1+3 dimensions by solving the equations of motion 
in the low energy effective theory:
\beq
&&\p_\alpha\p^\alpha \Theta +(m^2-\p_\alpha\Phi\p^\alpha \Phi) 
\sin\Theta\cos\Theta = 0,
\label{eq:eom_eff_1}\\
&&\p^\alpha\left(\sin^2\Theta\p_\alpha \Phi\right) = 0.
\label{eq:eom_eff_2}
\eeq
Eq.~(\ref{eq:eom_eff_2}) admits a constant solution for $\Phi$, 
say $\Phi=\eta$. 
Throughout this work, we focus our attention to this class 
of solutions. Then 
the equation of motion reduces to
the sine-Gordon equation. 
For static solutions it becomes 
\beq
- \Theta''  + m^2 \sin\Theta\cos\Theta = 0,
\eeq
where the prime stands for the derivative in terms of $x^3$.
The sine-Gordon model admits non-trivial topological 
excitations, 
kinks. 
The kinks interpolating $\Theta = 0$ and $\Theta = \pi$ are given by 
\beq
\Theta = 2 \arctan \exp\left(\pm m(x^3-X_m)\right).
\label{eq:sol_kink}
\eeq
The solution with the plus sign is the kink connecting $\Theta = 0$ 
at  
$x^3 \to -\infty$ and $\Theta = \pi$ 
at  
$x^3 \to + \infty$, while that with minus sign is the 
anti-kink which connects 
$\Theta = \pi$ at $x^3 \to - \infty$ and $0$ at $x^3 \to + \infty$,  
see Fig.~\ref{fig:sg_kinks}.

As is explained at the end of Sec.~\ref{sec:fomalism}, 
$\Theta = 0$ corresponds to the N-vortex and 
$\Theta = \pi$ corresponds to the S-vortex. 
Since the sine-Gordon kinks connect two different 
vortices, it should be a magnetic monopole. 
Now, we are ready to obtain the solution to
the full $1+3$ dimensional equations 
of motion in Eqs.~(\ref{eq:eom_H})--(\ref{eq:eom_sigma}) 
to the first order in $\epsilon$ according to the prescription 
given in the previous subsection.  
Firstly, the kink solution in terms of $\phi$ coordinate is 
\beq
\phi = - \exp\left(\pm m (x^3-X_m)-i\eta\right), 
\label{eq:moduli_kink}
\eeq
where $X_m$ and $\eta$ are the moduli representing the position 
and phase of the monopole.\footnote{
The moduli space of a monopole in the Higgs phase is 
$\mathbb{R}\times S^1$.
In contrast, the moduli space of the usual $SU(2)$ 
't Hooft-Polyakov monopole in the Coulomb 
phase is $\mathbb{R}^3 \times S^1$. The exponent $D$ of 
$\mathbb{R}^D$ corresponds to the spatial dimensions where 
the monopole can freely move. Including the position moduli of 
the host vortex string $\mathbb{R}^2$, the total moduli spaces 
of the monopole in the both phases coincide.
} 
From now on, we consider $X_m=\eta=0$ case for simplicity. 
Combining this with Eqs.~(\ref{eq:mm_vor}), 
(\ref{eq:H_0_zeroth}), (\ref{eq:S_zeroth}), and (\ref{eq:flavor_rot}),
we find 
\beq
H^{(0)}(x^1, x^2, \phi( x^3)) 
&=& v U^\dagger(x^3)
\left(
\begin{array}{cc}
z e^{-\frac{\psi}{2}} & 0 \\
0 & 1
\end{array}
\right)
U(x^3),\\
\bar A^{(0)}(x^1, x^2, \phi(x^3)) &=& 
U^\dagger(x^3)
\left(
\begin{array}{cc}
-\frac{i}{2}\bar\p\psi & 0 \\
0 & 0
\end{array}
\right)
U(x^3),
\label{eq:A_0thorder}
\eeq
with
\beq
U(x^3) = 
\frac{1}{\sqrt{1+|\phi(x^3)|^2}}
\left(
\begin{array}{cc}
1 & \bar\phi(x^3)\\
-\phi(x^3) & 1
\end{array}
\right).
\eeq
Comparing these with the approximate solutions  
$H=H_{\rm mono}$ and $\bar A=\bar A_{\rm mono}$ in 
Eqs.~(\ref{eq:sol_BPS1}) and (\ref{eq:sol_BPS2}) 
obtained through the Ansatz  in Eq.~(\ref{eq:CFsolution1}), 
we find that 
$\{H_{\rm mono},\bar A_{\rm mono}\}$ and $\{H^{(0)},\bar A^{(0)}\}$ 
are identical (apart from sign choices in Eq.~(\ref{eq:moduli_kink}) 
and the moduli parameters being set as $X_{\rm m}=0$ and $\eta=0$). 
Furthermore, plugging $S^{(0)}$
\beq
S^{(0)} 
= \left(
\begin{array}{cc}
\frac{e^{\frac{\psi}{2}} + z |\phi|^2}{1+|\phi|^2} 
& \frac{(e^{\frac{\psi}{2}}-z)\bar\phi}{1+|\phi|^2}\\
- \phi & 1
\end{array}
\right),\qquad \phi = -\exp(\pm mx^3),
\eeq
into the solutions $A_\alpha^{(1)}$ and $\Sigma^{(1)}$ given 
in Eqs.~(\ref{eq:a0_1st}) and
(\ref{eq:sigma_1st}),  
we obtain the induced  
fields 
\beq
A_3 &\simeq& \pm\frac{im}{2}\sech  mx^3\,
U^{\dagger}(x^3)
\left(\begin{array}{cc}
0 & 1-ze^{-\frac{\psi}{2}} \\
\bar{z}e^{-\frac{\psi}{2}}-1 & 0 \\
\end{array}\right) U(x^3),\\
\Sigma &\simeq& \frac{m}{2}\sech mx^3\,
U^{\dagger}(x^3)
\left(\begin{array}{cc}
\mp\sinh mx^3 & ze^{-\frac{\psi}{2}}\\
 \bar{z}e^{-\frac{\psi}{2}} & \pm\sinh mx^3 \\
\end{array}\right) U(x^3).
\eeq 
The configurations with the upper sign are exactly identical 
to those given in Eqs.~(\ref{eq:A_CF}) and (\ref{eq:sigma_CF}). 
In this way, we can interpret the approximate solution of the 
slender monopole given in Sec.~\ref{sec:monopole_in_higgs_phase} 
as identical to the solution in the rigid-body approximation. 
The electric fields are the same as before: 
\beq
F^0_{12} \simeq - \p\bar\p\psi,\quad F^0_{23} \simeq 0,\quad F^0_{31} \simeq 0,
\eeq
and  
the sign of the  magnetic fields depends on the sign choices of 
the moduli field in Eq.~(\ref{eq:moduli_kink}) 
\beq
B^\Sigma_3 &=& F^\Sigma_{12} \simeq\pm \p\bar\p\psi \tanh m x^3,\\
B^\Sigma_1 &=& F^\Sigma_{23} \simeq 
\pm\frac{m}{4} \p_1(r^2e^{-\psi}) \sech^2mx^3 ,\\
B^\Sigma_2 &=& F^\Sigma_{31} \simeq
\pm\frac{m}{4} \p_2 (r^2e^{-\psi}) \sech^2mx^3.
\eeq
 
\begin{figure}[ht]
\begin{center}
\includegraphics[width=16cm]{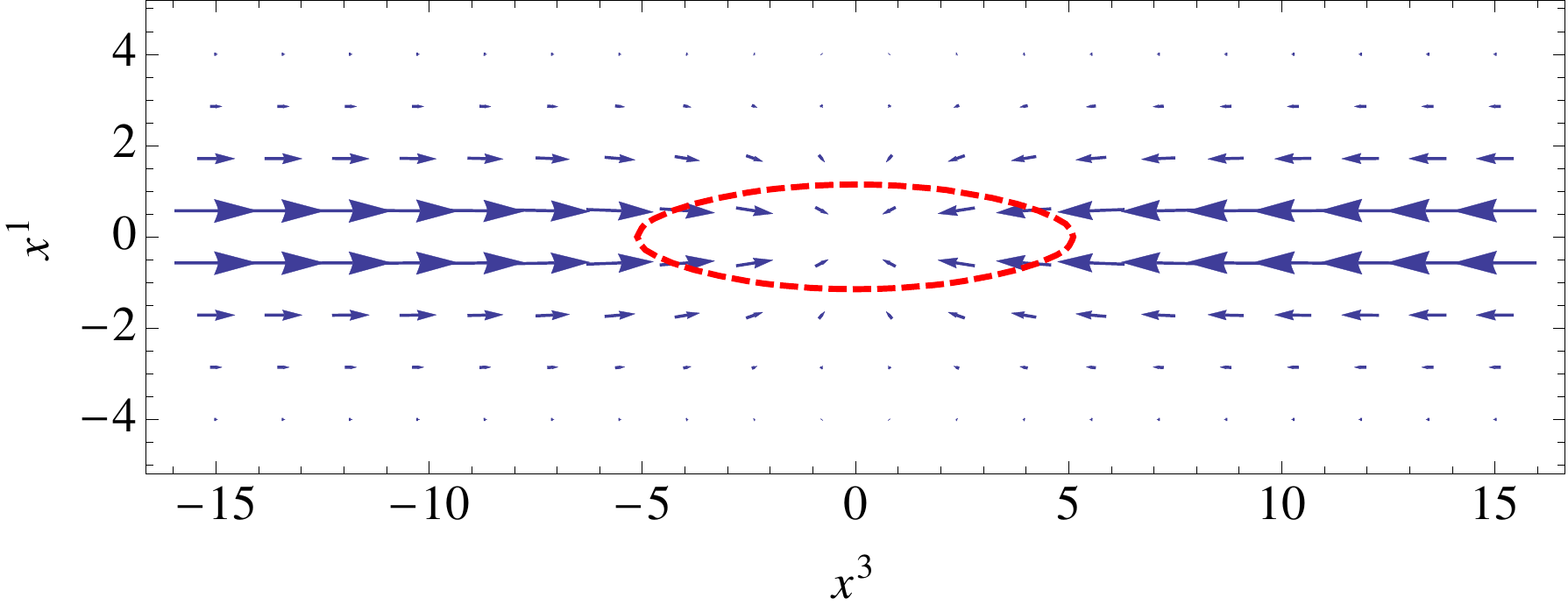}
\caption{\capfont The slender magnetic anti-monopole in the Higgs phase. 
The parameters are set as
$gv = 1$ and $m=1/5$. The red dashed 
curve shows a contour of 
the topological charge density 
${\cal Q}_m = -0.05$. 
The vectors show the magnetic field of 
$(F^\Sigma_{12},F^\Sigma_{23})$. 
Length of the vector is proportional to norm of the magnetic field.}
\label{fig:antimonopole}
\end{center}
\end{figure}

The new solution with the lower sign connects  
the N-vortex as $x^3 \to +\infty$ and the S-vortex 
as $x^3 \to -\infty$, which is opposite to the configuration 
with upper sign.
The corresponding monopole  
has the magnetic field $F^\Sigma_{ij}$ 
pointing toward monopole,  
namely it is an anti-monopole in the 
Higgs phase. 
We show the configuration in Fig.~\ref{fig:antimonopole}.

Magnetic charges of the above solutions can be easily calculated
\beq
Q_{\rm m} = \frac{1}{g} \int d^3x~ {\rm div} \vec B^\Sigma 
= \frac{1}{g} \left[
\int_{x^3\to\infty} \!\!\!\!\!\!\!\! dx^1dx^2~(\pm \p\bar\p\psi)
- \int_{x^3\to-\infty} \!\!\!\!\!\!\!\! dx^1dx^2~(\mp \p\bar\p\psi) 
\right]
= \pm \frac{2\pi}{g},
\eeq
where we used $\int dx^1dx^2~\p\bar\p\psi = \pi$ and $r^2e^{-\psi} \to 1$ as $r \to \infty$.
Here the factor $1/g$ is needed due to our notation that the gauge coupling is absorbed in the gauge field,
see Eqs.~(\ref{eq:coder1}) -- (\ref{eq:coder3}). This magnetic charge precisely coincides with one
of the 't Hooft-Polyakov monopole in the Coulomb phase \cite{Tong:2003pz}.

\section{Dynamics of slender monopoles and anti-monopoles}
\label{sec:dynamics}

\subsection{Brief summary of the rigid body approximation}

So far, we have studied the static single slender (anti-)monopole 
in the rigid-body approximation. 
Now we come to the main point of this paper,
dynamics of the slender monopoles and anti-monopoles. 
To utilize the rigid-body approximation fully, 
let us summarize the result of the approximation 
in a compact form
\beq
H & \approx & H^{(0)} = v U^\dagger(x^0,x^3)
\left(
\begin{array}{cc}
z e^{-\frac{\psi}{2}} & 0 \\
0 & 1
\end{array}
\right)
U(x^0,x^3),\\
\bar A & \approx & \bar A^{(0)} = 
U^\dagger(x^0,x^3)
\left(
\begin{array}{cc}
-\frac{i}{2}\bar\p\psi & 0 \\
0 & 0
\end{array}
\right)
U(x^0,x^3),\\
A_\alpha & \approx & A_\alpha^{(1)} 
= i \left[ (\delta_\alpha S^{(0)\dagger})S^{(0)\dagger -1} -
S^{(0)-1}\delta_\alpha^\dagger S^{(0)}\right],\\
\Sigma & \approx & \Sigma^{(1)} 
=  M + i\left[ (\delta_\phi S^{(0)\dagger})S^{(0)\dagger-1} 
- S^{(0)-1}\delta_\phi^\dagger S^{(0)}\right],
\eeq
with
\beq
U(x^0,x^3) &=& 
\frac{1}{\sqrt{1+|\phi(x^0,x^3)|^2}}
\left(
\begin{array}{cc}
1 & \bar\phi(x^0,x^3)\\
-\phi(x^0,x^3) & 1
\end{array}
\right),\\
S^{(0)} &=& \left(
\begin{array}{cc}
\dfrac{e^{\frac{\psi }{2}} + z |\phi(x^0,x^3) |^2}
{1+|\phi(x^0,x^3) |^2} 
& \dfrac{(e^{\frac{\psi }{2}}-z)\bar\phi(x^0,x^3) }
{1+|\phi(x^0,x^3) |^2}\\
- \phi(x^0,x^3)  & 1
\end{array}
\right).
\eeq
The moduli field $\phi(x^0,x^3)$ 
\beq
\phi(x^0,x^3) = -e^{i\Phi(x^0,x^3)}\tan\frac{\Theta(x^0,x^3)}{2},
\eeq
should be a solution of the equations of motion
\beq
&&\p_\alpha\p^\alpha \Theta +(m^2-\p_\alpha\Phi\p^\alpha \Phi) \sin\Theta\cos\Theta = 0,\\
&&\p^\alpha\left(\sin^2\Theta\p_\alpha \Phi\right) = 0.
\eeq

\subsection{A dictionary: mapping onto the sine-Gordon model}

In the following, we fully make use of the similarity between 
our system and the sine-Gordon model. 
Let us denote another choice of the range of angles as 
\beq
\tilde\Theta \in \mathbb{R}\ (\text{mod}\ 2\pi),\qquad
\tilde\Phi \in [0,\pi) , 
\eeq
to parametrize the $\mathbb{C}P^1$ moduli $\phi$ 
\beq
\phi(x^0,x^3) = -e^{i\tilde\Phi(x^0,x^3)}\tan\frac{\tilde\Theta(x^0,x^3)}{2}.
\eeq
The equations of motion for $\tilde \Theta,\tilde \Phi$ are 
the  
same as those for $\Theta,\Phi$. 
Therefore, $\tilde \Phi = {\rm const.}$ is a solution, to which 
we restrict ourselves in the following. 
Without loss of generality, the value of the constant $\Phi$ 
can be chosen as
\beq
\tilde \Phi =0.
\eeq
Then the equation of motion for $\tilde \Theta$ 
is reduced to 
\beq
\p_\alpha\p^\alpha \tilde \Theta 
+ m^2 \sin\tilde\Theta\cos\tilde\Theta = 0,\qquad
\tilde\Theta \in \mathbb{R}\ (\text{mod}\ 2\pi).
\label{eq:sG_eq}
\eeq
This is nothing but the sine-Gordon equation with a periodicity 
$\pi$ in 1+1 dimensions. 

Now we can compute all field configurations in $1+3$ 
dimensions with the help of 
the sine-Gordon field $\tilde \Theta$ 
\beq
F^0_{12} = -\p\bar\p\psi,\quad
F^0_{23} = F^0_{31} = F^0_{01} = F^0_{02} = F^0_{03} = 0,
\eeq
and
\beq
F^\Sigma_{12} &=& -\p\bar\p\psi \cos \tilde \Theta,\\
F^\Sigma_{23} &=& 
\frac{1}{4} \p_1(r^2e^{-\psi})
\p_3\tilde\Theta \sin \tilde \Theta,\\
F^\Sigma_{31}&=& 
\frac{1}{4} \p_2(r^2e^{-\psi})
\p_3\tilde\Theta \sin \tilde \Theta,\\
F^\Sigma_{01} &=& 
\frac{1}{4} \p_2(r^2e^{-\psi})
\p_0\tilde\Theta \sin \tilde \Theta,\\ 
F^\Sigma_{02} &=& -
\frac{1}{4} \p_1(r^2e^{-\psi})
\p_0\tilde\Theta \sin \tilde \Theta,\\
F^\Sigma_{03} &=& 0.
\eeq
Here we define Abelian and non-Abelian electric fields in the
same sprit as in Eq.~(\ref{eq:abel_nonabel_flux})
\beq
F_{0i}^0 = \Tr\left[F_{0i}\frac{{\bf 1}_2}{2}\right],\quad
F_{0i}^\Sigma = \Tr\left[F_{0i}\frac{\Sigma}{m}\right].
\eeq
Note that the electric field and magnetic field are orthogonal 
\beq
\epsilon^{ijk}F^\Sigma_{ij}F^\Sigma_{0k} = 0.
\eeq
Therefore, there is no energy dissipation.
This is the $1+3$-dimensional manifestation 
of the well-known special feature of the sine-Gordon dynamics: 
There is no dissipation of energy in multi-kink and anti-kink dynamics 
because of the infinite number of conservation law associated 
with the integrable sine-Gordon theory in $1+1$ dimensions.

The Hamiltonian density is decomposed into two parts: 
the energy density of the rigid vortex-string ${\cal H}_{\rm vortex}^{(0)}$ 
and that of the dressed monopole ${\cal H}_{\rm dress}^{(2)}$ 
\beq
{\cal H} = {\cal H}_{\rm vortex}^{(0)} 
+ {\cal H}_{\rm dress}^{(2)} + {\cal O}(\epsilon^4).
\eeq
Note that the monopole contribution in our regime  is of order ${\cal O}(\epsilon^2)$ as explained in Eq.~(\ref{eq:background_energy}).
The rigid vortex-string Hamiltonian density does not depend 
on $x^0$ and $x^3$ 
\beq
{\cal H}_{\rm vortex}^{(0)} &=&
\Tr\left[\frac{1}{g^2}(F_{12}^{(0)})^2 + (D_iH)^{(0)}(D_iH)^{(0)\dagger} 
+ \frac{g^2}{4}\left(H^{(0)}H^{(0)\dagger} - v^2{\bf 1}_2\right)^2
\right]
\non
&=& 
\Tr\left[
\frac{1}{g^2}\left(F_{12}^{(0)} 
- \frac{g^2}{2}\left(H^{(0)}H^{(0)\dagger} - v^2{\bf 1}_2\right)\right)^2
+ 4(\bar D H)^{(0)} (\bar DH)^{(0)\dagger}
\right.\non
&& \left. - v^2 F_{12}^{(0)} + i \left\{\p_1(H^{(0)}(D_2H)^{(0)\dagger})
-\p_2(H^{(0)}(D_1H)^{(0)\dagger})
\right\}\right]\non
&=& 2v^2 {\cal V}.
\label{eq:vortex_energy_density}
\eeq
where we used the master equation (\ref{eq:master_ANO}) to reach the last expression and
defined a vortex energy density
\beq
{\cal V}(x^1,x^2) = 
\p\bar\p\psi - \frac{4}{g^2v^2} (\p\bar\p)^2\psi
\eeq
The first term in the 
right-hand side
is the topological term 
while the second one is a total derivative which does not 
contribute to the 
total vortex energy. 
In deriving Eq.~(\ref{eq:vortex_energy_density}), 
we have used the 
same identities in Eqs. (\ref{id1})--(\ref{id4}).
The dressed Hamiltonian density which depends on $x^0$ and 
$x^3$ is given by 
\beq
{\cal H}_{\rm dress}^{(2)} &=&
\Tr\bigg[\frac{1}{g^2}\left\{
(F_{23}^{(1)})^2+(F_{31}^{(1)})^2
+(F_{01}^{(1)})^2+(F_{02}^{(1)})^2+(D_1\Sigma^{(1)})^2
+(D_2\Sigma^{(1)})^2
\right\}  \non
&&
+ D_0H^{(0)}(D_0H^{(0)})^\dagger+ D_3H^{(0)}(D_3H^{(0)})^\dagger
+ (\Sigma^{(1)}H^{(0)}-H^{(0)}M)(\Sigma^{(1)}H^{(0)}-H^{(0)}M)^\dagger
\bigg] \non
&=& \frac{{\cal V}}{g^2}
\left((\p_0\tilde\Theta)^2+ (\p_3\tilde\Theta)^2+m^2 \sin ^2\tilde\Theta\right),
\eeq
where we used the master equation (\ref{eq:master_ANO}).

Let us also add the topological charge density (\ref{tcd}) 
in our dictionary
\beq
{\cal Q}_{\rm m} 
= \frac{{\cal V}}{g}\p_3\tilde\Theta \sin\tilde\Theta.
\label{eq:mag_ene_dens}
\eeq
From this expression, one can easily compute the magnetic charge as
\beq
Q_{\rm m} = \int d^3x~{\cal Q}_{\rm m} 
= \frac{1}{g} \int dx^1dx^2~
{\cal V}~~\int dx^3~\p_3\tilde \Theta \sin\tilde\Theta 
= \frac{\pi}{g}
\left[-\cos\tilde\Theta\right]^{x^3=+\infty}_{x^3=-\infty}.
\eeq
Here we used $\int dx^1dx^2 \p\bar\p \psi = \pi$.
As a check, one can compute the energy of the magnetic 
monopoles for the solutions given in Eq.~(\ref{eq:sol_kink})
\beq
Q_{\rm m} = \pm \frac{2\pi}{g}.
\eeq
Similarly, one may introduce a electric charge density by
\beq
{\cal Q}_{\rm e} = \frac{1}{g}\p^iF_{0i}^\Sigma.
\label{eq:EC_m_am}
\eeq
But this is identically zero  for any $\tilde \Theta(x^0,x^3)$.
This matches with a naive intuition that the fixed azimuthal angle $\tilde \Phi$ does not generate
any electric charges. Note, however, that this does not mean the electric fields themselves are zeros.
One can easily find that rotation of $\vec E^\Sigma = (F_{10}^\Sigma,F_{20}^\Sigma,F_{30}^\Sigma)$ are non-zero.
\beq
(\vec \nabla \times \vec E^\Sigma)_3 = 
- \frac{4}{g^2v^2}(\p\bar\p)^2\psi~\p_0\tilde\Theta\sin\tilde\Theta.
\eeq
The other components are of higher order, so we ignore them.

\subsection{Two different species of slender monopoles}

\begin{figure}[ht]
\begin{center}
\includegraphics[height=5.5cm]{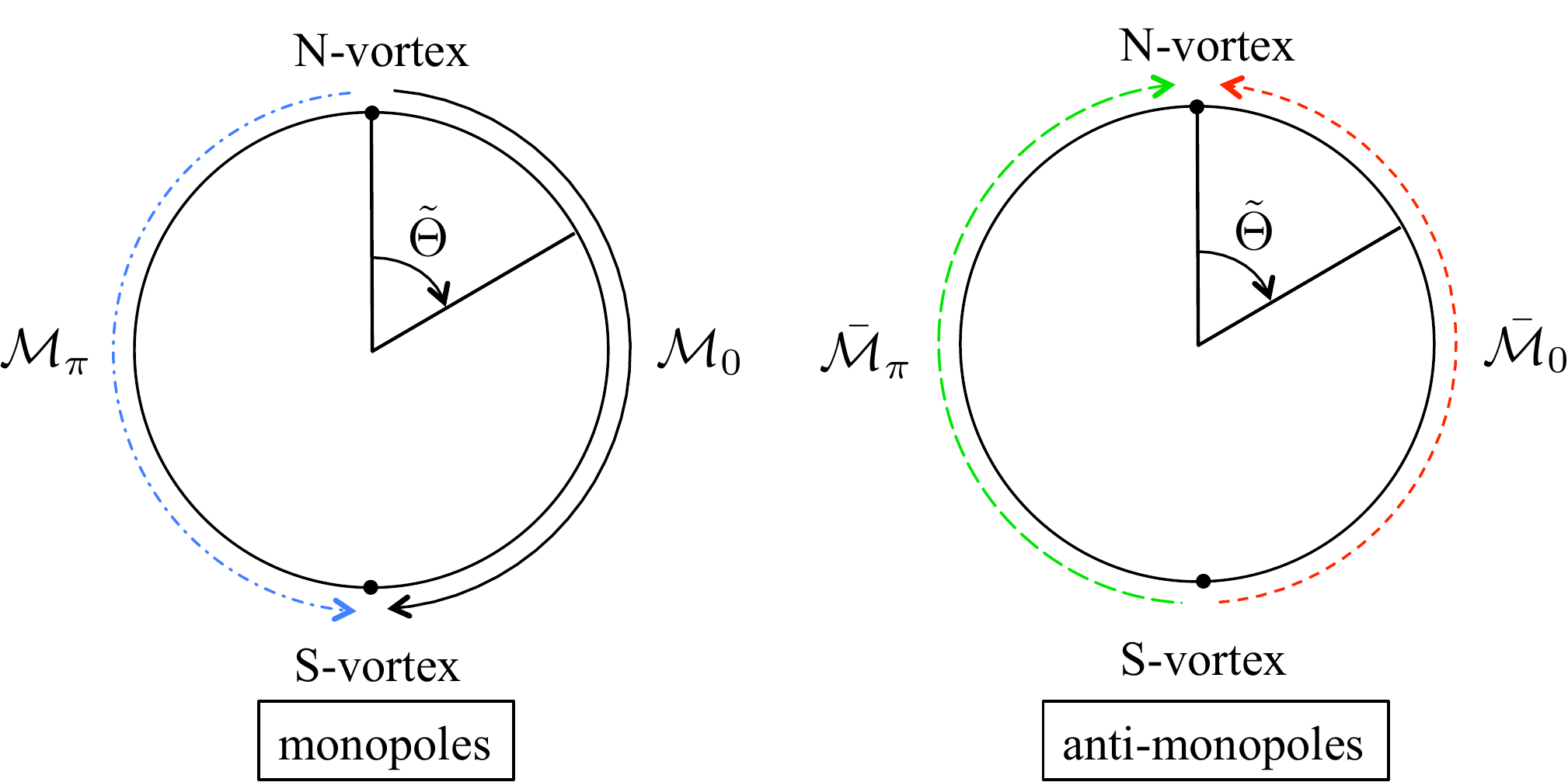}\\\ \\
\includegraphics[height=5.5cm]{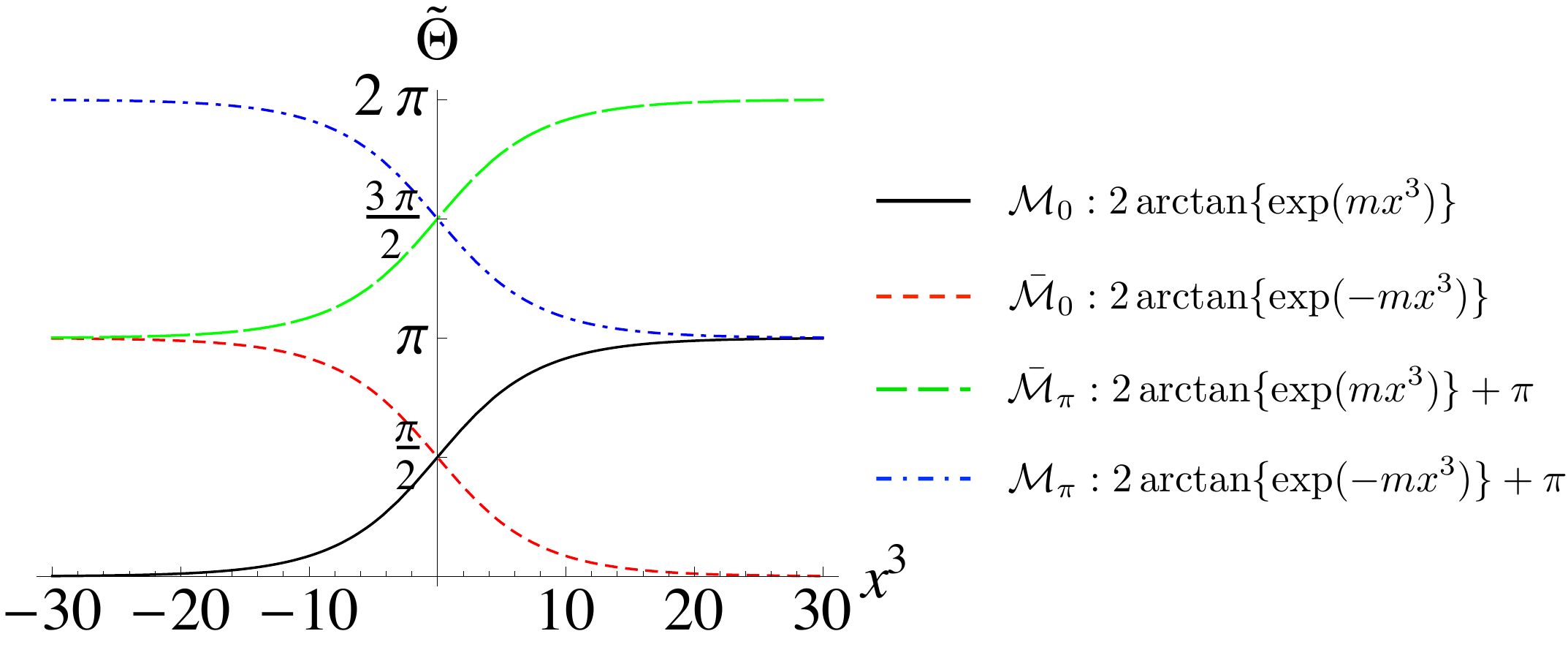}
\caption{\capfont The slender monopole ${\cal M}_0$ 
(black solid line) corresponds to the sine-Gordon kink 
($0$ to $\pi$), and another kind of the slender monopole 
${\cal M}_\pi$ (blue dash-dotted line) to the sine-Gordon 
anti-kink ($2\pi$ to $\pi$). 
The slender anti-monopole $\bar{\cal M}_0$ 
(red dotted line) corresponds to the sine-Gordon anti-kink 
($\pi$ to $0$), and another kind of slender 
anti-monopole $\bar{\cal M}_\pi$ (green dashed line) 
to the sine-Gordon kink ($\pi$ to $2\pi$). 
}
\label{fig:rel_kink_mono}
\end{center}
\end{figure}

The zenith angle $\tilde \Theta$ takes values between $0$ 
and $2\pi$ ($=\mathbb{R}\ {\rm mod}~2\pi$), 
and 
the sine-Gordon equation (\ref{eq:sG_eq}) is periodic 
with  
a period $\pi$. 
Therefore, there exist two
sine-Gordon kinks: the one interpolates from $0$ to $\pi$ as 
$x^3 =-\infty \to +\infty$, 
and the other interpolates from $\pi$ to $2\pi$ as $x^3 =-\infty \to +\infty$.
Here we need to pay some attention to our terminology 
in translating 
the sine-Gordon kinks into monopoles in $1+3$ dimensions.
Although these two configurations 
are both to be called kinks in the 
sense of the sine-Gordon model, the former connects the N-vortex 
and S-vortex from left to right, while 
the latter connects them from right to left. 
Namely, the former kink is the  monopole (denoted as
${\cal M}_0$) and 
the latter kink is the  anti-monopole ($\bar{\cal M}_\pi$). 
Similarly, the anti-kink interpolating from $\pi$ to $0$ as 
$x^3 = -\infty \to +\infty$ is the anti-monopole ($\bar{\cal M}_0$), 
while the other anti-kink interpolating from $2\pi$ to $\pi$ 
as $x^3 = -\infty \to +\infty$ 
is the monopole (${\cal M}_\pi$).
Correspondence between the sine-Gordon (anti-)kinks and the 
slender (anti-)monopoles 
are depicted in Fig.~\ref{fig:rel_kink_mono}.
The configurations are given by
\beq
{\cal M}_0&:&\ \tilde \Theta = 2\arctan\exp (mx^3) + 2n\pi,\\
\bar{\cal M}_0&:&\ \tilde \Theta = 2\arctan\exp (-mx^3) + 2n\pi,\\
{\cal M}_\pi&:&\ \tilde \Theta = 2\arctan\exp (-mx^3)+(2n+1)\pi,\\
\bar{\cal M}_\pi&:&\ \tilde \Theta = 2\arctan\exp (mx^3)+(2n+1)\pi,
\eeq
with $n$ being an integer.

An advantage of mapping the slender monopoles in the 
non-Abelian superconductor onto the sine-Gordon system 
is that the interactions between the slender monopoles and 
anti-monopoles can be easily found. 
For instance, forces between the sine-Gordon 
kink and anti-kink have been obtained \cite{Perring:1962vs,Manton:2004tk} assuming large separation $R$.
We just need to translate them to the interactions of slender 
monopoles in  
the non-Abelian gauge theory. 
The slender monopole and anti-monopole of the same kind 
(${\cal M}_0$ and $\bar{\cal M}_0$, or ${\cal M}_\pi$ and 
$\bar{\cal M}_\pi$) at  
large separations $R$ exert an attractive force each other  
\beq
{\cal F}({\cal M}_0:\bar{\cal M}_0) = {\cal F}({\cal M}_\pi:\bar{\cal M}_\pi) 
= - 4 m^2 \exp(-mR).
\label{eq:attractive_force}
\eeq
On the other hand, the interaction between the monopole and 
anti-monopole of the different kind 
(${\cal M}_0$ and $\bar{\cal M}_\pi$ 
or ${\cal M}_\pi$ and $\bar{\cal M}_0$) 
at large separations $R$ is a  
repulsion 
\beq
{\cal F}({\cal M}_0:\bar{\cal M}_\pi) = {\cal F}({\cal M}_\pi:\bar{\cal M}_0) 
= + 4 m^2 \exp(-mR).
\label{eq:repulsive_force}
\eeq 
This is because both of ${\cal M}_0$ and $\bar{\cal M}_\pi$ 
(${\cal M}_\pi$ and $\bar{\cal M}_0$) 
are kinks (anti-kinks) with half windings 
($\Delta \Theta=\pi$) as solitons in the sine-Gordon system. 
The  
exponential force is in sharp contrast to the Coulomb 
force between monopoles in the unbroken phase. 
Since the gauge fields become massive in the superconducting phase,  
the interaction between monopoles becomes the Yukawa type 
which decreases exponentially with the Compton wave length $1/m$ 
of the massive particle. 

\subsection{Magnetic meson}

It is well-known that the sine-Gordon model admits a bound 
state of kink and anti-kink,  
 the so-called  
breather solution. 
In our case, 
it is nothing but 
a  
bound state of the slender monopole and anti-monopole, which we call 
the magnetic meson. 
The configuration is given by 
\beq
\tilde \Theta(x^0,x^3) = 2\arctan\left(\frac{\eta\sin\omega x^0}
{\cosh\eta\omega x^3}\right),\quad
\eta = \sqrt{\frac{m^2}{\omega^2}-1},\quad \omega < m,
\eeq
where $\omega$ is  
the  
frequency and $(\eta\omega)^{-1} = 1/\sqrt{m^2-\omega^2}$ is  
the  
typical size 
of the magnetic meson.
The mass of the meson depends on $\omega$ as 
\beq
M_{\rm meson} = 2 M_{\rm mono} \times \sqrt{1-\frac{\omega^2}{m^2}} 
< 2 M_{\rm mono}.
\eeq
The mass of the mesonic bound state is smaller than 
the sum  
of the masses of isolated 
monopole and anti-monopole.

\begin{figure}
\begin{center}
    \begin{tabular}{cc}
      \begin{minipage}{9cm}
        \begin{center}
          \includegraphics[clip, height=1.2cm]{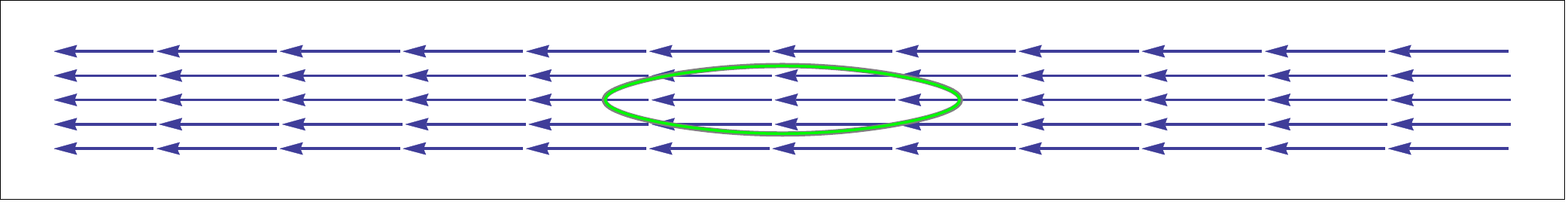}\\
          \includegraphics[clip, height=1.2cm]{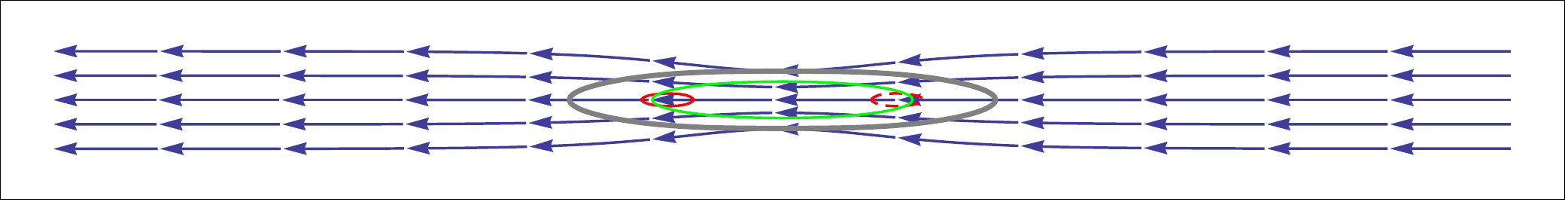}\\
          \includegraphics[clip, height=1.2cm]{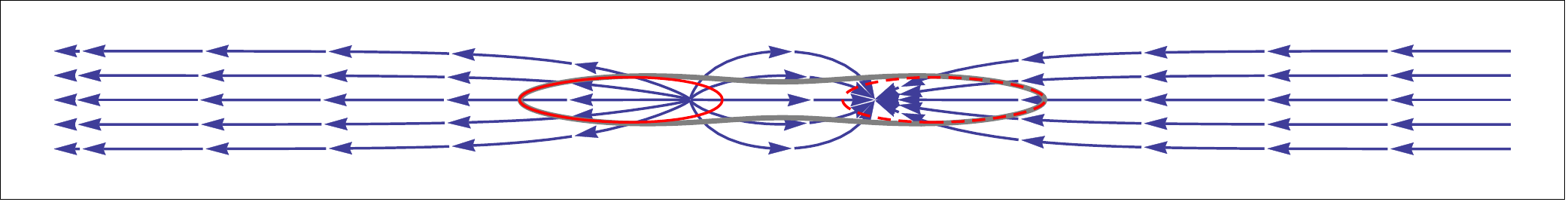}\\
          \includegraphics[clip, height=1.2cm]{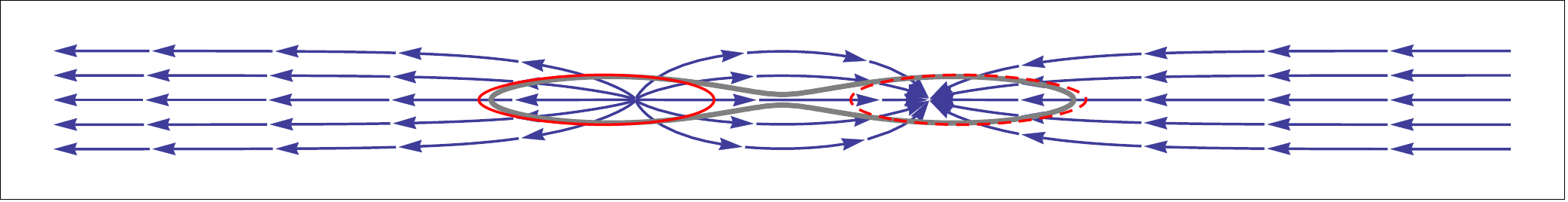}\\
          \includegraphics[clip, height=1.2cm]{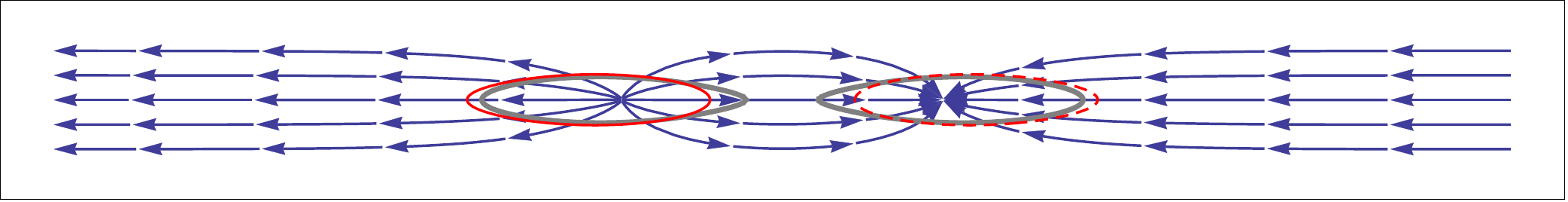}\\
          \includegraphics[clip, height=1.2cm]{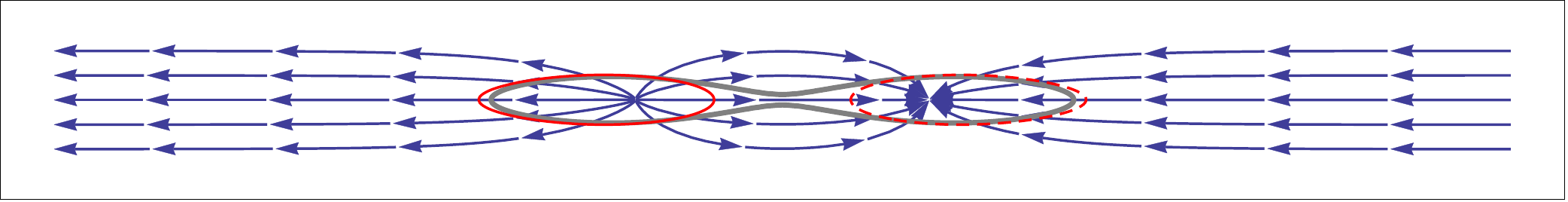}\\
          \includegraphics[clip, height=1.2cm]{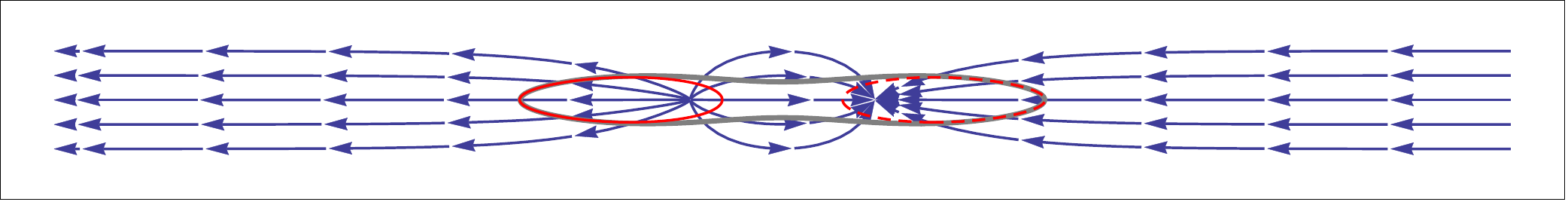}\\
          \includegraphics[clip, height=1.2cm]{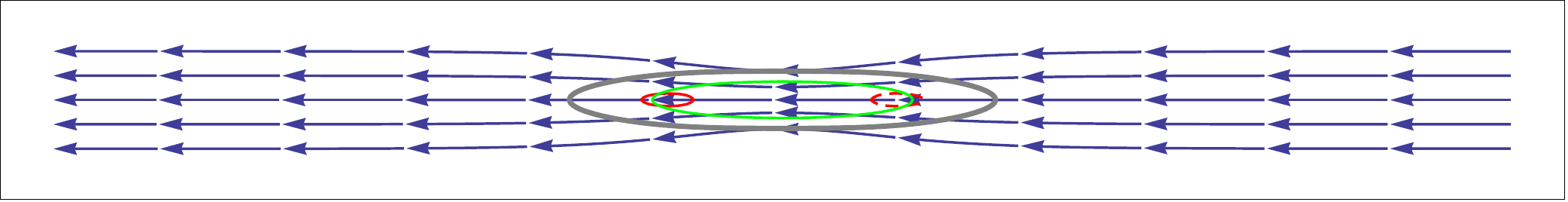}\\
          \includegraphics[clip, height=1.2cm]{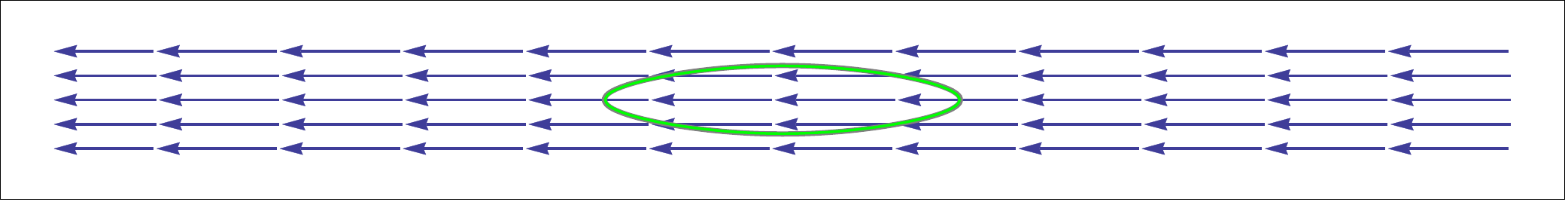}\\
          \includegraphics[clip, height=1.2cm]{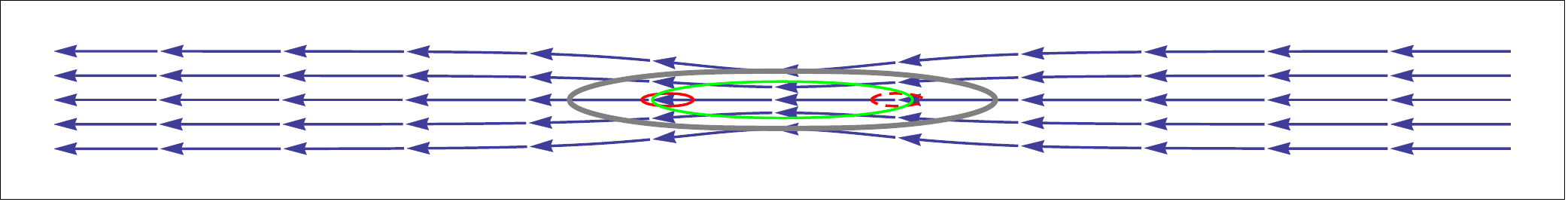}\\
          \includegraphics[clip, height=1.2cm]{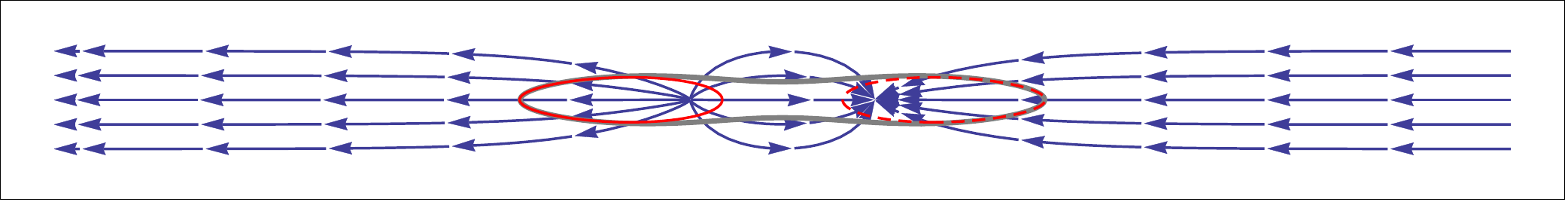}\\
          \includegraphics[clip, height=1.2cm]{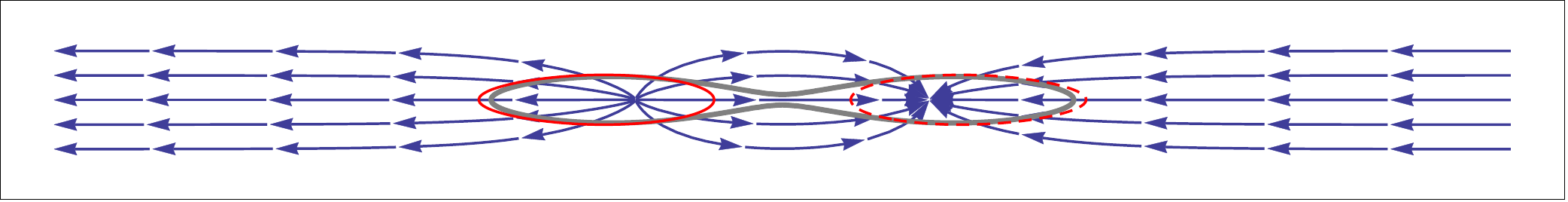}\\
          \includegraphics[clip, height=1.2cm]{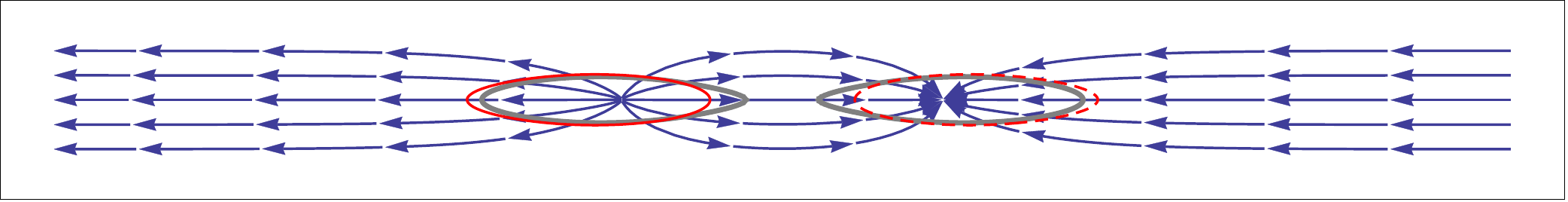}\\
          \includegraphics[clip, height=1.2cm]{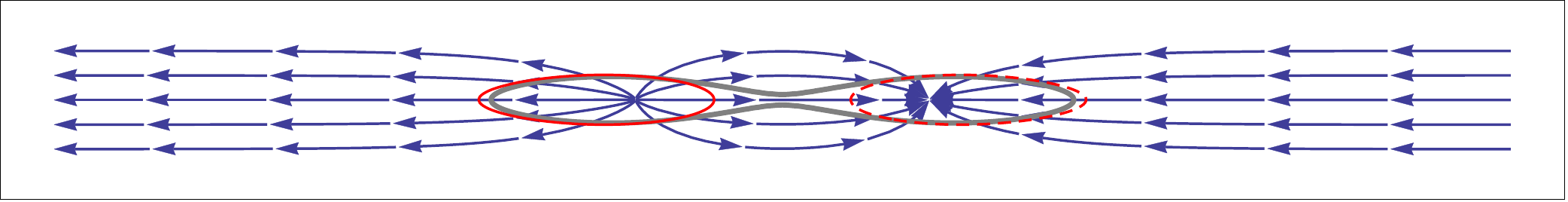}\\
          \includegraphics[clip, height=1.2cm]{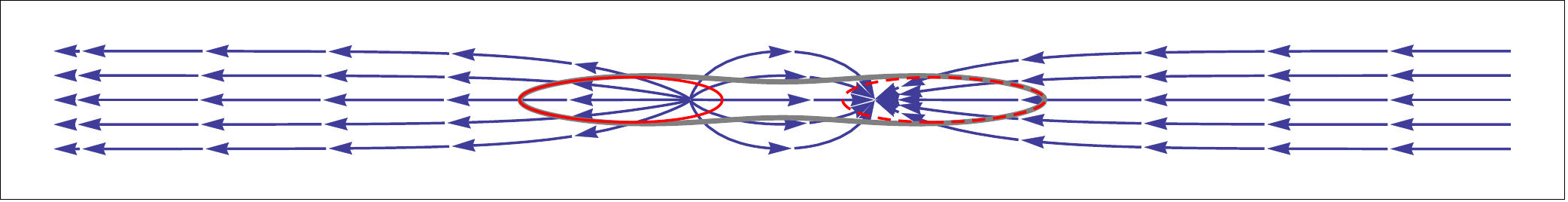}\\
          \includegraphics[clip, height=1.2cm]{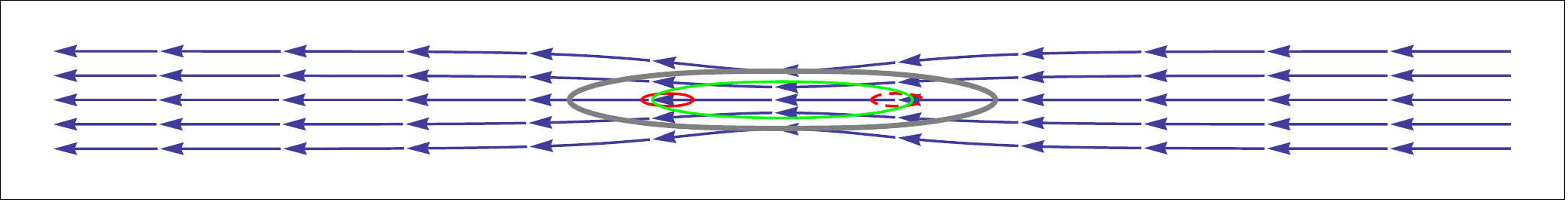}
         \end{center}
      \end{minipage}
      &
      \begin{minipage}{7cm}
        \begin{center}
          \includegraphics[clip, height=1.2cm]{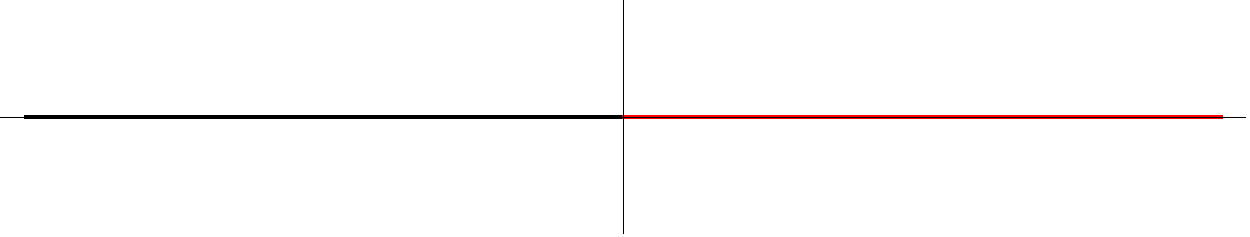}\\
          \includegraphics[clip, height=1.2cm]{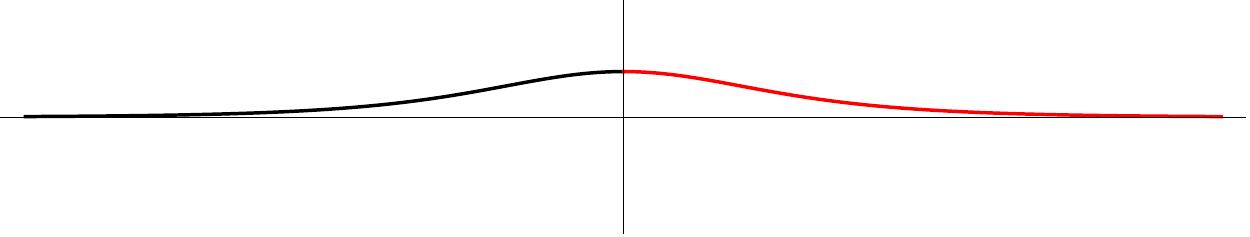}\\
          \includegraphics[clip, height=1.2cm]{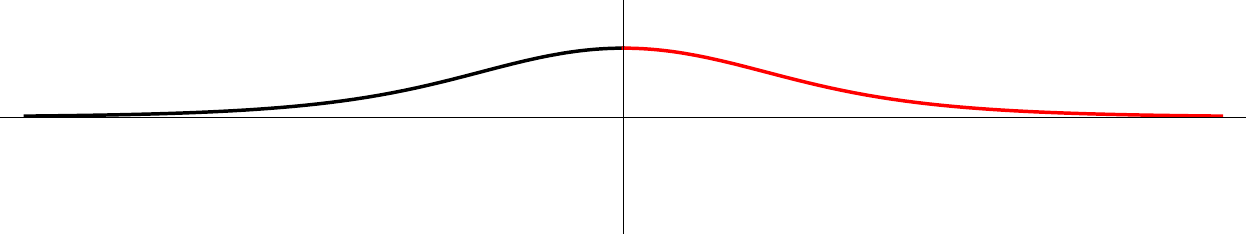}\\
          \includegraphics[clip, height=1.2cm]{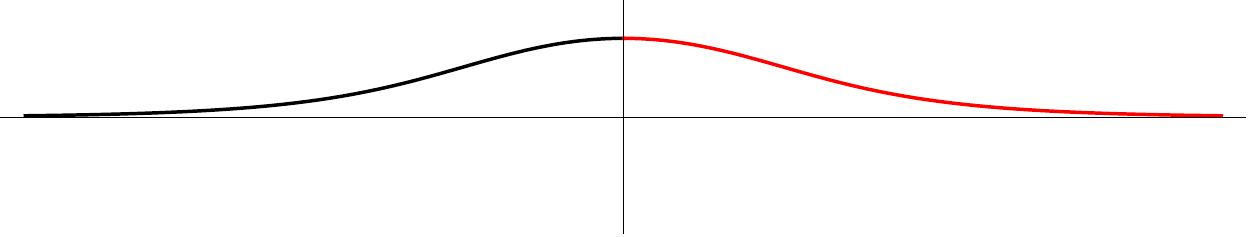}\\
          \includegraphics[clip, height=1.2cm]{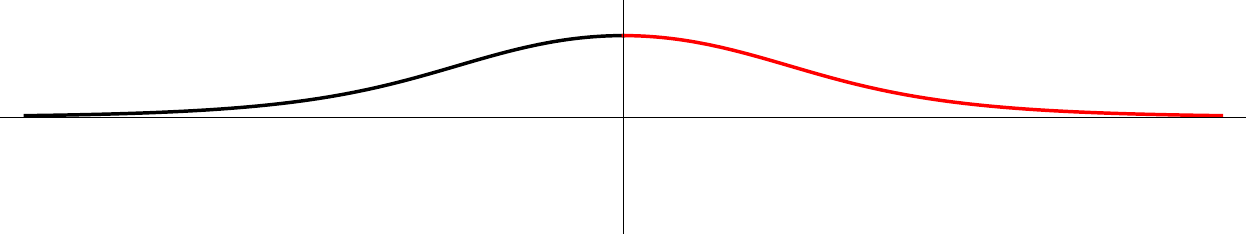}\\
          \includegraphics[clip, height=1.2cm]{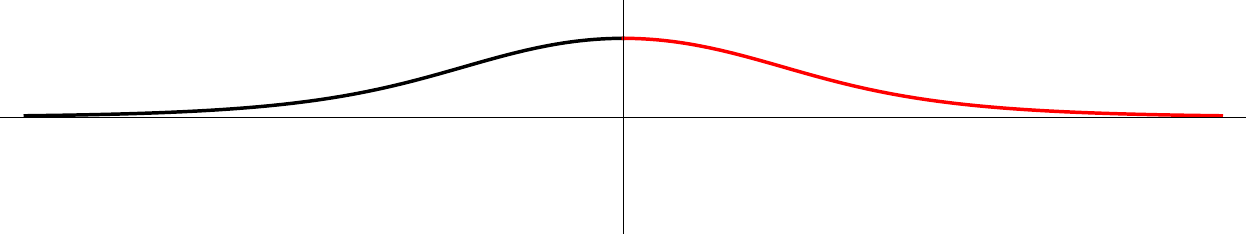}\\
          \includegraphics[clip, height=1.2cm]{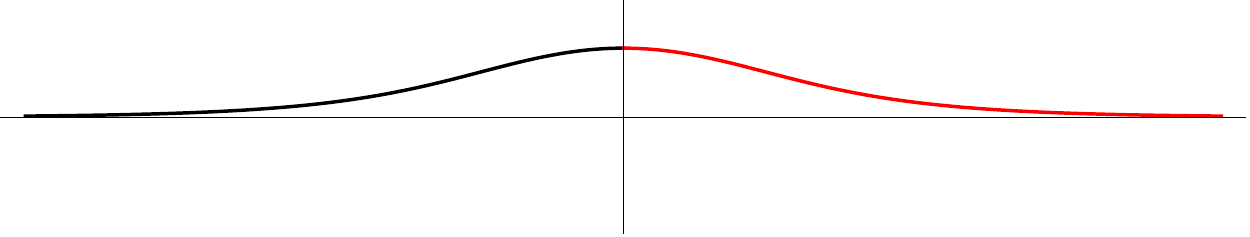}\\
          \includegraphics[clip, height=1.2cm]{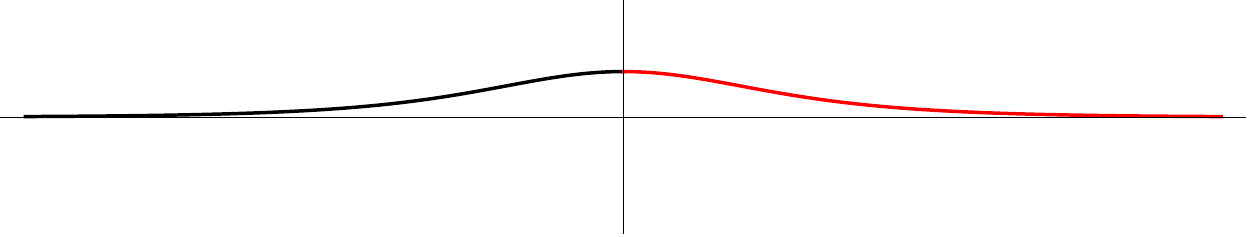}\\
          \includegraphics[clip, height=1.2cm]{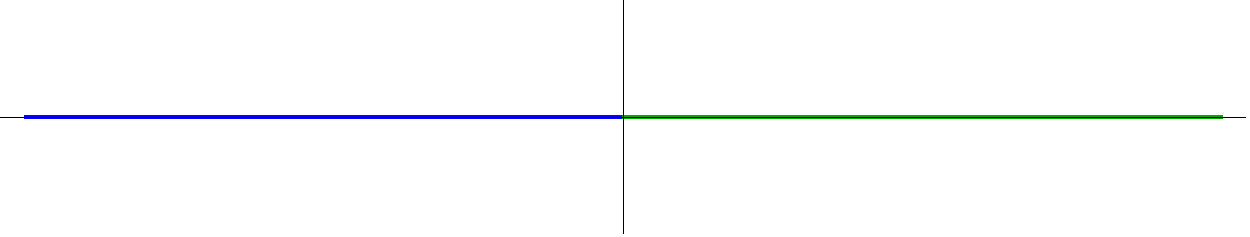}\\
          \includegraphics[clip, height=1.2cm]{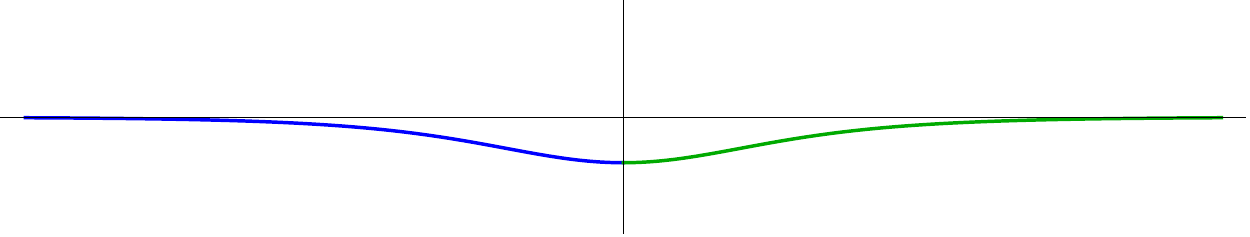}\\
          \includegraphics[clip, height=1.2cm]{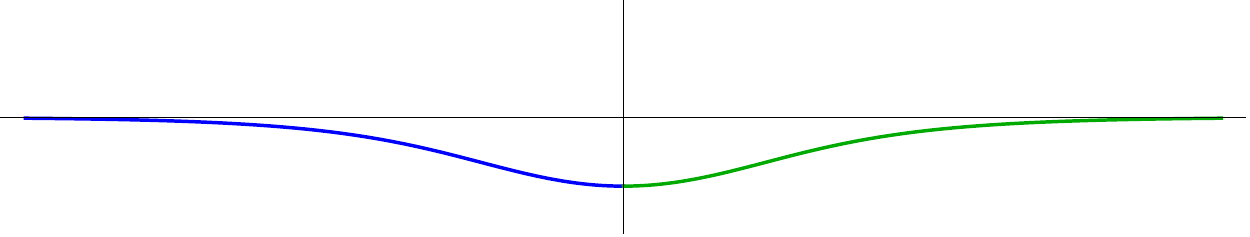}\\
          \includegraphics[clip, height=1.2cm]{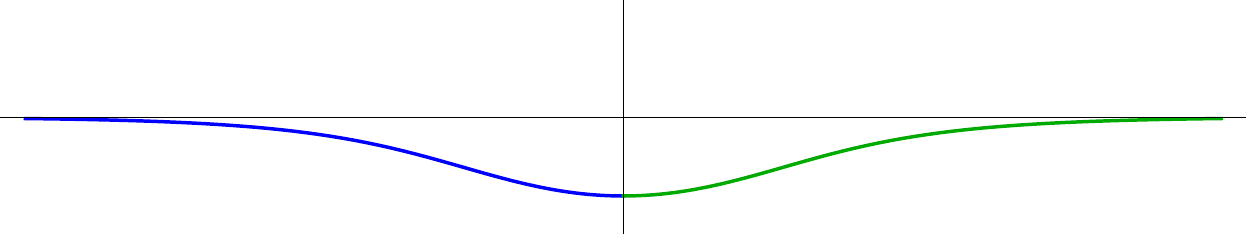}\\
          \includegraphics[clip, height=1.2cm]{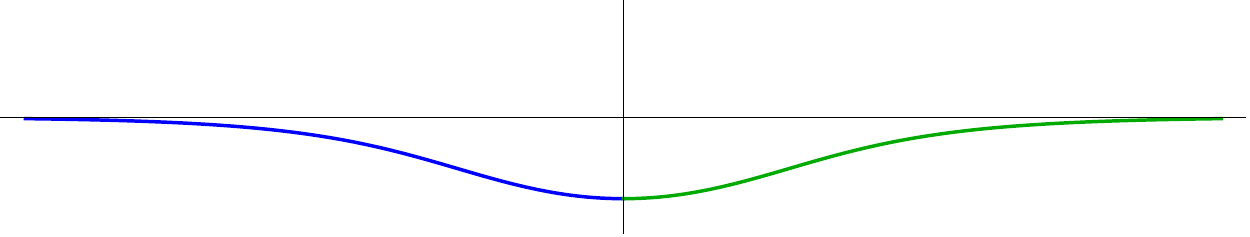}\\
          \includegraphics[clip, height=1.2cm]{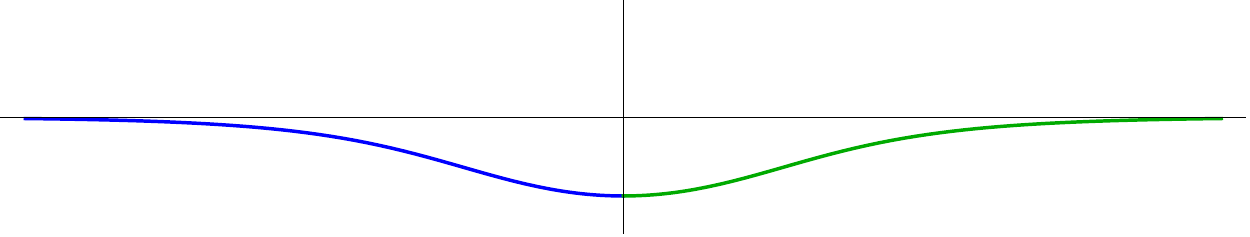}\\
          \includegraphics[clip, height=1.2cm]{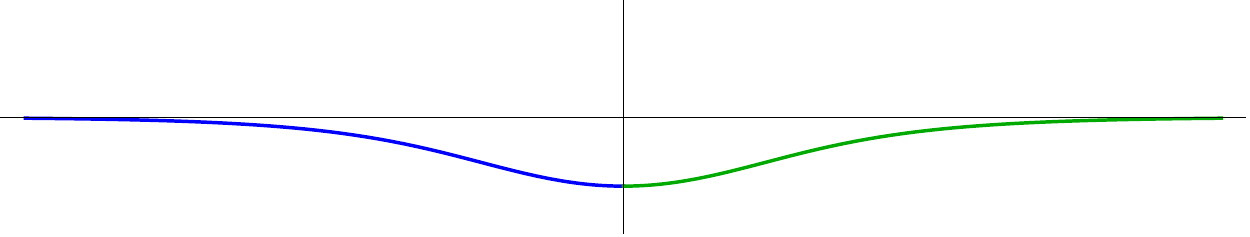}\\
          \includegraphics[clip, height=1.2cm]{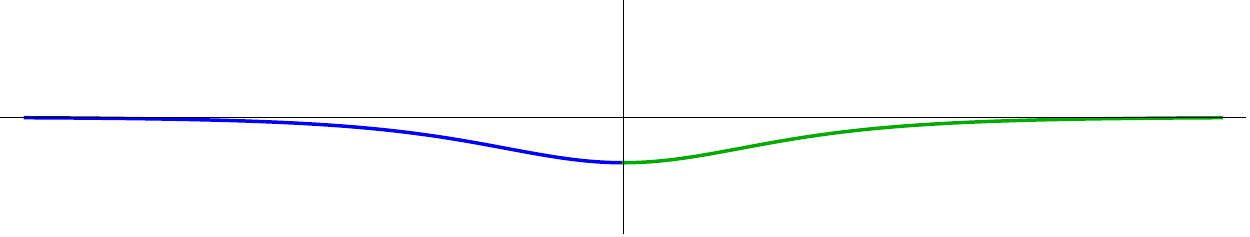}
         \end{center}
      \end{minipage}
     \end{tabular} 
\caption{\capfont Snapshots of a single period of the magnetic meson. 
The top is at $t=0$ and the bottom is at 
$t=T-\delta t$
with $\delta t = T/16$.
The left panel shows the
magnetic field $(F^\Sigma_{12},F^\Sigma_{23})$ by blue 
streamlines and the 
topological charge densities,
${\cal M}=\pm 0.017$,
electric energy density, 
${\cal E}=0.012$, 
and the dressed energy density 
${\cal H}_{\rm dress}=0.02$ 
by red/green/grey contours.
In the right figures, $\tilde \Theta(x^3,t)$ is plotted. 
The curves are piecewise colored by black, red, blue and 
green for ${\cal M}_0$, $\bar {\cal M}_0$,
${\cal M}_\pi$ and $\bar{\cal M}_\pi$, respectively. We set $gv=1$,
$m=1/5$ and $\omega=1/10$. $x^1\in[-3,3]$ and $x^3\in[-30,30]$.}
\label{fig:2d_breather}
\end{center}
\end{figure}

We show how the magnetic meson  
varies  
in one period $T = \frac{2\pi}{\omega}$ 
in Fig.~\ref{fig:2d_breather}. 
The sources of  
outgoing magnetic field are identified as slender monopoles 
and those of incoming magnetic field as slender anti-monopoles. 
It is interesting to observe that the meson is made of 
${\cal M}_0$ and $\bar{\cal M}_0$ at an instance 
(for example $t = T/4$), and that 
it transforms into a different meson made of ${\cal M}_\pi$ and 
$\bar{\cal M}_\pi$ at another instance (for example $t =3T/4$).  
In Fig.~\ref{fig:2d_breather}, we also show 
the topological charge density ${\cal Q}_{\rm m}$ 
given in Eq.~(\ref{eq:mag_ene_dens}) together with the energy density of the electric field
\beq
{\cal E} = \frac{1}{g^2}\Tr\left[(F_{01})^2 + (F_{02})^2 \right] 
= \frac{1}{g^2}\left|1-z\p\psi\right|^2e^{-\psi}(\p_t\tilde\Theta)^2.
\eeq
As the monopole and anti-monopole approach each other, the magnetic energy 
density ${\cal M}$ decreases and the electric field energy density ${\cal E}$ 
grows. At the very instance of collision, the 
magnetic energy disappears and is transferred into the 
electric energy completely.  
The electric field is generated by the time variation 
(decrease) of the magnetic field as monopole and 
anti-monopole merge.

\subsection{Scattering of the slender monopole and anti-monopole}

Let us next study the head-on collision of the slender monopole 
and anti-monopole. 
There are two 
types of collisions: one type is the collision  
 of ${\cal M}_0$ and $\bar {\cal M}_0$, 
(${\cal M}_\pi$ and $\bar {\cal M}_\pi$) 
and the other
type 
is that of ${\cal M}_0$ and $\bar{\cal M}_\pi$ 
(${\cal M}_\pi$ and $\bar{\cal M}_0$). 

\subsubsection*{Scattering of ${\cal M}_0$ and $\bar {\cal M}_0$ (${\cal M}_\pi$ and $\bar {\cal M}_\pi$)}

\begin{figure}
\begin{center}
    \begin{tabular}{cc}
      \begin{minipage}{10cm}
        \begin{center}
          \includegraphics[clip, height=1.3cm]{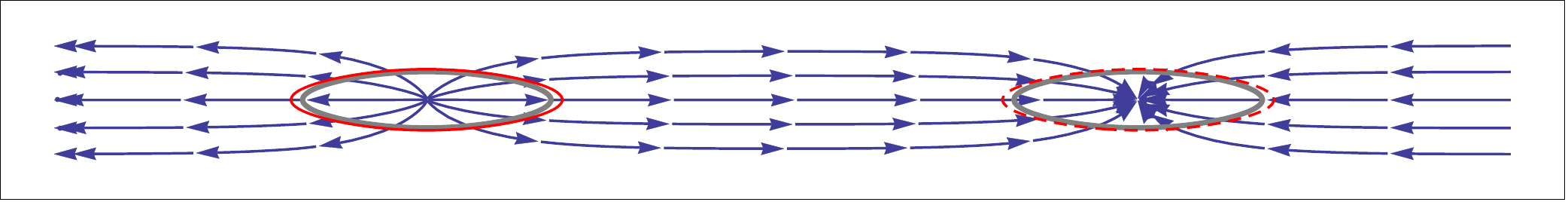}\\
          \includegraphics[clip, height=1.3cm]{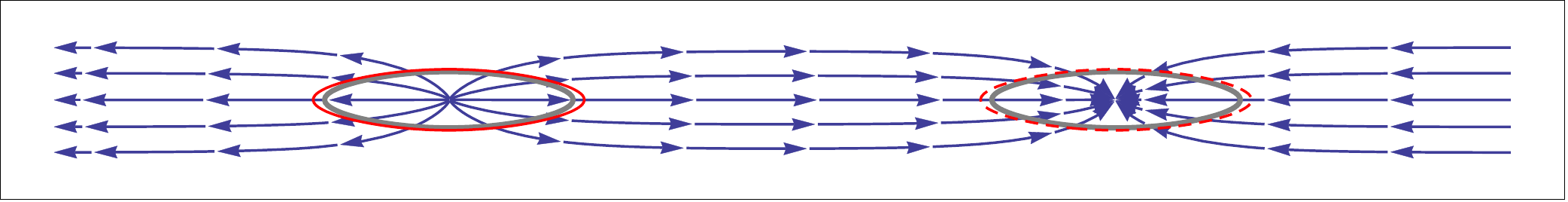}\\
          \includegraphics[clip, height=1.3cm]{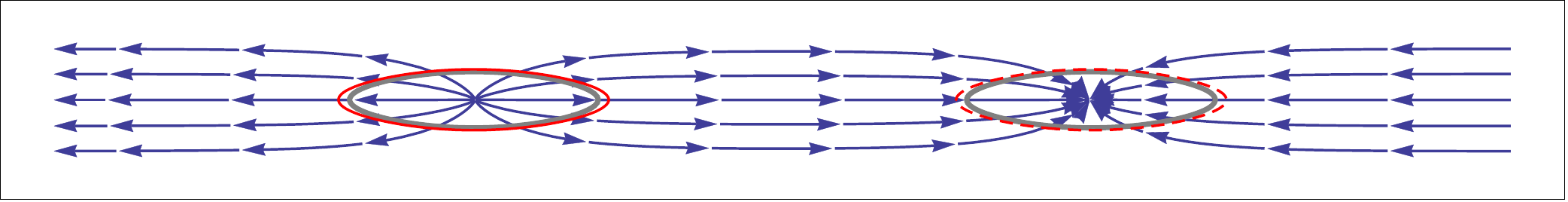}\\
          \includegraphics[clip, height=1.3cm]{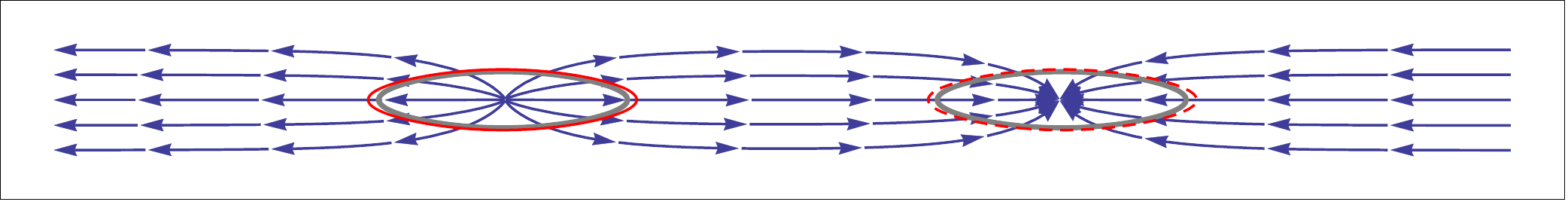}\\
          \includegraphics[clip, height=1.3cm]{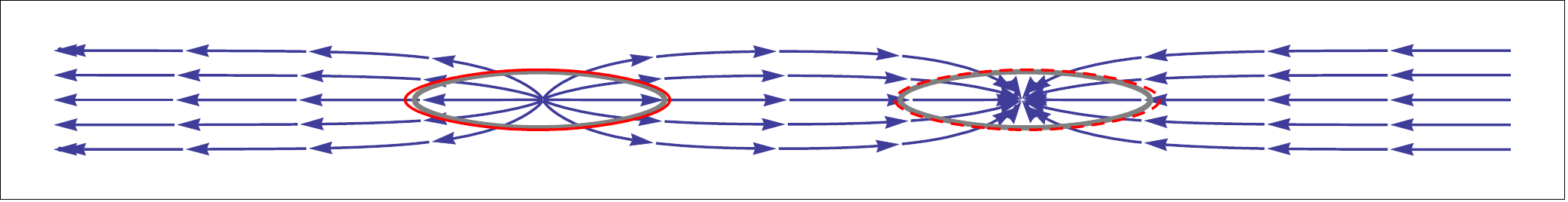}\\
          \includegraphics[clip, height=1.3cm]{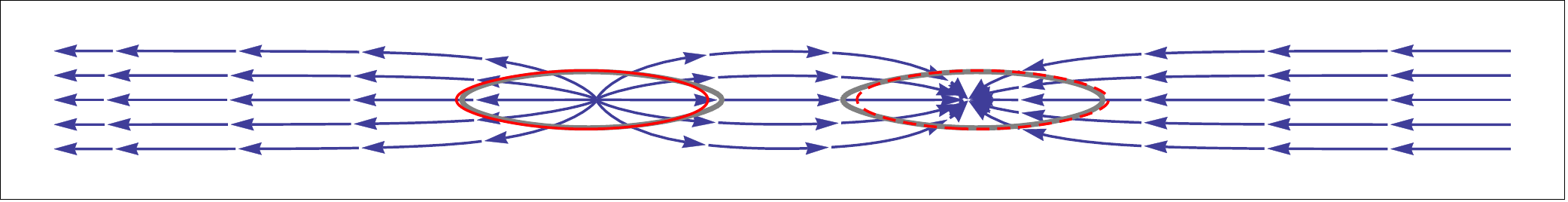}\\
          \includegraphics[clip, height=1.3cm]{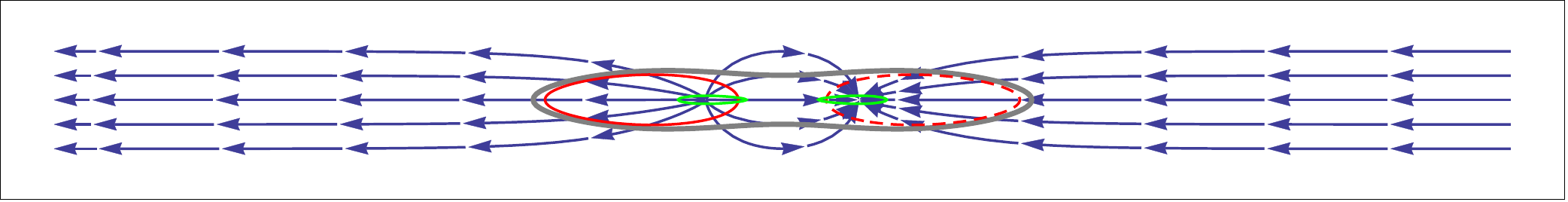}\\
          \includegraphics[clip, height=1.3cm]{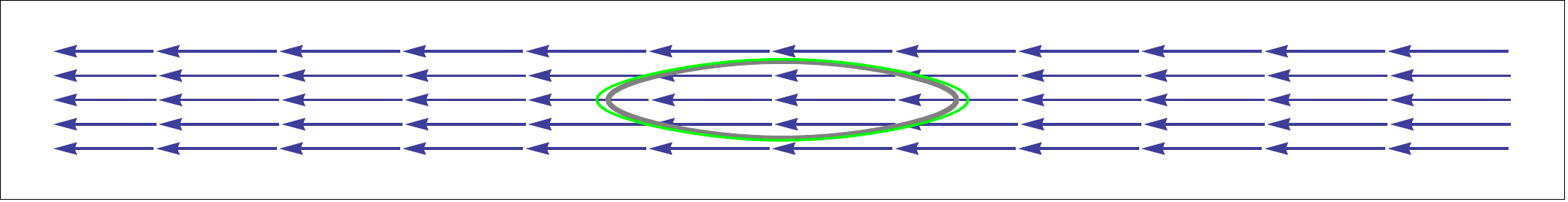}\\
          \includegraphics[clip, height=1.3cm]{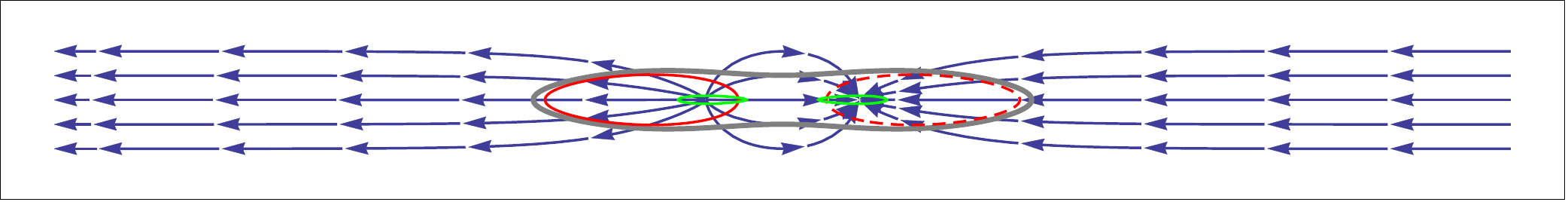}\\
          \includegraphics[clip, height=1.3cm]{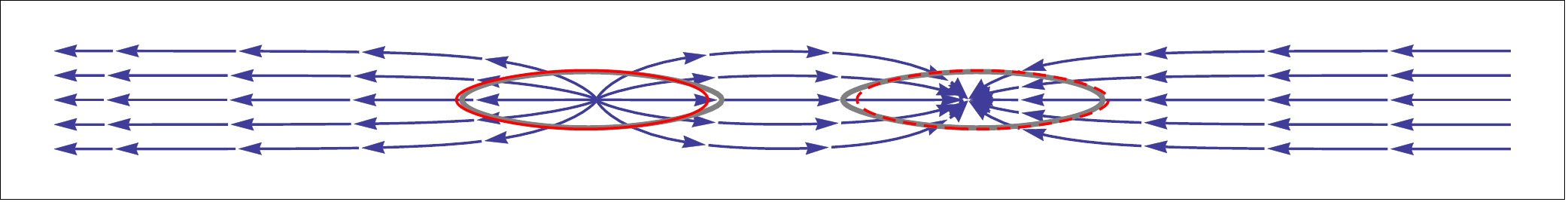}\\
          \includegraphics[clip, height=1.3cm]{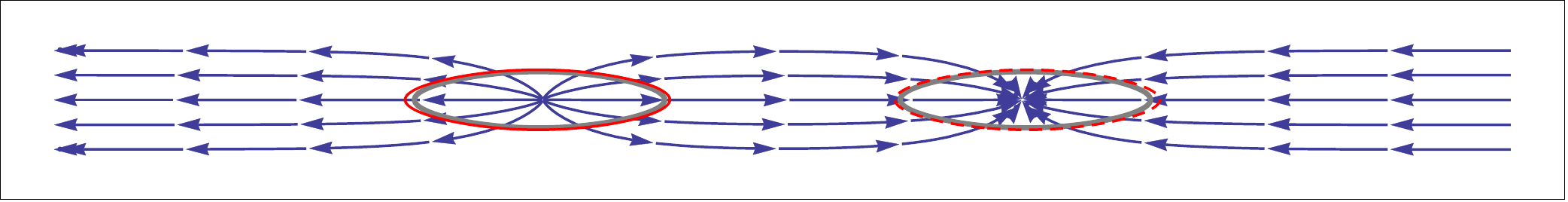}\\
          \includegraphics[clip, height=1.3cm]{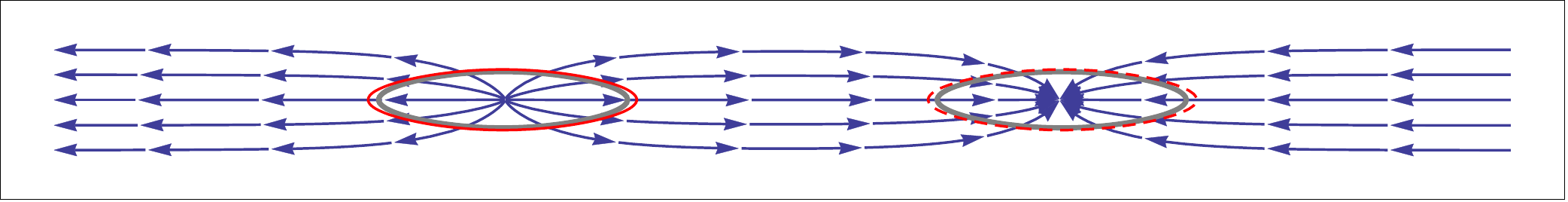}\\
          \includegraphics[clip, height=1.3cm]{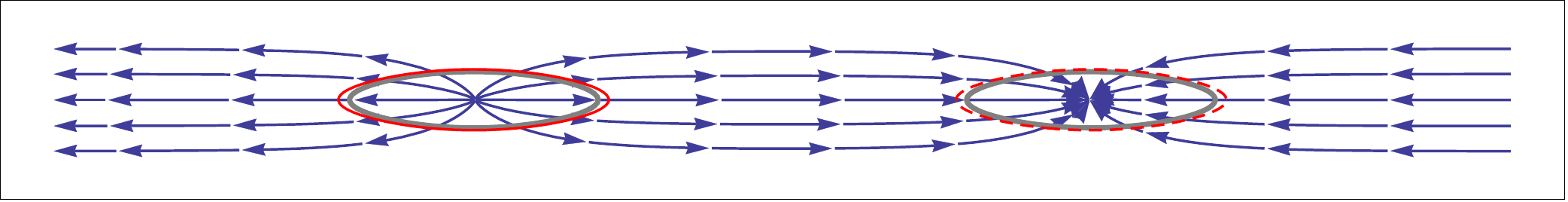}\\
          \includegraphics[clip, height=1.3cm]{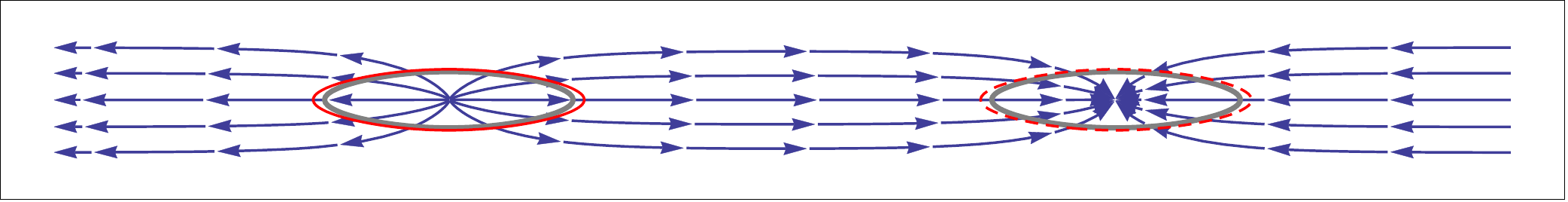}\\
          \includegraphics[clip, height=1.3cm]{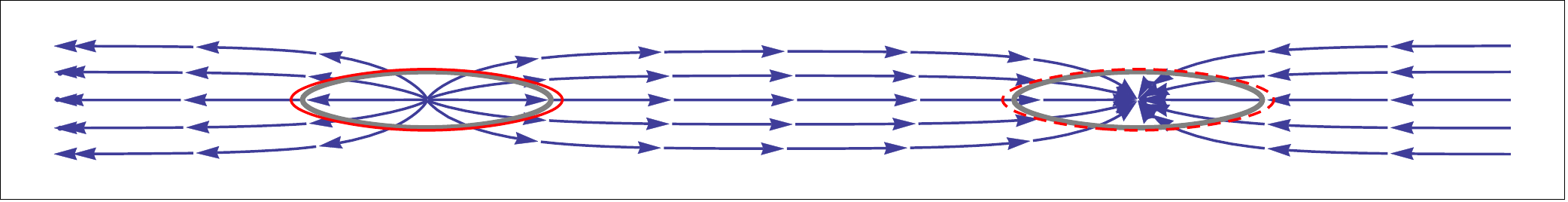}
         \end{center}
      \end{minipage}
      &
      \begin{minipage}{6.5cm}
        \begin{center}
          \includegraphics[clip, height=1.3cm]{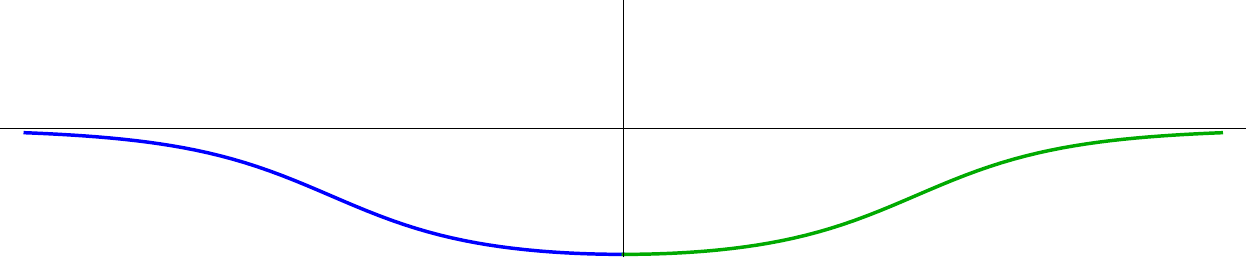}\\
          \includegraphics[clip, height=1.3cm]{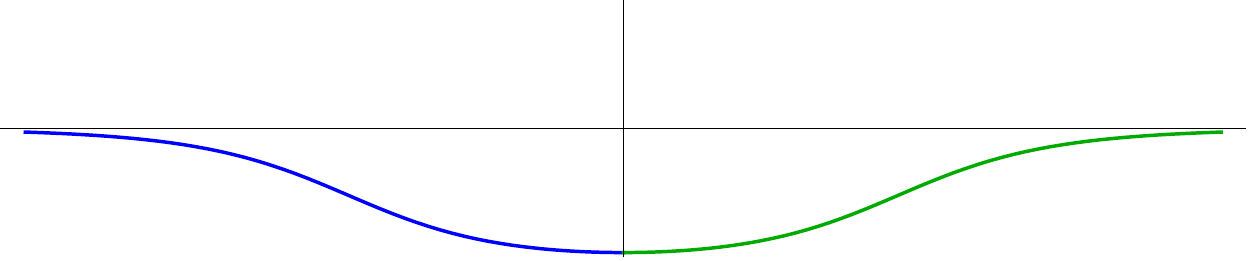}\\
          \includegraphics[clip, height=1.3cm]{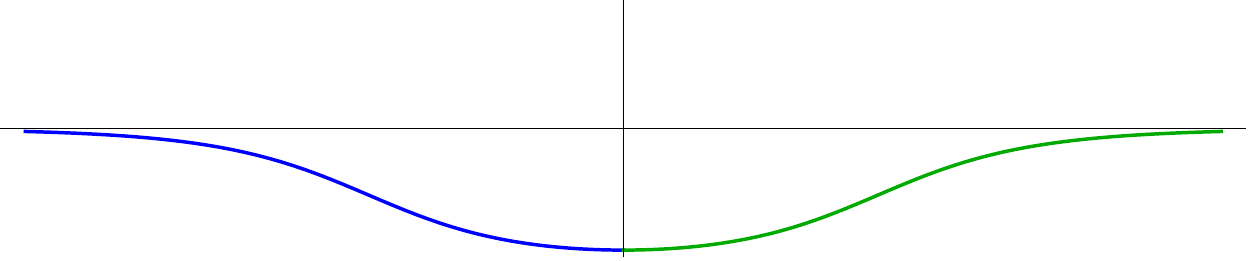}\\
          \includegraphics[clip, height=1.3cm]{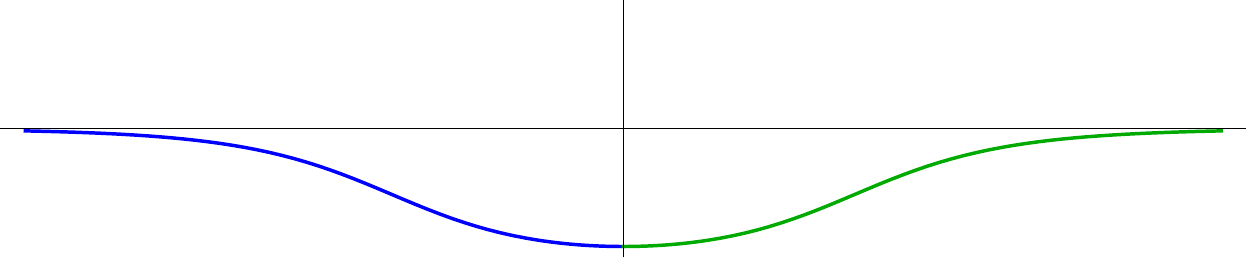}\\
          \includegraphics[clip, height=1.3cm]{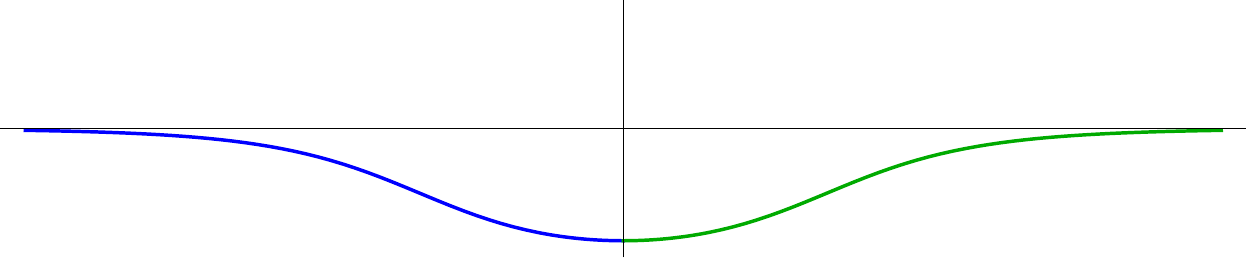}\\
          \includegraphics[clip, height=1.3cm]{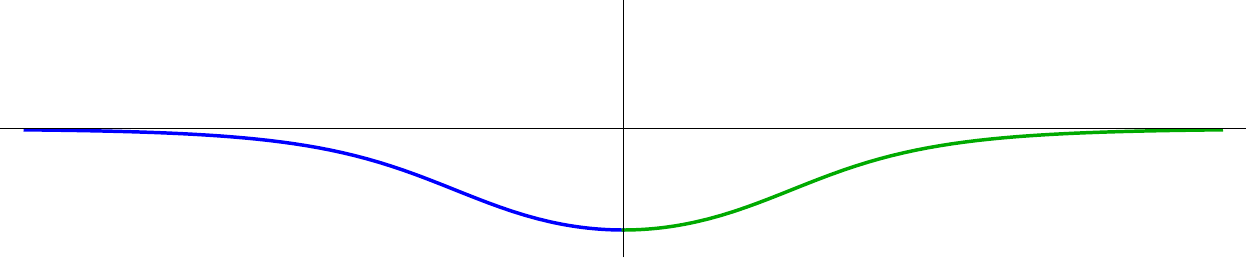}\\
          \includegraphics[clip, height=1.3cm]{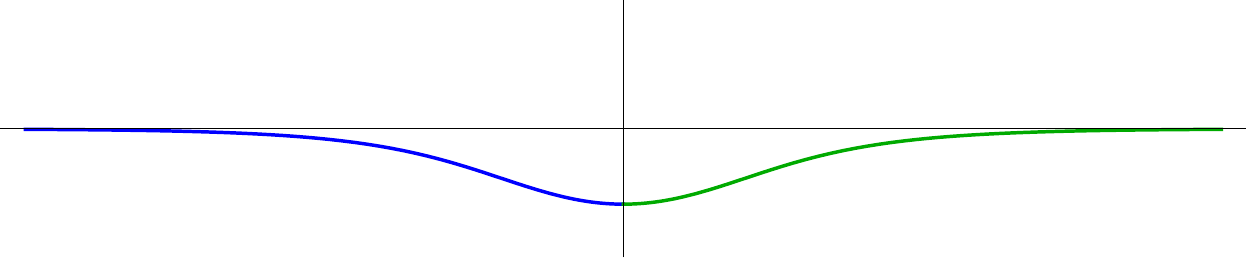}\\
          \includegraphics[clip, height=1.3cm]{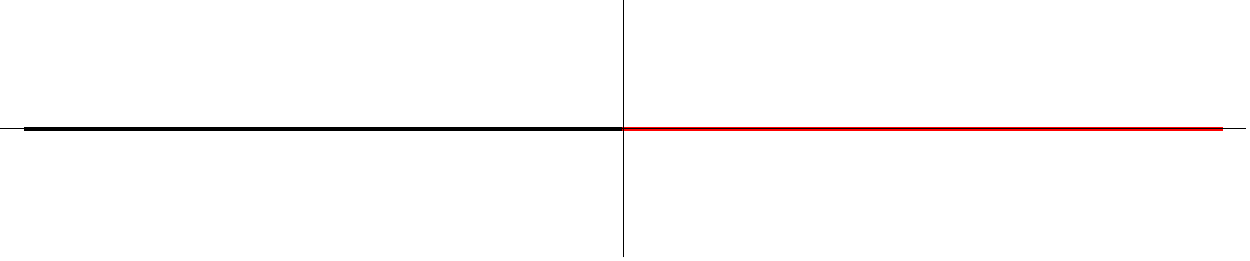}\\
          \includegraphics[clip, height=1.3cm]{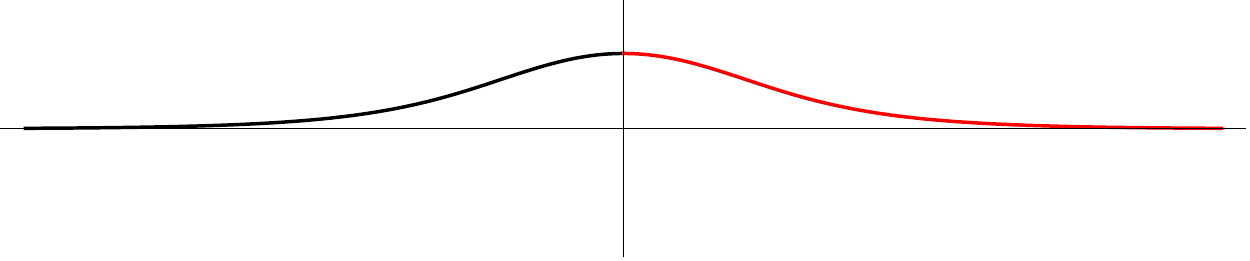}\\
          \includegraphics[clip, height=1.3cm]{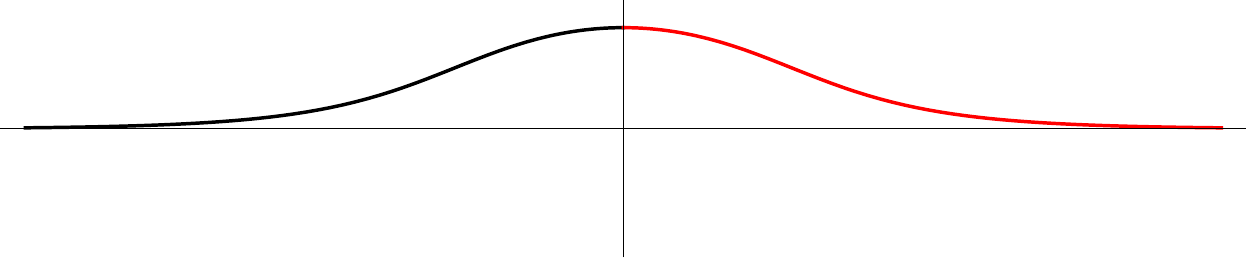}\\
          \includegraphics[clip, height=1.3cm]{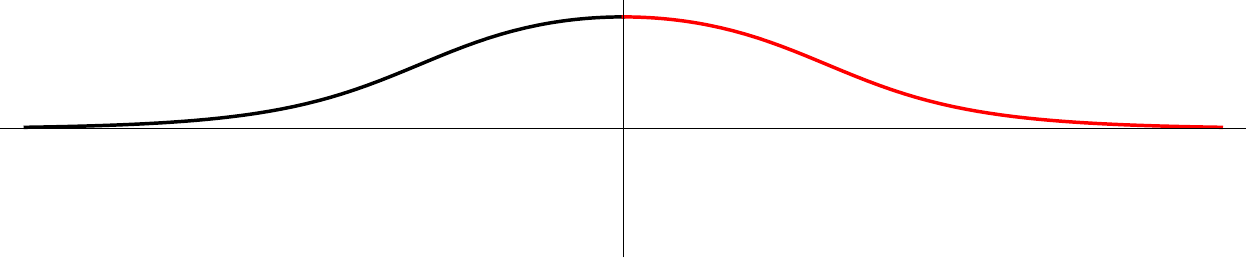}\\
          \includegraphics[clip, height=1.3cm]{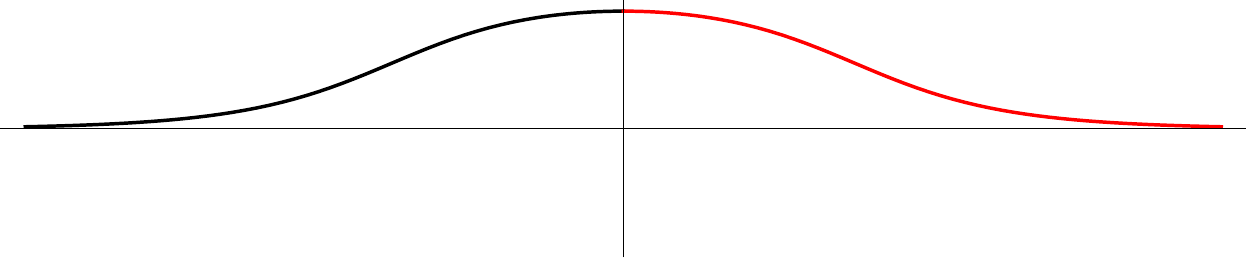}\\
          \includegraphics[clip, height=1.3cm]{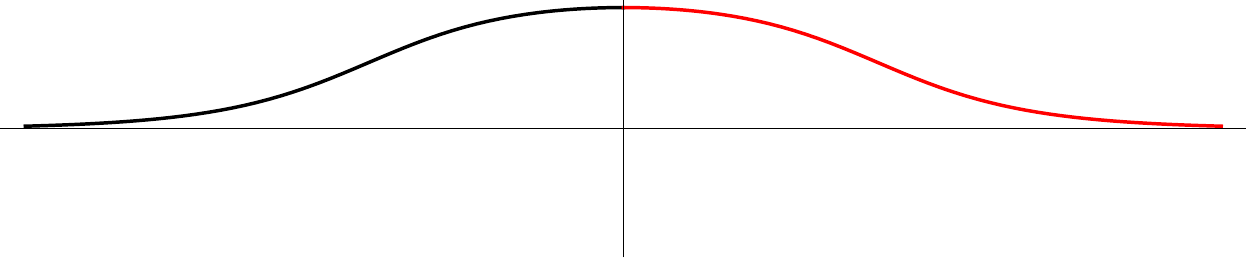}\\
          \includegraphics[clip, height=1.3cm]{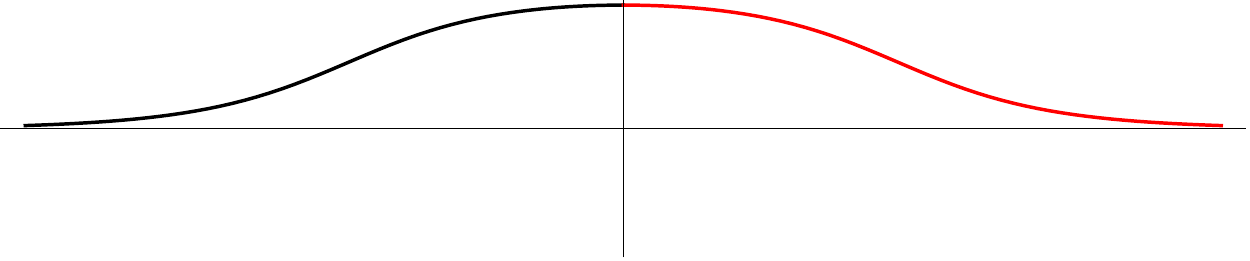}\\
          \includegraphics[clip, height=1.3cm]{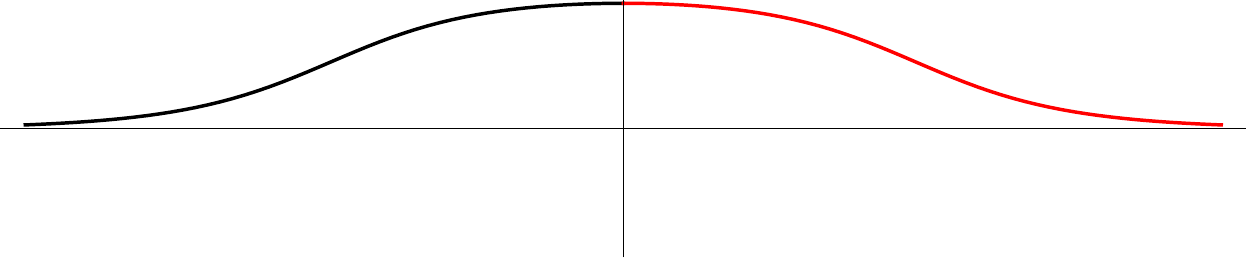}
         \end{center}
      \end{minipage}
     \end{tabular} 
\caption{
\capfont Snapshots (from $t=-42$ to $t=42$ with 
$\delta t =6$ interval) 
of scattering of the slender monopole and anti-monopole.
The red/green/gray contours 
are $({\cal M},{\cal E},{\cal H}_{\rm dress})=(\pm 0.017,0.01,0.02)$,
see the caption of Fig.~\ref{fig:2d_breather} for explanation. We set $gv=1$,
$m=1/5$ and $u=1/10$. $x^1\in[-3,3]$ and $x^3\in[-30,30]$.
}
\label{fig:2d_scatter1}
\end{center}
\end{figure}

The  
exact solution for the moduli field for the collision of a 
monopole and anti-monopole of the same species 
(${\cal M}_0$ or ${\cal M}_\pi$)  
is given by 
\beq
\tilde \Theta =
2 \arctan\left(\frac{\sinh mu\gamma x^0}
{u\cosh m\gamma x^3}\right),\quad
\gamma = \frac{1}{\sqrt{1-u^2}}.
\eeq
The parameter $u$ corresponds to 
the 
velocity of the monopole.
However, we should keep in mind that our approximation holds 
only for small velocities, that is 
\beq \label{eq:restr}
u \ll1\qquad (\gamma \simeq 1).
\eeq
Since we are using the rigid-body approximation we cannot 
faithfully describe Lorentz boosted monopoles. 
Thus even though we can solve the 1+1-dimensional effective 
dynamics for arbitrary velocities, the full 1+3-dimensional 
dynamics is correctly represented only within the restriction 
of Eq.~(\ref{eq:restr}). 
A typical configuration is shown in Fig.~\ref{fig:2d_scatter1}.
The slender magnetic monopole ${\cal M}_\pi$ comes from the 
left infinity and the anti-monopole $\bar{\cal M}_\pi$ comes 
from the right infinity. 
As they approach to the 
collision point, the magnetic energy decreases 
while the electric energy grows. 
After the collision, the magnetic energy grows as the monopole 
${\cal M}_0$ (anti-monopole $\bar{\cal M}_0$) goes toward 
the left (right) infinity. 
Thus we find that the species of the monopole and the 
anti-monopole changes after the collision.  
The attractive force 
given in Eq.~(\ref{eq:attractive_force}) gives rise to a 
negative time delay 
\beq
\delta t = \frac{1-u^2}{u}\log u < 0.
\eeq

\subsubsection*{Scattering of ${\cal M}_0$ and $\bar {\cal M}_\pi$ (or ${\cal M}_\pi$ and $\bar {\cal M}_0$)}

\begin{figure}
\begin{center}
    \begin{tabular}{cc}
      \begin{minipage}{10cm}
        \begin{center}
          \includegraphics[clip, height=1.3cm]{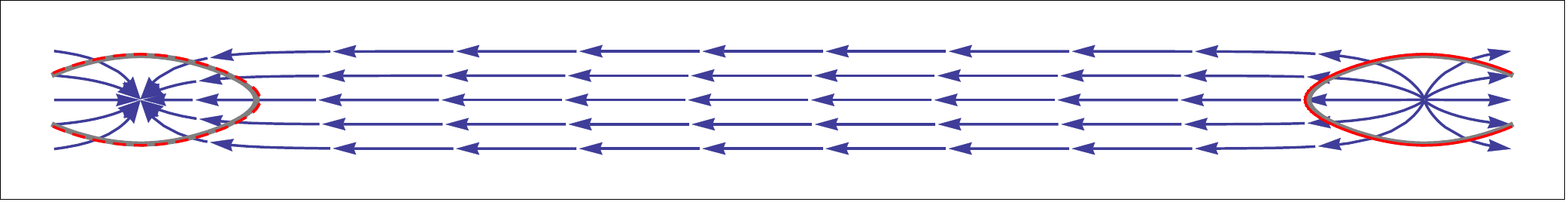}\\
          \includegraphics[clip, height=1.3cm]{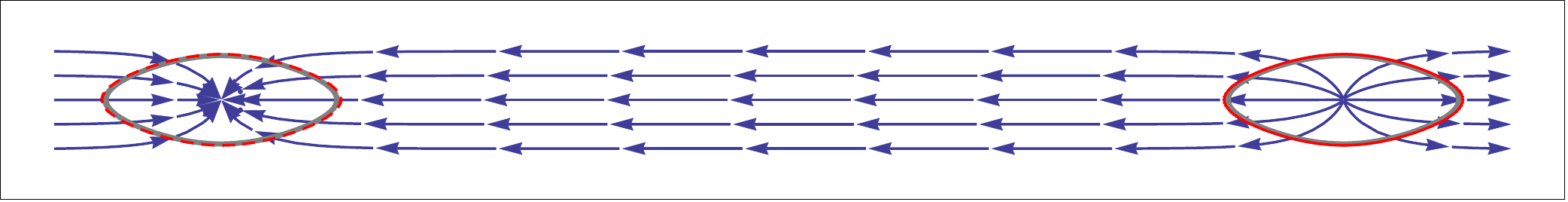}\\
          \includegraphics[clip, height=1.3cm]{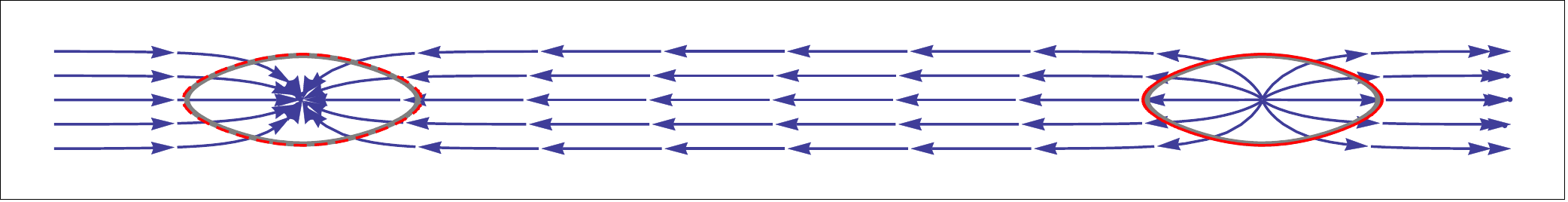}\\
          \includegraphics[clip, height=1.3cm]{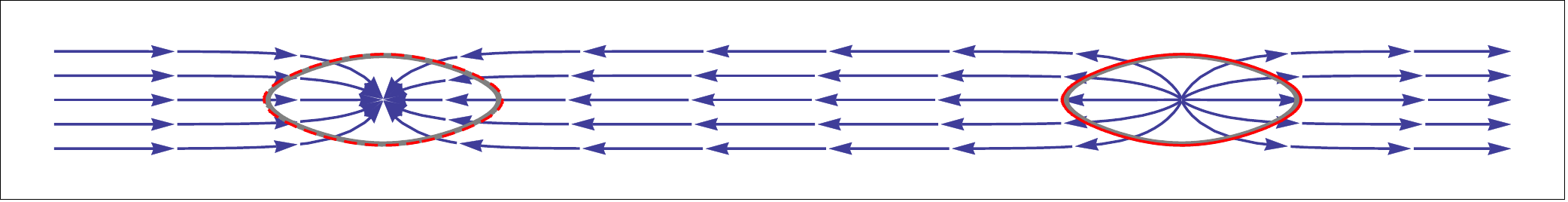}\\
          \includegraphics[clip, height=1.3cm]{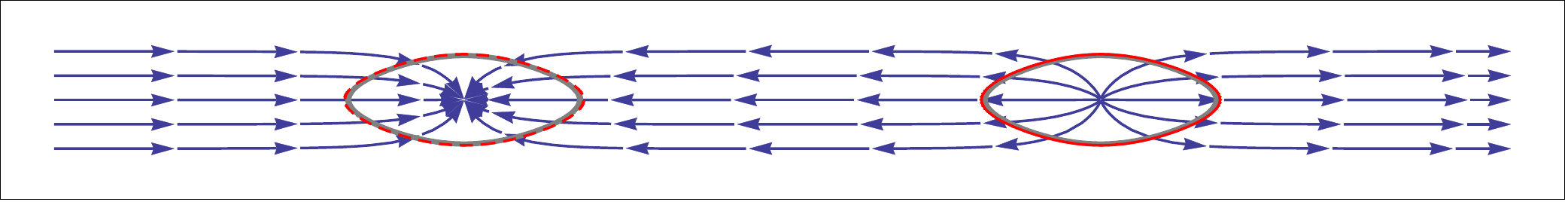}\\
          \includegraphics[clip, height=1.3cm]{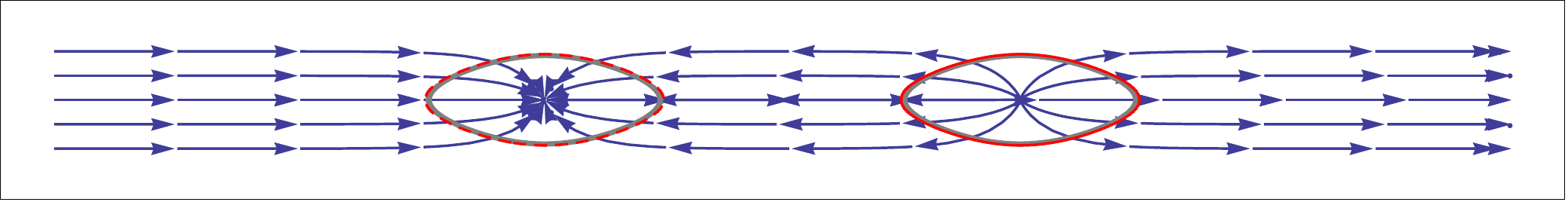}\\
          \includegraphics[clip, height=1.3cm]{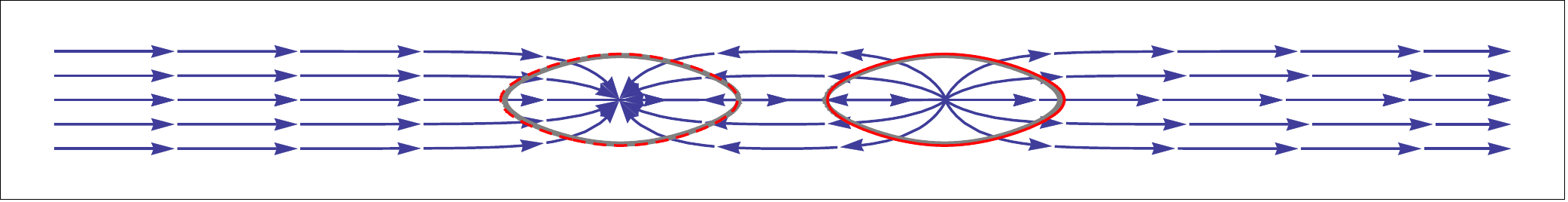}\\
          \includegraphics[clip, height=1.3cm]{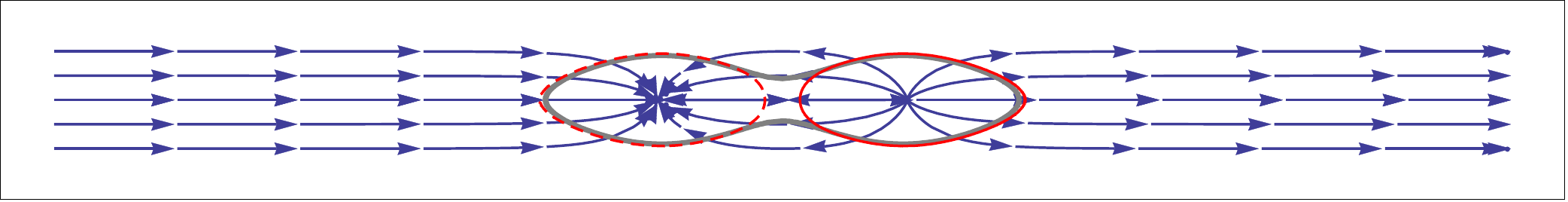}\\
          \includegraphics[clip, height=1.3cm]{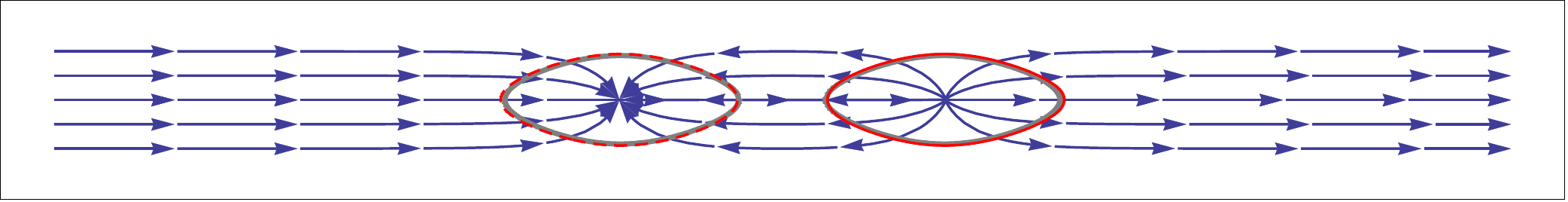}\\
          \includegraphics[clip, height=1.3cm]{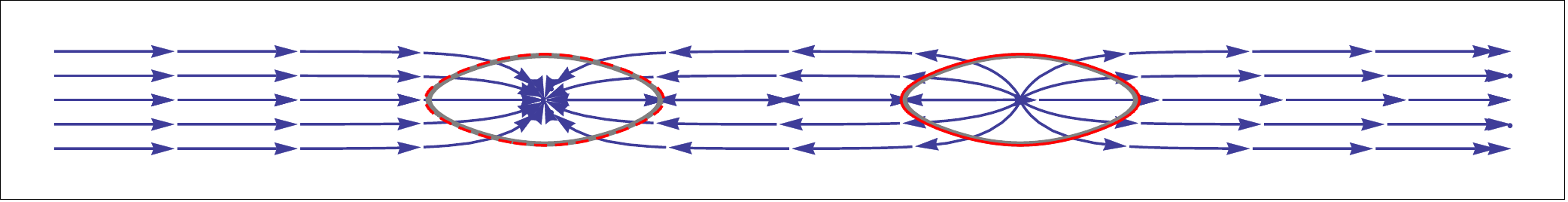}\\
          \includegraphics[clip, height=1.3cm]{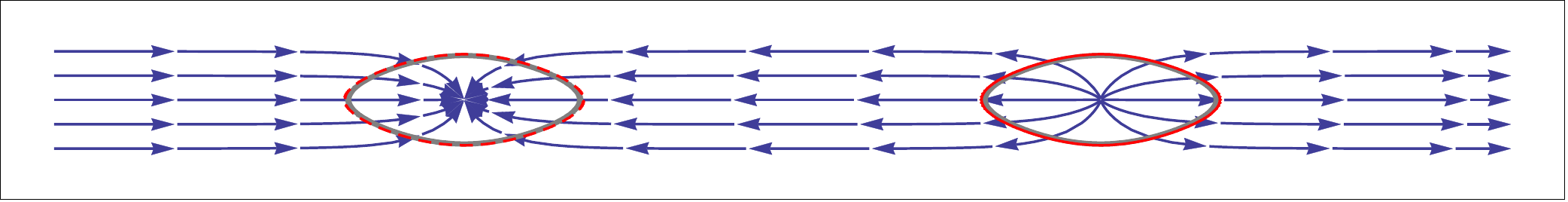}\\
          \includegraphics[clip, height=1.3cm]{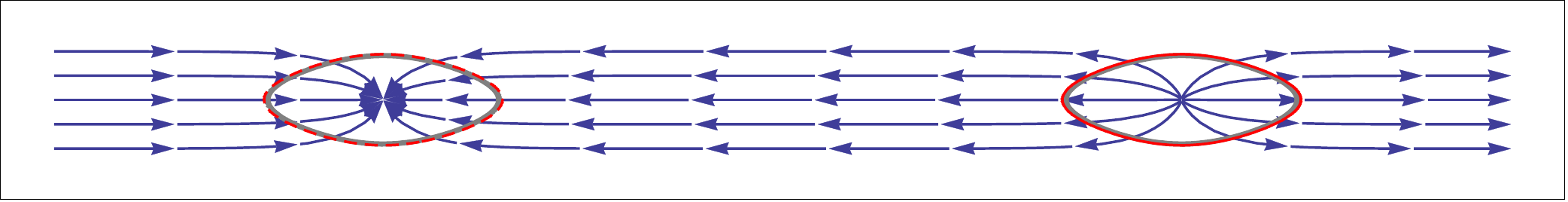}\\
          \includegraphics[clip, height=1.3cm]{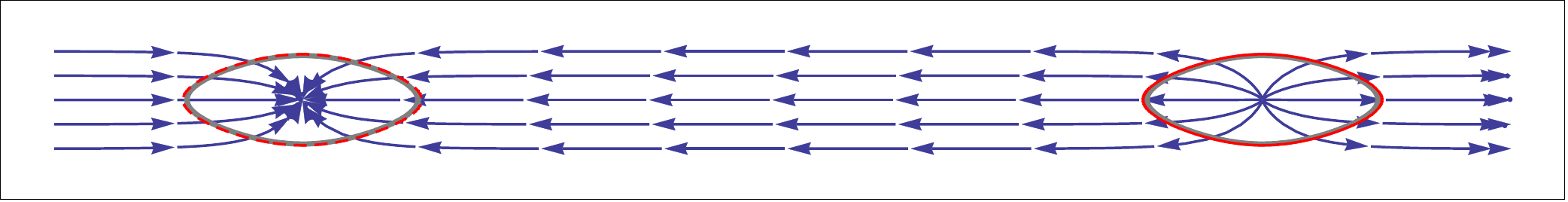}\\
          \includegraphics[clip, height=1.3cm]{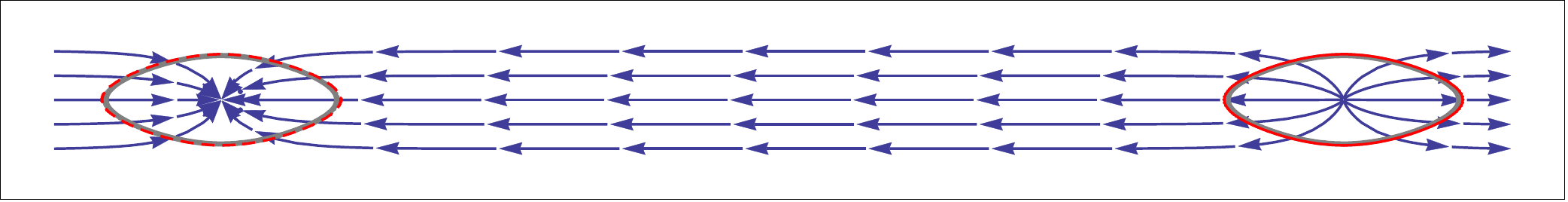}\\
          \includegraphics[clip, height=1.3cm]{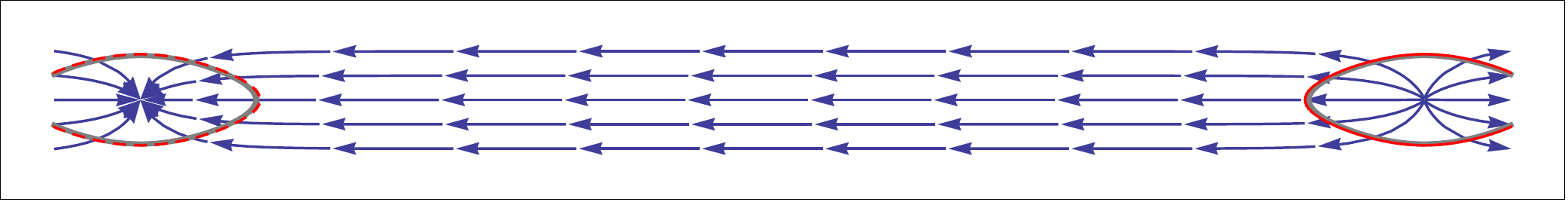}
         \end{center}
      \end{minipage}
      &
      \begin{minipage}{6.5cm}
        \begin{center}
          \includegraphics[clip, height=1.3cm]{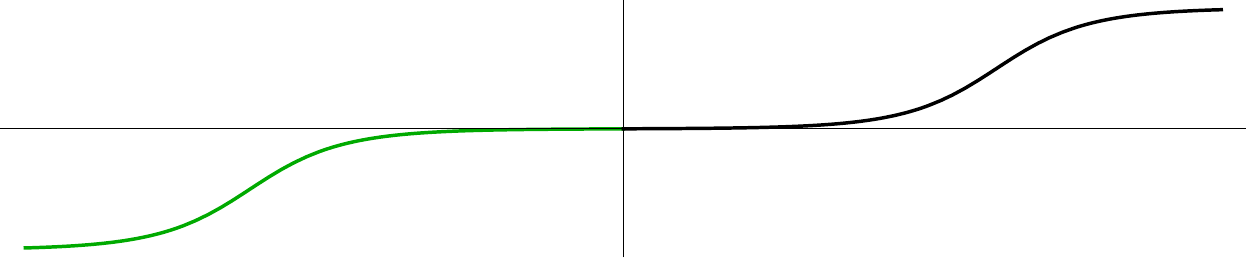}\\
          \includegraphics[clip, height=1.3cm]{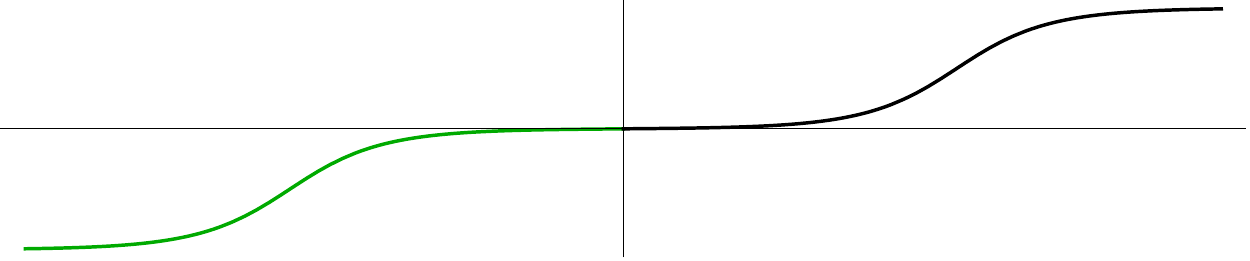}\\
          \includegraphics[clip, height=1.3cm]{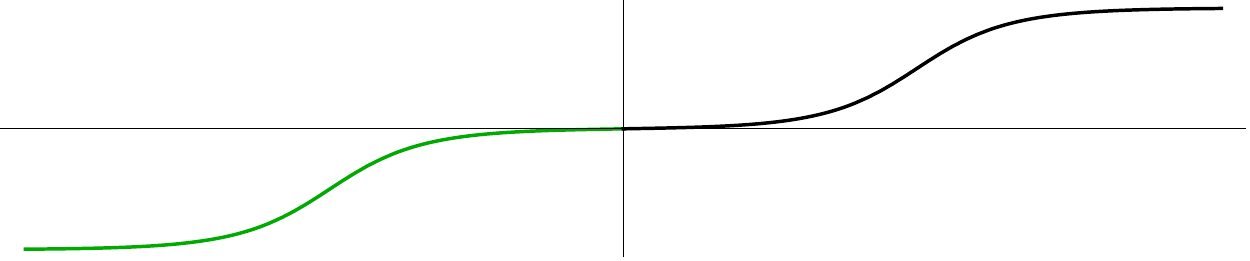}\\
          \includegraphics[clip, height=1.3cm]{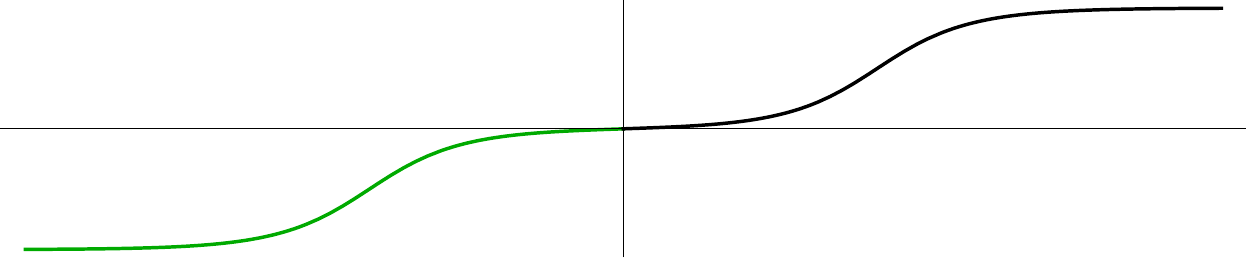}\\
          \includegraphics[clip, height=1.3cm]{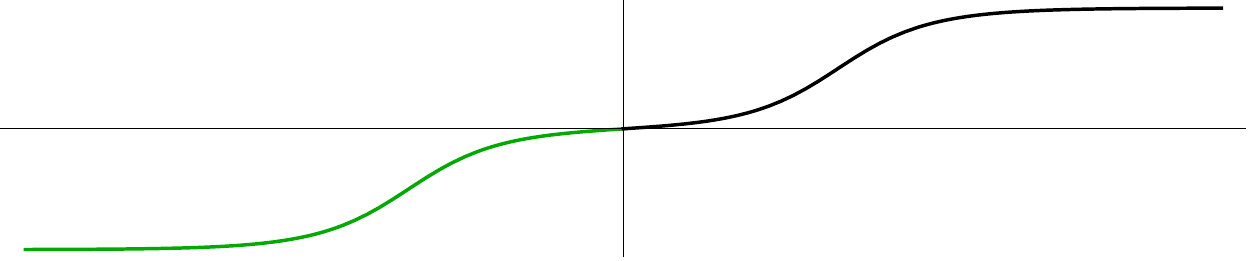}\\
          \includegraphics[clip, height=1.3cm]{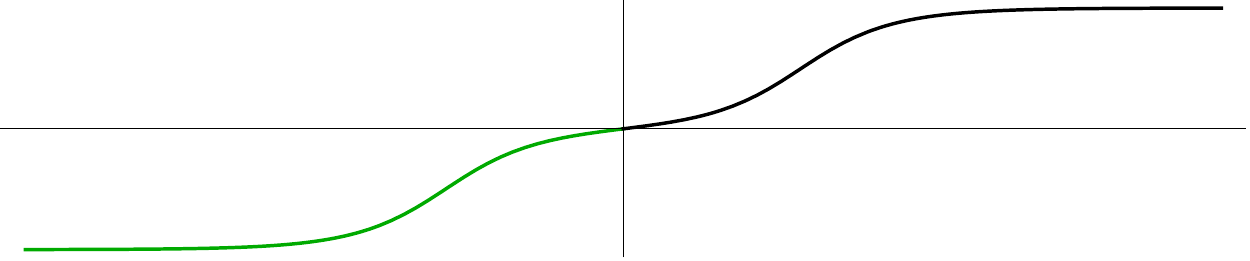}\\
          \includegraphics[clip, height=1.3cm]{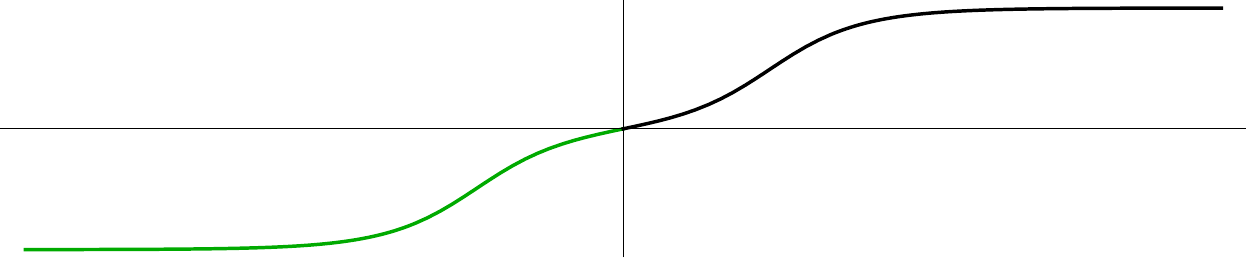}\\
          \includegraphics[clip, height=1.3cm]{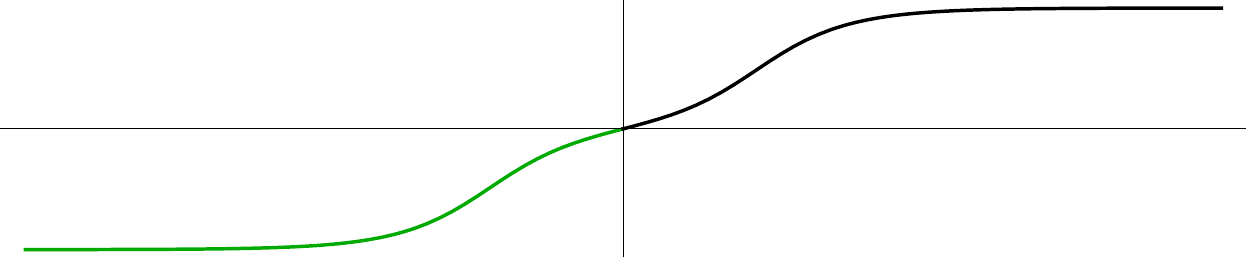}\\
          \includegraphics[clip, height=1.3cm]{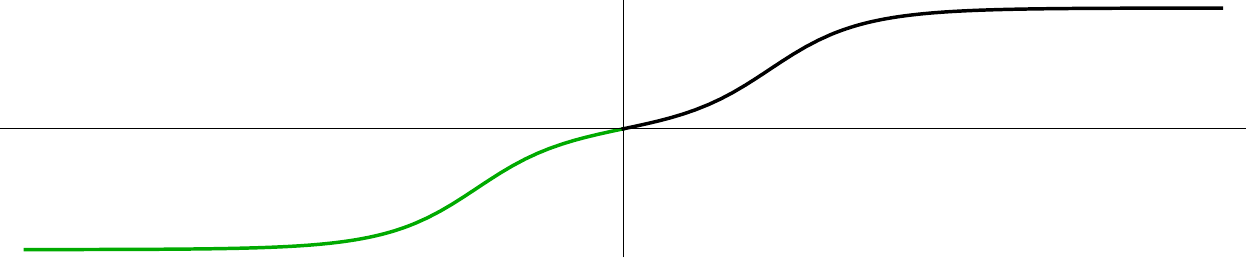}\\
          \includegraphics[clip, height=1.3cm]{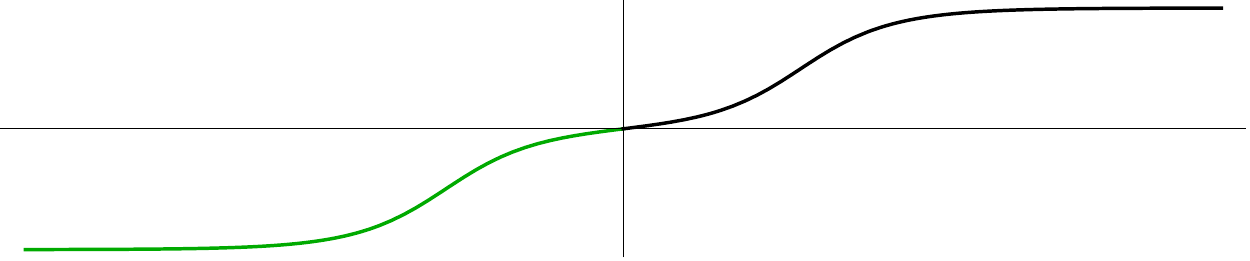}\\
          \includegraphics[clip, height=1.3cm]{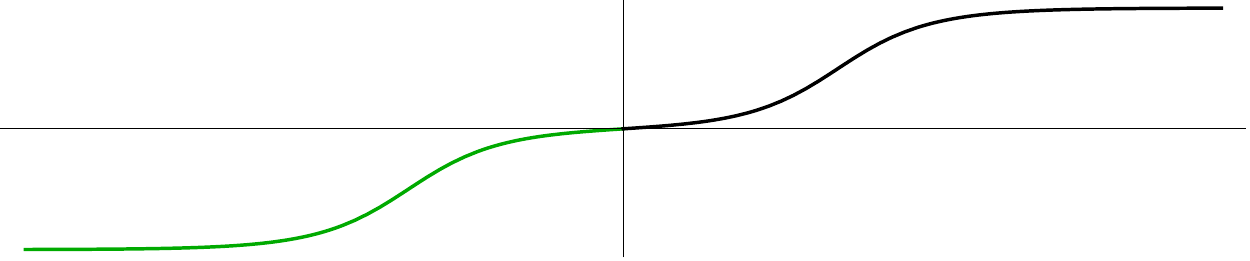}\\
          \includegraphics[clip, height=1.3cm]{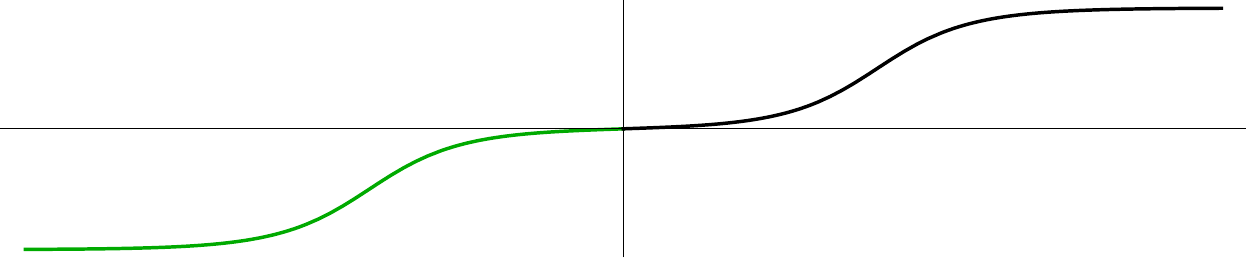}\\
          \includegraphics[clip, height=1.3cm]{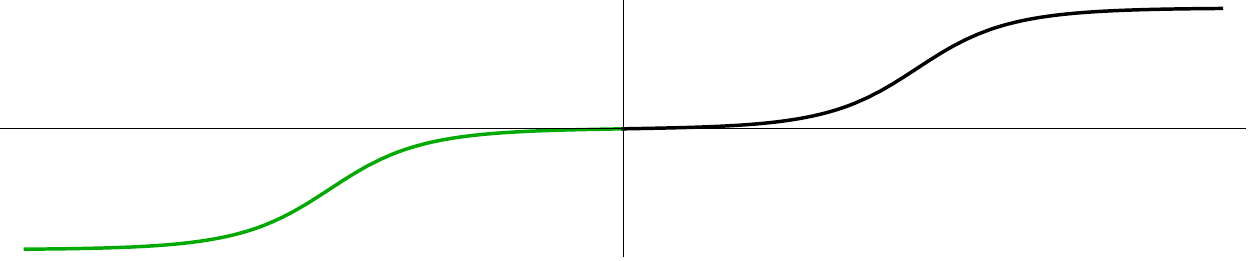}\\
          \includegraphics[clip, height=1.3cm]{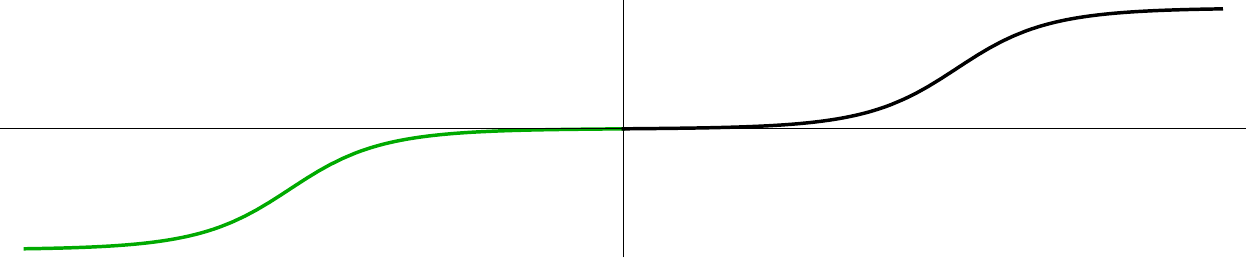}\\
          \includegraphics[clip, height=1.3cm]{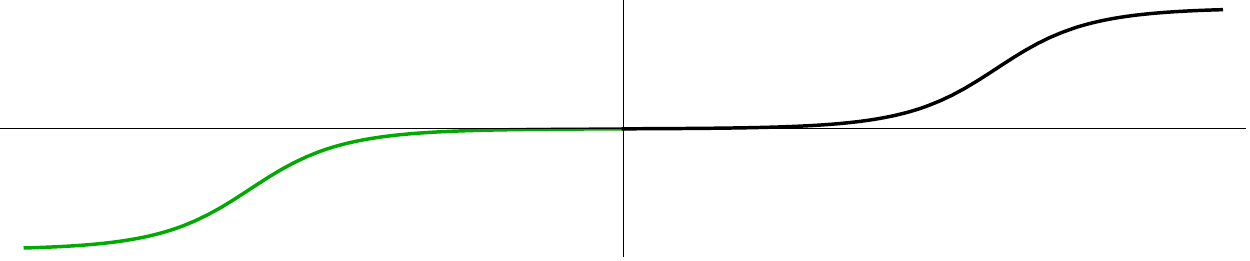}
         \end{center}
      \end{minipage}
     \end{tabular} 
\caption{\capfont Snapshots (from $t=-70$ to $t=70$ with 
$\delta t =10$ interval) 
of scattering of the slender monopole and anti-monopole.
The red/gray contours are $({\cal M},{\cal H}_{\rm dress})=(\pm 0.017,0.02)$.
See the caption of Fig.~\ref{fig:2d_breather} for explanation. We set $gv=1$,
$m=1/3$ and $u=1/3$. $x^1\in[-3,3]$ and $x^3\in[-30,30]$.}
\label{fig:2d_scatter2}
\end{center}
\end{figure}

The solution for the scattering of a monopole and an 
anti-monopole of different species  
is given by 
\beq
\tilde \Theta =
2 \arctan\left(\frac{u \sinh m\gamma x^3}{\cosh m u\gamma x^0}\right).
\eeq
A typical configuration is shown in Fig.~\ref{fig:2d_scatter2}. 
In contrast to
the previous type of scattering 
the species of monopoles 
(${\cal M}_0$ or ${\cal M}_\pi$) do not change into different species 
during the collision. 
As~shown in Fig.~\ref{fig:2d_scatter2}, the anti-monopole 
$\bar{\cal M}_\pi$ comes from the left infinity and reflects back 
toward the left infinity, while the monopole ${\cal M}_0$ 
comes from the right infinity and reflects back toward the 
right infinity.

\clearpage

\clearpage

\clearpage

\subsection{Lifting the zero mode}    

So far, we have considered only solutions 
with a constant azimuthal angle $\tilde\Phi ={\rm const.}$, where
all the solutions 
are mapped onto trajectories on  a great circle of $\mathbb{C}P^1$.
Since the fundamental homotopy group of the total target 
space $\mathbb{C}P^1\simeq S^2$ of the $U(2)$ vortex is 
trivial, a small fluctuation of $\tilde \Phi$ 
may destabilize our solutions with constant $\tilde \Phi$ 
and can convert the magnetic mesons from bound states 
 (of monopole and anti-monopole) into resonances. 
This observation appears plausible, but we would 
like to emphasize that our configurations 
with constant $\tilde \Phi$ are all solutions 
of the full equations of motion (within our approximation). 
Therefore, our statement that the monopole and anti-monopole 
do not always decay into radiation but can make bound states 
or resonances is still valid. 
Moreover, one should note that these magnetic mesons 
play a significant role in understanding dynamics of monopole 
and anti-monopole system, irrespective of whether they are 
genuine bound states or resonances.

If one desires, one can introduce a small  scalar potential
such as 
\beq
V_{\rm add} = \lambda g^2m^2
\left(\Tr\left[H\sigma_1H^\dagger\right]\right)^2.
\label{eq:V_add}
\eeq
This potential respects the gauge symmetry $U(2)_{\rm C}$ while 
the flavor symmetry $U(1)_{\rm F3}$ is no longer genuine symmetry of the system.
It is the approximate symmetry, and the zero 
mode of the azimuthal angle $\tilde \Phi$ is weakly lifted.
Let us see how this additional potential gives a finite mass to $\tilde \Phi$.
First of all, we assume that $V_{\rm add}$ is sufficiently small in such a way that the 
rigid-body
approximation works. Namely, the zero-th order equations (\ref{eq:zeroth1}) and (\ref{eq:zeroth2}), and
the first order equations (\ref{eq:eom_NLO_1}) and (\ref{eq:eom_NLO_2}) are intact.
Thus, to the first order of the expansion in $\epsilon$, 
solutions are the same as those for no additional potential given in
Eqs.~(\ref{eq:0thorder_moduli_field}), (\ref{eq:a0_1st}), and (\ref{eq:sigma_1st}).
On the other hand,
the quasi zero mode $\phi(x^0,x^3)$ may receive a correction from $V_{\rm add}$.
Plugging the zero-th order configuration $H^{(0)}(x^1,x^2;\phi(x^0,x^3))$ into Eq.~(\ref{eq:V_add}), 
we 
obtain the following expression
\beq
V_{\rm add}=  
\frac{\lambda m^2}{g^2}F^2 \times \left(\frac{\phi + \bar\phi}{1+|\phi|^2}\right)^2,
\eeq
where $F$ is defined in Eq.~(\ref{eq:def_F}).
This potential should be think of as a correction of order $\epsilon^2$ for the quasi zero mode.
Indeed, it is possible if we choose $\lambda$ to be a parameter of  ${\cal O}(1)$. After integrating
in $x^1$ and $x^2$, one 
obtains
\beq
{\cal L}_{\rm eff + add} = 
\frac{4\pi}{g^2} 
\frac{|\p_\alpha\phi|^2 
- m^2 |\phi|^2 - \tilde m^2(\phi+\bar\phi)^2}{(1+|\phi|^2)^2},
\eeq
with
\beq
\tilde m^2 = m^2 \times \frac{\lambda }{4\pi}\int F^2 d^2x \sim {\cal O}(m^2).
\eeq
All the terms here are of order $\epsilon^2$ compared to the leading term $2\pi v^2$, as we
desired.
In this way, the new mass $\tilde m$ appears, which sets the phase of $\phi$ zero or $\pi$.
In summary, the stability of the solution with $\tilde \Phi = 0$ is assured by introducing 
$V_{\rm add}$.

\section{Dyon in the Higgs phase}

Another non-trivial application of the rigid-body approximation is 
a dyonic solution  
in the non-Abelian superconductor. 
The corresponding solution in 3+1 dimensions are known 
as 1/4 BPS state \cite{Eto:2004rz,Kim:2006ee,Eto:2005sw}.
A time-dependent stationary solution of the equations of motion
(\ref{eq:eom_eff_1}) and (\ref{eq:eom_eff_2}) is given by
\beq
\Theta(x^0,x^3) = 2 \arctan\left(e^{\sqrt{m^2-\omega^2}\, 
x^3}\right),\quad
\Phi(x^0,x^3) = \omega x^0,
\label{eq:dyon_moduli}
\eeq
with $\omega < m$. Here, we again use the coordinates $\Theta$ and $\Phi$.
This is called a Q-kink solution\cite{Abraham:1992vb}, 
which carries both topological (magnetic) and Noether (electric) charges.
Let us see how the above solution can be derived through a standard Bogomol'nyi technique.
The 1+1 dimensional Hamiltonian corresponding to the effective Lagrangian (\ref{eq:eff_lag}) 
can be cast into the following perfect square form as
\beq
{\cal H}_{\rm eff} &=& \frac{\pi}{g^2}
\left[
\dot\Theta^2 + \Theta'{}^2 + \sin^2\Theta
\left(\dot \Phi^2+\Phi'{}^2\right) + m^2 \sin^2\Theta
\right]\non
&=& \frac{\pi}{g^2}
\bigg[\dot\Theta^2+\sin^2\Theta\Phi'{}^2
+ \left(\Theta' - m\cos\alpha\sin\Theta\right)^2
+ \sin^2\Theta\left(\dot\Phi - m\sin\alpha\right)^2\non
&+& 2m\Theta'\cos\alpha \sin\Theta+2m\dot\Phi\sin\alpha\sin^2\Theta\bigg] \non
&\ge& \frac{2\pi m}{g^2} \left(\Theta'\cos\alpha \sin\Theta + \dot\Phi\sin\alpha\sin^2\Theta\right).
\eeq
Here $\alpha$ is an arbitrary constant.
Integrating this over the $x^3$ direction, one 
obtains the following inequality 
\beq
\int dx^3~{\cal H}_{\rm eff} \ge \frac{4\pi m}{g^2} \left(T \cos\alpha + N \sin \alpha\right),
\eeq
where we defined the topological charge and the Noether charge by
\beq
T &=& \int dx^3~\frac{1}{2}\Theta'\sin\Theta = - \frac{1}{2}\big[\cos\Theta\big]^{x^3\to\infty}_{x^3\to-\infty},\\
N &=& \int dx^3~ \frac{1}{2} \dot \Phi \sin^2\Theta. \label{noether}
\eeq
Since the parameter $\alpha$ is arbitrary, the strictest bound for given $T$ and $N$ is
obtained when it holds
\beq
\tan\alpha = \frac{N}{T}.
\eeq
The bound is saturated for solutions for the following first order equations 
\beq
\Theta' = \frac{mT}{\sqrt{T^2+N^2}} \sin \Theta,\quad
\dot\Phi = \frac{m N}{\sqrt{T^2+N^2}},
\eeq
with the energy
\beq
\int x^3~{\cal H}_{\rm eff} = \frac{4\pi m}{g^2}\sqrt{T^2 + N^2}.
\eeq
The solution given in Eq.~(\ref{eq:dyon_moduli}) corresponds to the solution with
the charges
\beq
T= 1,\quad N = \frac{\omega}{\sqrt{m^2-\omega^2}}.
\eeq
Thus, the energy is given by
\beq
\int dx^3~{\cal H}_{\rm eff} = \frac{4\pi m}{g^2}\frac{m}{\sqrt{m^2-\omega^2}}.
\eeq

\begin{figure}[ht]
\begin{center}
\begin{tabular}{cc}
\includegraphics[height=4.5cm]{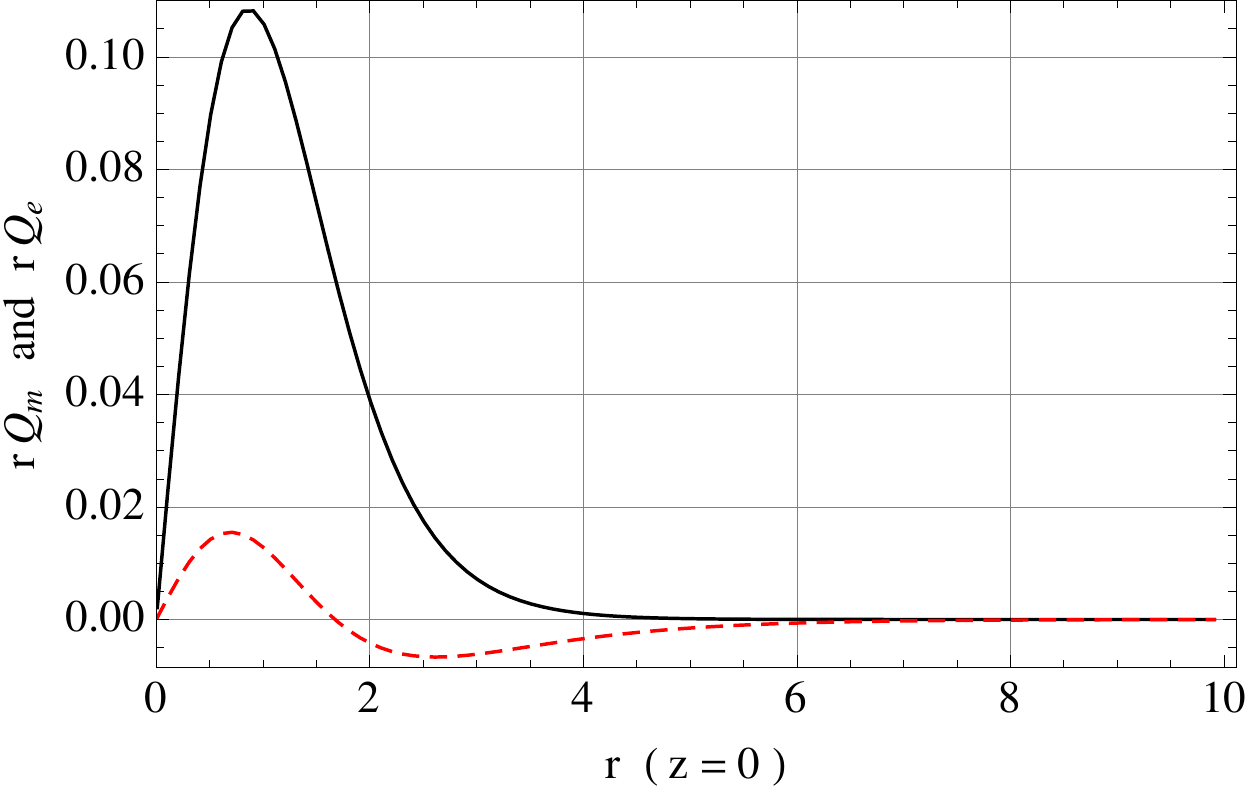}
&
\includegraphics[height=4.5cm]{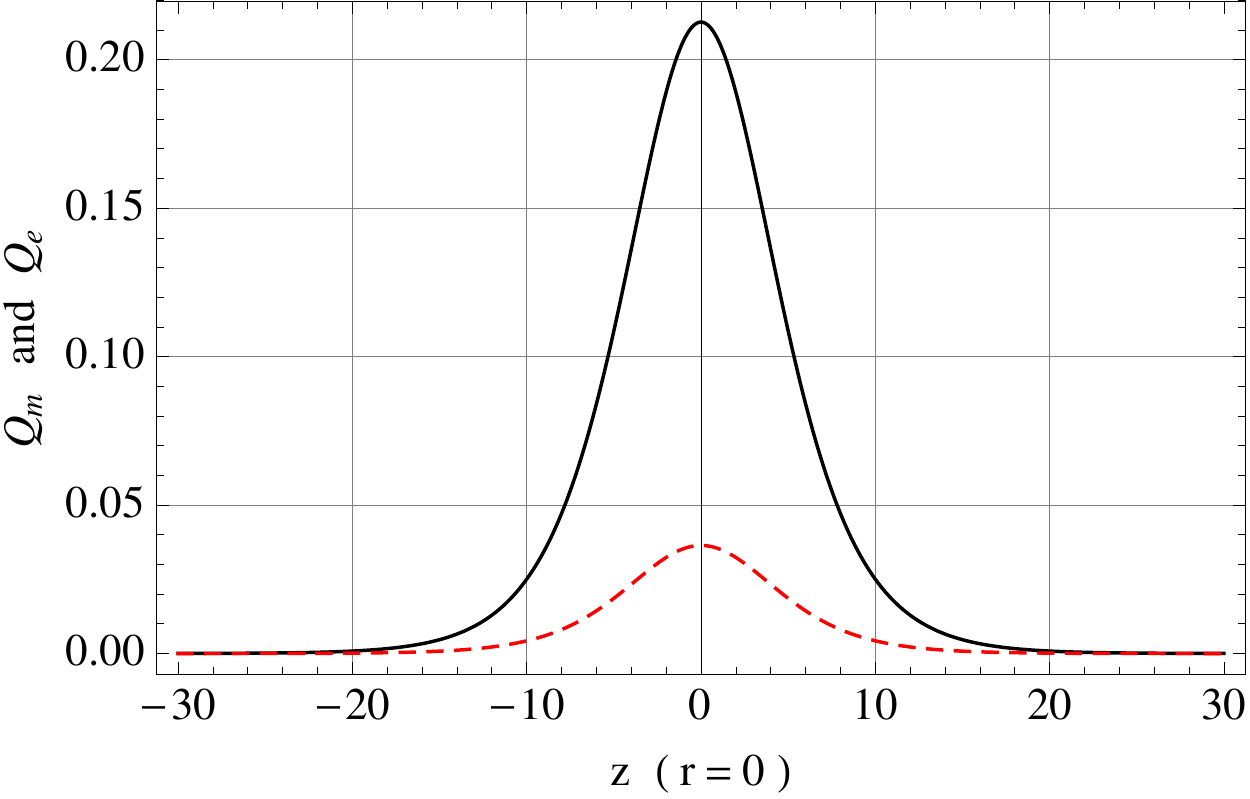}
\end{tabular}
\caption{\capfont The electric and magnetic charge distribution 
of a slender dyonic monopole in the Higgs phase. 
The left panel shows the magnetic (black solid line) 
$r{\cal Q}_{\rm m}$ and electric (red dashed line) $r{\cal Q}_{\rm e}$ 
charge densities multiplied by $r$ at $z=0$ as a function of $r$. 
The right panel shows those at $r=0$ as a function of $z$. 
The parameters are set as $gv = 1$, $m=1/5$ and $\omega = 1/10$. 
}
\label{fig:dyon}
\end{center}
\end{figure}

Let us see this configuration in 1+1 dimensions from the 3+1 dimensional perspective.
The corresponding electric and magnetic  
fields can be easily obtained from $\bar A^{(0)}$ in 
Eq.~(\ref{eq:A_0thorder}) and $A_\alpha^{(1)}$ in Eq.~(\ref{eq:a0_1st}) 
by noting that $x^0, x^3$ dependence resides only in $\Theta$ 
and $\Phi$ in Eq.~(\ref{eq:dyon_moduli}) which appears through 
$U(x^0, x^3)$ in Eqs.~(\ref{eq:flavor_rot}), (\ref{eq:cp1_angle}). 
The Abelian electric and magnetic fields are given as 
\beq
F^0_{12} \simeq - \p\bar\p\psi,\quad F^0_{23} \simeq 0,
\quad F^0_{31} \simeq 0,
\quad F^0_{01} = F^0_{02} = F^0_{03} = 0,
\eeq
and their non-Abelian counterparts as 
\beq
B_3^\Sigma &=& F^\Sigma_{12} \simeq \p\bar\p\psi \tanh 
\left(\sqrt{m^2-\omega^2}~ x^3\right),\\
B_1^\Sigma &=& F^\Sigma_{23} \simeq 
\frac{\sqrt{m^2-\omega^2}}{4} \p_1(r^2e^{-\psi}) \sech^2 
\left(\sqrt{m^2-\omega^2}~ x^3\right),\\
B_2^\Sigma &=& F^\Sigma_{31} \simeq
\frac{\sqrt{m^2-\omega^2}}{4} \p_2(r^2e^{-\psi}) \sech^2 
\left(\sqrt{m^2-\omega^2}~ x^3\right),\\
E_1^\Sigma &=& F^\Sigma_{01} \simeq  
\frac{\omega}{4} \p_1(r^2e^{-\psi}) \sech^2 
\left(\sqrt{m^2-\omega^2}~ x^3\right),\\
E_2^\Sigma &=& F^\Sigma_{02} \simeq
\frac{\omega}{4} \p_2(r^2e^{-\psi}) \sech^2 
\left(\sqrt{m^2-\omega^2}~ x^3\right),\\
E_3^\Sigma &=& F^\Sigma_{03} \simeq 0.
\eeq
$F_{03}$ is of order ${\cal O}(\epsilon^2)$ by definition.
Then we can define 
topological and electric charge densities as
\beq
{\cal Q}_{\rm m} &=& \frac{1}{g}{\rm div}\vec{B}^\Sigma = \frac{1}{g}\sqrt{m^2-\omega^2}\sech^2 
\left(\sqrt{m^2-\omega^2}~ x^3\right)
\left(\p\bar\p - \frac{4}{g^2v^2}(\p\bar\p)^2\right)\psi,\\
{\cal Q}_{\rm e} &=& \frac{1}{g}{\rm div}\vec{E}^\Sigma = \frac{1}{g}\omega \sech^2 
\left(\sqrt{m^2-\omega^2}~ x^3\right)\left(- \frac{4}{g^2v^2}(\p\bar\p)^2\psi\right).
\eeq
Note that the factor $\omega \sech^2(\sqrt{m^2-\omega^2}~x^3)$ appearing
in ${\cal Q}_{\rm e}$ corresponds to the integrand of the Noether charge $N$ of $Q$-kink, 
Eq. (\ref{noether}), as
\beq
\dot\Phi \sin^2\Theta = \omega  \sech^2(\sqrt{m^2-\omega^2}~x^3).
\eeq
In this way, one realizes that 
the electric charge density ${\cal Q}_{\rm e}$ is a direct manifestation of the Noether charge
density of the $Q$-kink, as should be expected.
Since the magnetic and electric fields except for $F^{\Sigma}_{12}$ go to zero
at spatial infinity, total 
topological
and electric charges are given by
\beq
Q_{\rm m} = \frac{2\pi}{g},\qquad
Q_{\rm e} = 0.
\eeq
Although the total electric charge is zero, its density is 
not equal to zero in contrast to the case of the monopole and 
anti-monopole scattering, see Eq.~(\ref{eq:EC_m_am}).
Similarly to a neutron, there is a non-trivial charge density 
distribution inside the monopole.
The electric and magnetic charge densities are shown in Fig.~\ref{fig:dyon}.

Let us compare the dyonic monopoles in the Coulomb phase and the Higgs phase.
In the Coulomb phase, the conserved Noether charge of the unbroken $U(1)$ {\it gauge}
symmetry is associated with the time dependent internal moduli $S^1$ of the static monopole.
Since there is no fields which would screen the electric charge, the total electric charge can be non
zero in the Coulomb phase \cite{Christ:1976cg}. 
On the other hand, the $U(1)$ symmetry is broken in the Higgs phase, but it is locked with
the flavor symmetry of the fundamental Higgs field $H$. Therefore, the non-zero Noether charge of
{\it flavor} $U(1)$ symmetry also exists. However, since $H$ is charged under the $U(1)$
symmetry and it is condensed in the Higgs phase, the electric charge of the monopole generated
by the rotation of the $U(1)$ phase is screened. Thus, the total electric charge
becomes zero  in the Higgs phase.

By flipping the 
sign of $\omega$, one can also construct the 
slender dyonic monopole with opposite electric 
charge density distribution. It is also easy to 
construct dyonic anti-monopoles by flipping the topological  charge.

\section{Conclusion and discussion}

In this paper we have investigated the low energy dynamics of 
monopoles and anti-monopoles in the non-Abelian superconductor.
We have restricted ourselves to 
the parameter region $m \ll gv$ where the monopoles are of slender 
ellipsoidal shape, confined on a vortex string, with the 
cross-section comparable to that of the monopole. 
For that reason, the scattering problem 
becomes essentially $1+1$ dimensional. 
Indeed, we have 
found that at least a part of the low energy dynamics 
is identical to the sine-Gordon system 
in $1+1$ dimensions up to the first  order of the expansion 
in $\epsilon=m/(gv)$, when 
$\{m,\p_0,\p_3\} \ll \{gv,\partial_1,\partial_2\}$ holds. 
This observation is very useful  
because the sine-Gordon system is solvable. 
In the literature, only the static kink was identified with the monopole.
In this paper, we have dealt with all the sine-Gordon solutions and 
have constructed the dictionary with which one can easily translate  
the dynamics of sine-Gordon kinks in 1+1 dimensions into the 
dynamics 
of monopoles in $1+3$ dimensions. 
A surprising fact is that the monopole and anti-monopole do 
not always decay into radiation when they make a head-on collision, 
although they are 
not protected by topology. 
We have studied  
three concrete 
examples: (1) the magnetic meson which is the bound state of 
the slender monopole and anti-monopole, 
(2) the scattering of the monopole and anti-monopole 
of the same species, and (3) the scattering of the monopole 
and anti-monopole of the different species. 
All these three examples show that the monopole 
and anti-monopole do not always annihilate. 
This observation may be counter-intuitive and 
remarkable.  

\begin{figure}
\begin{center}
    \begin{tabular}{cc}
      \begin{minipage}{5cm}
        \begin{center}
          \includegraphics[clip, height=1.2cm]{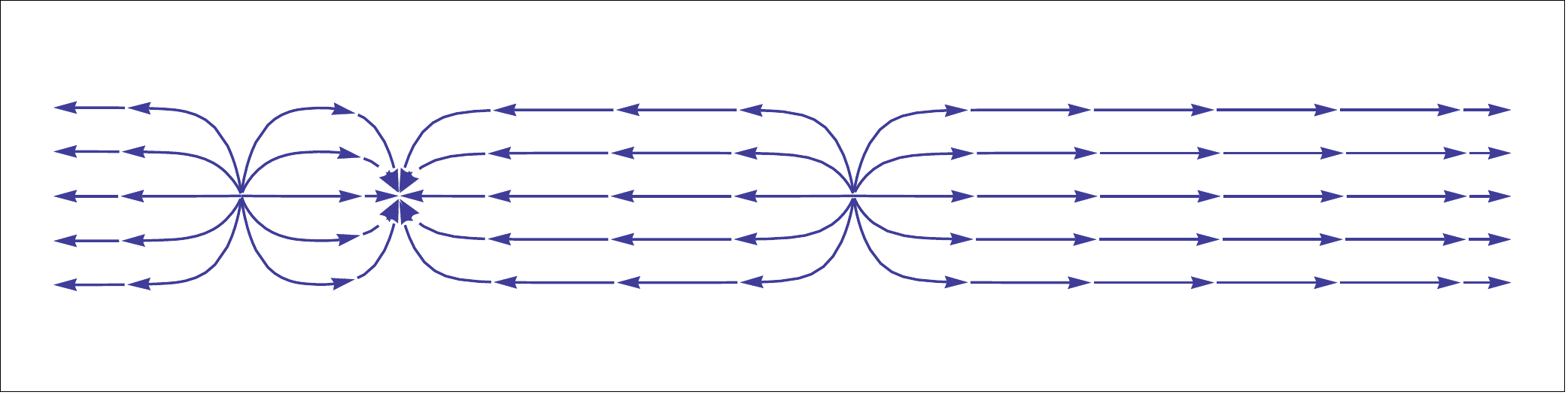}\\
          \includegraphics[clip, height=1.2cm]{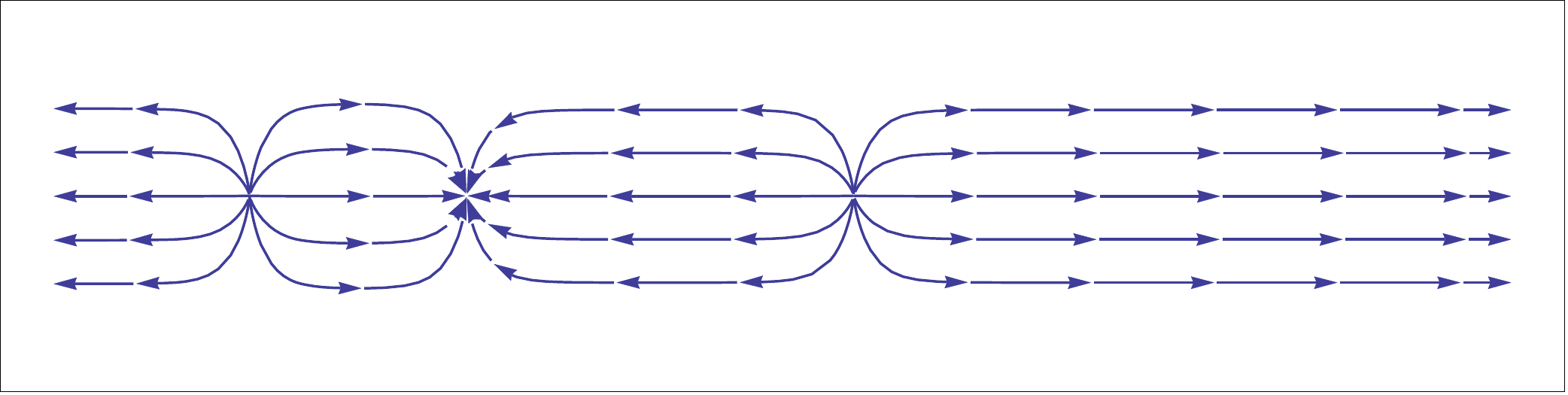}\\
          \includegraphics[clip, height=1.2cm]{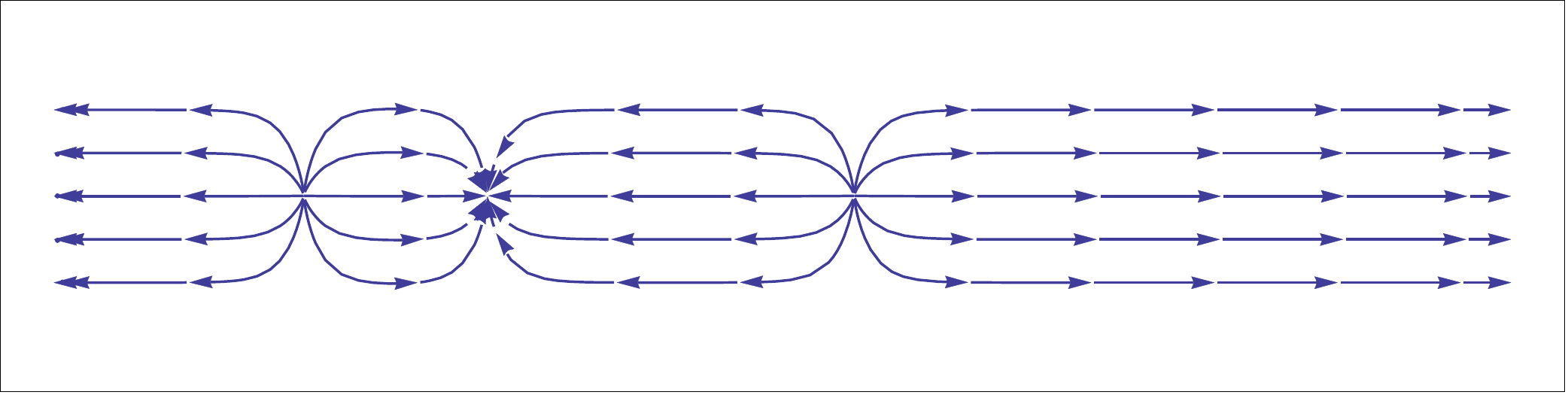}\\
          \includegraphics[clip, height=1.2cm]{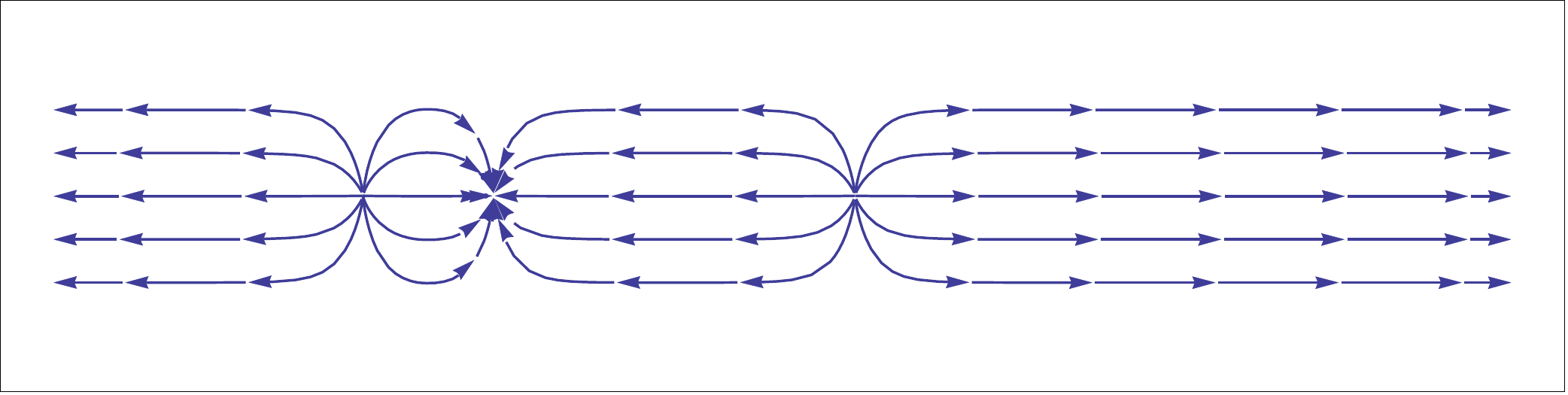}\\
          \includegraphics[clip, height=1.2cm]{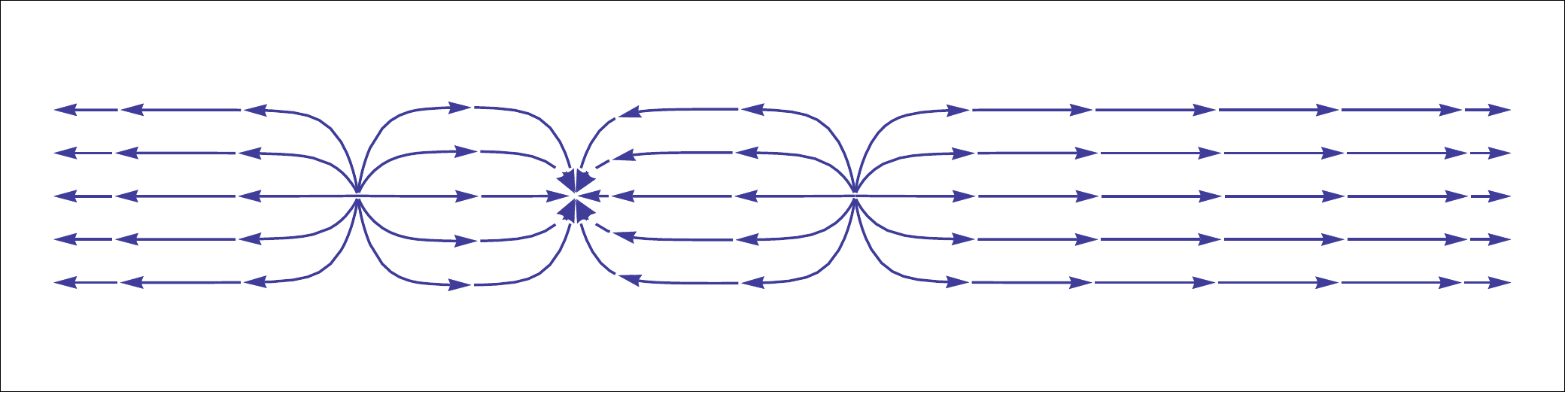}\\
          \includegraphics[clip, height=1.2cm]{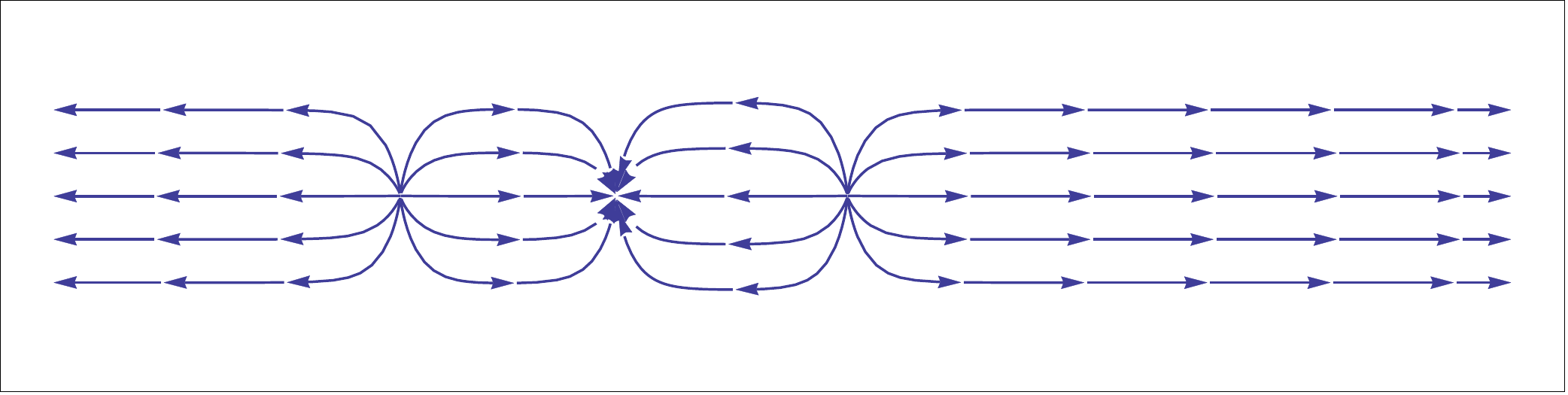}\\
          \includegraphics[clip, height=1.2cm]{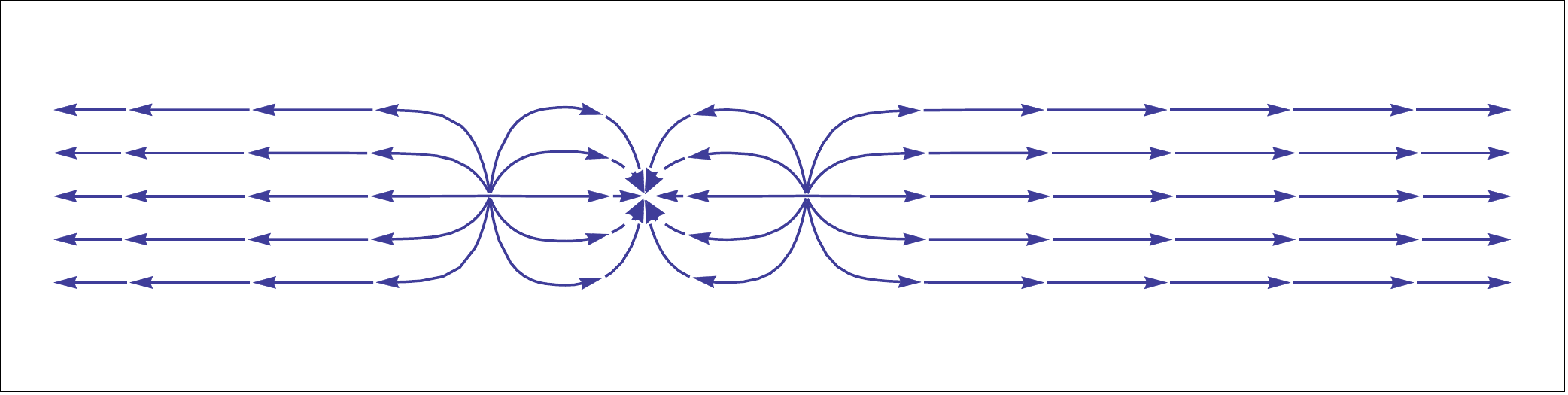}\\
          \includegraphics[clip, height=1.2cm]{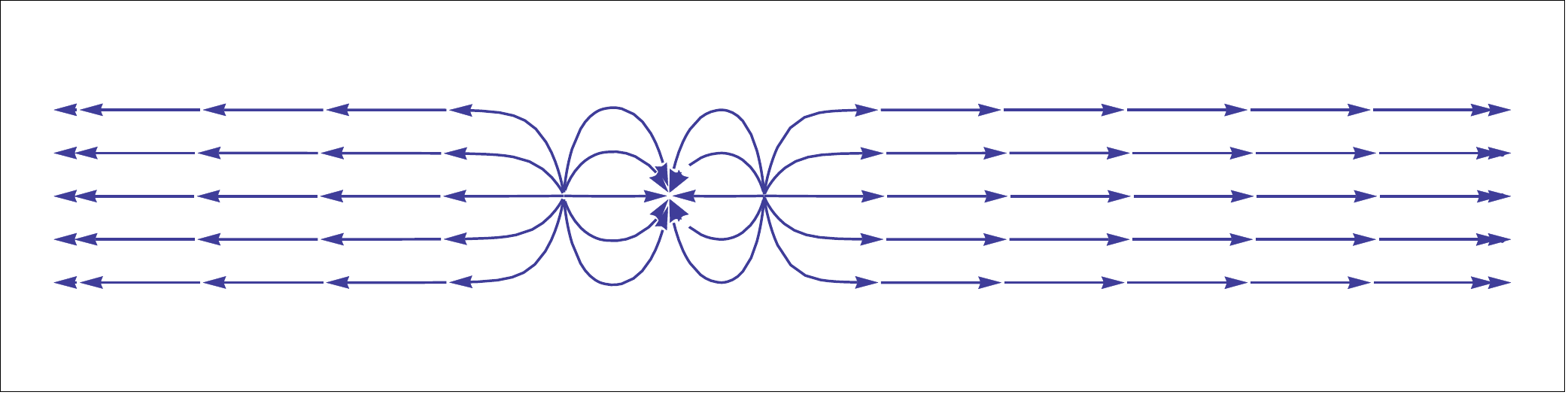}\\
          \includegraphics[clip, height=1.2cm]{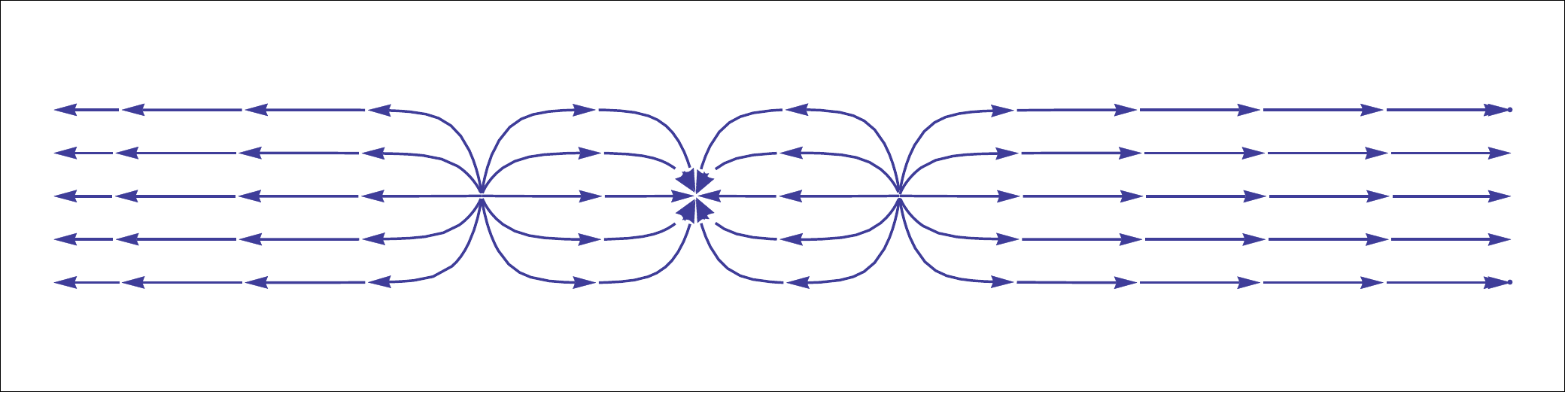}\\
         \includegraphics[clip, height=1.2cm]{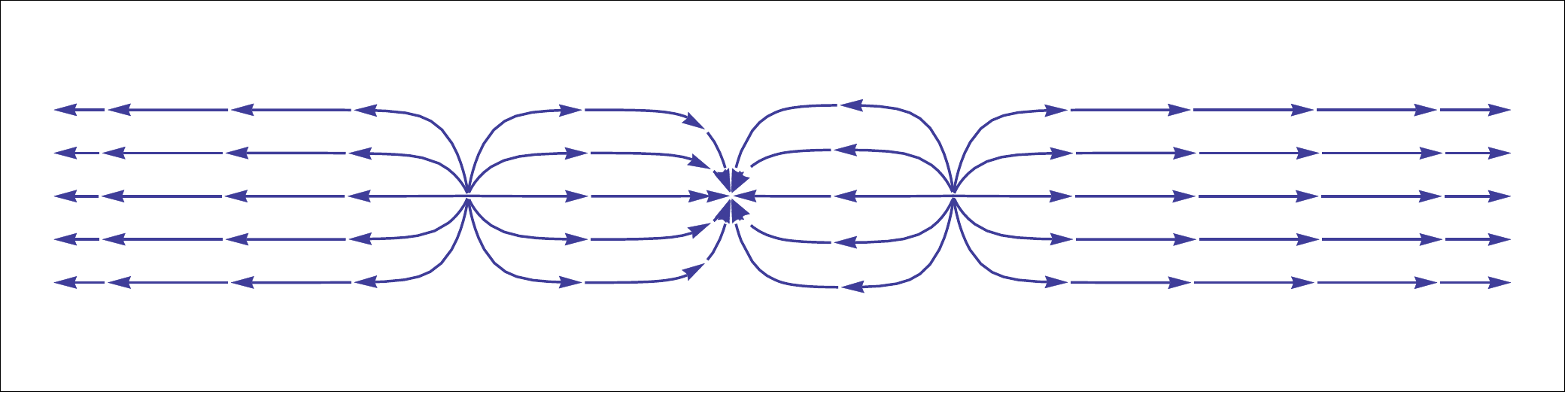}\\
         \includegraphics[clip, height=1.2cm]{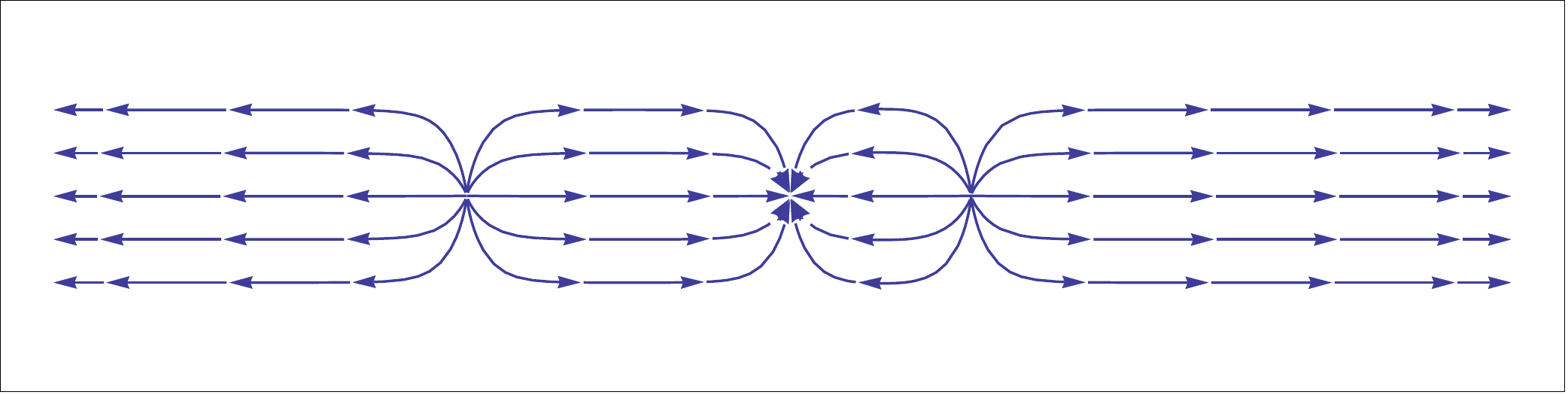}\\
         \includegraphics[clip, height=1.2cm]{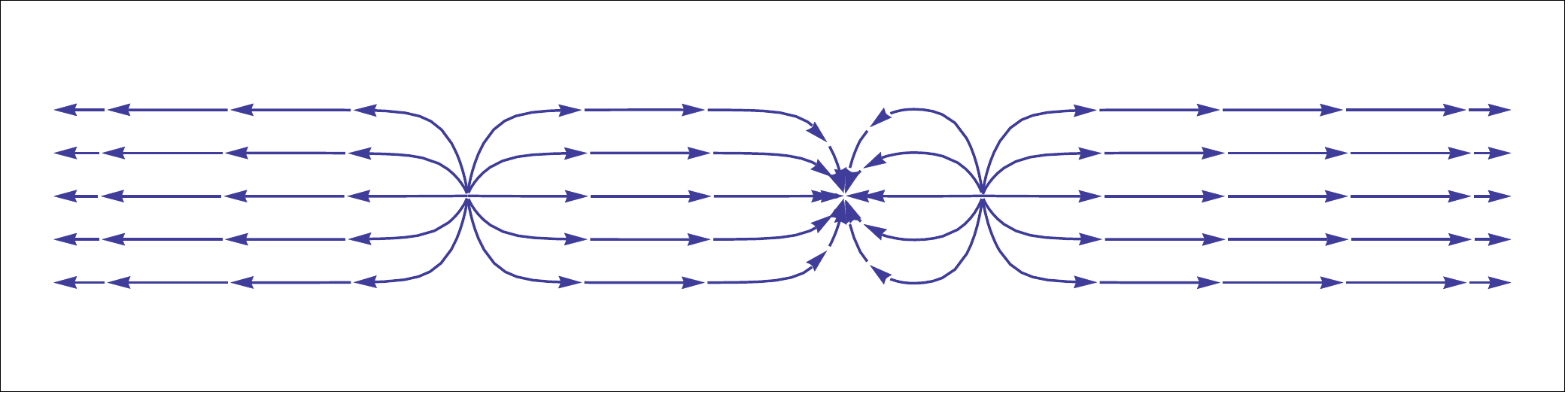}\\
         \includegraphics[clip, height=1.2cm]{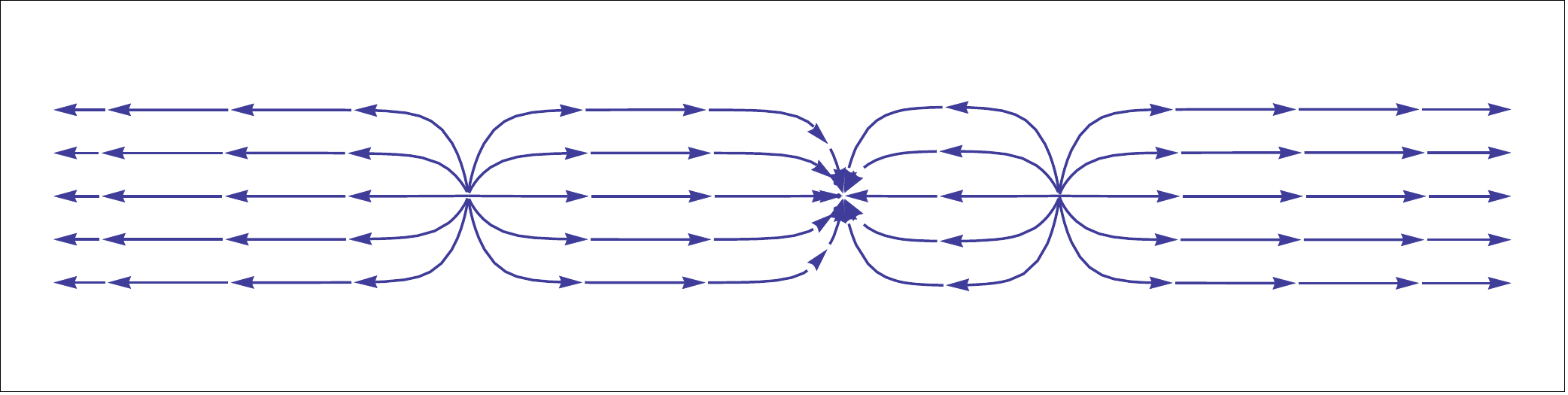}\\
         \includegraphics[clip, height=1.2cm]{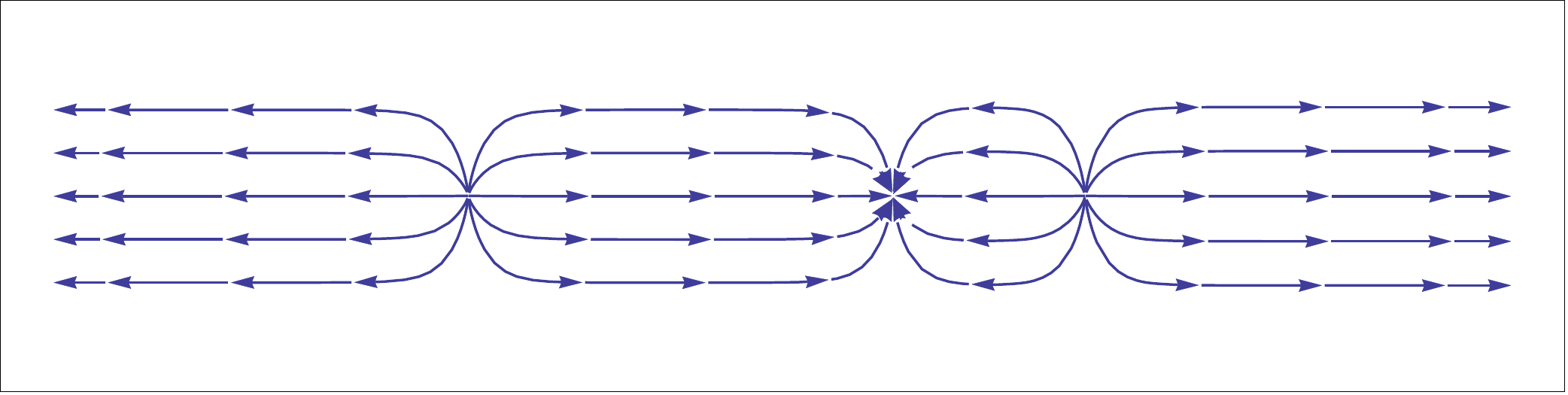}\\
         \includegraphics[clip, height=1.2cm]{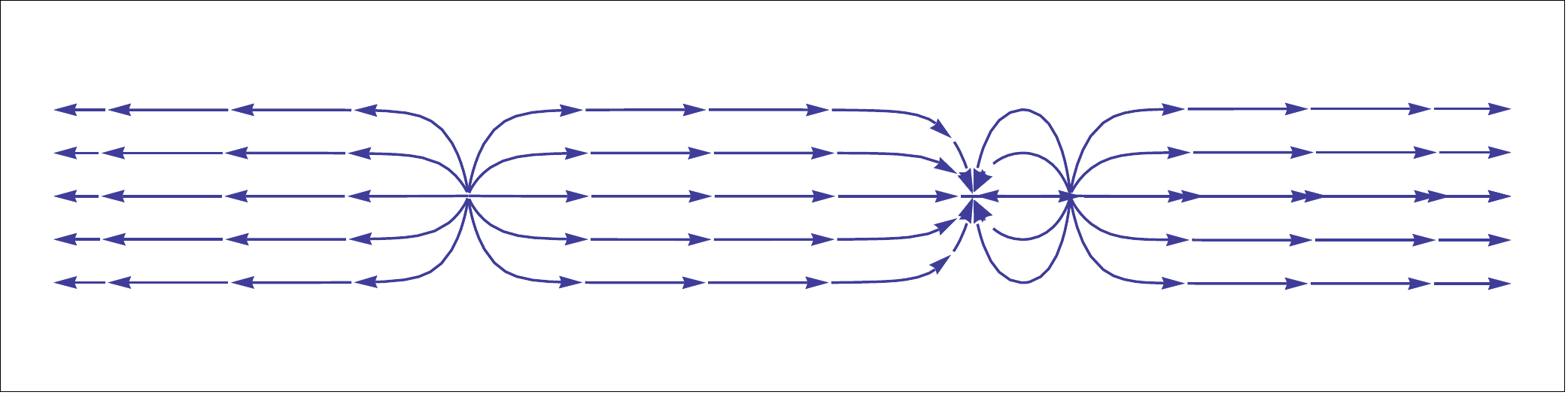}\\
         \includegraphics[clip, height=1.2cm]{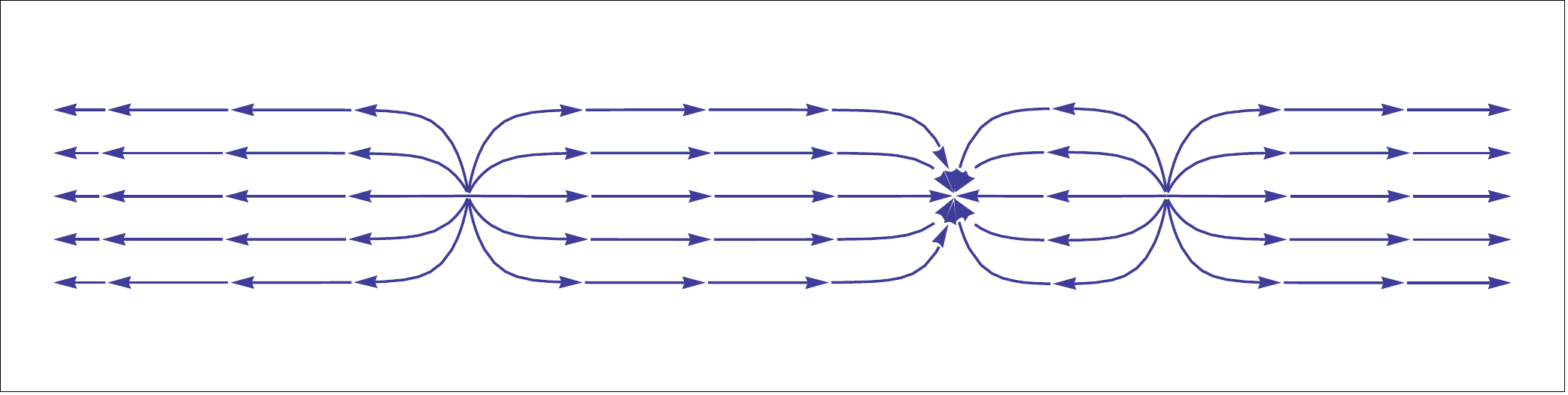}
         \end{center}
      \end{minipage}
      &
      \begin{minipage}{5cm}
        \begin{center}
         \includegraphics[clip, height=1.2cm]{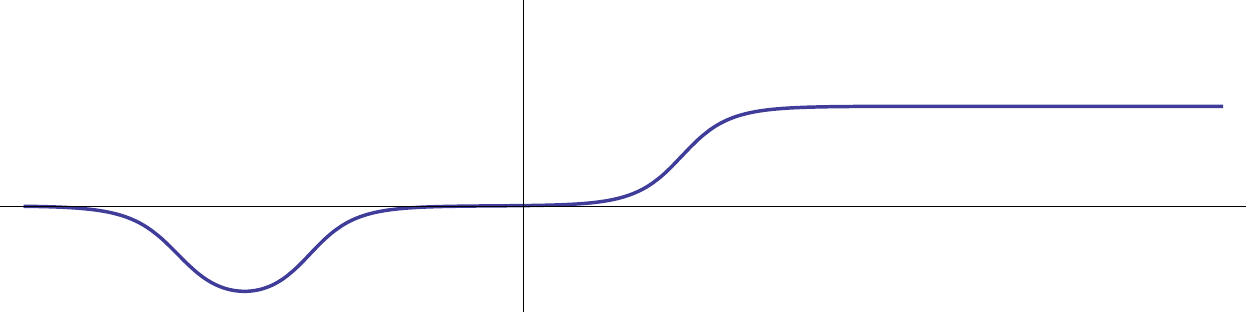}\\
         \includegraphics[clip, height=1.2cm]{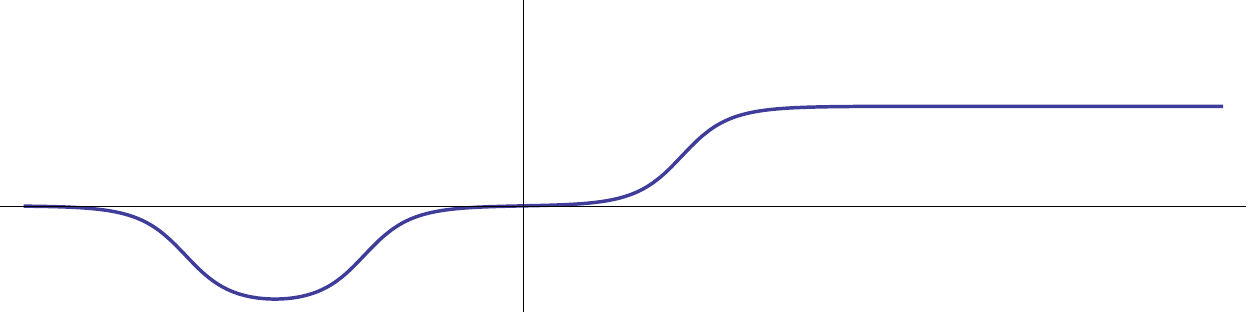}\\
         \includegraphics[clip, height=1.2cm]{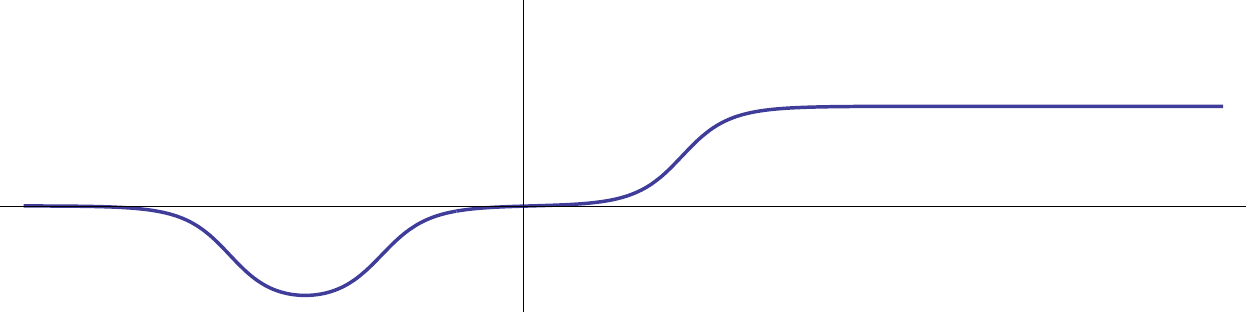}\\
         \includegraphics[clip, height=1.2cm]{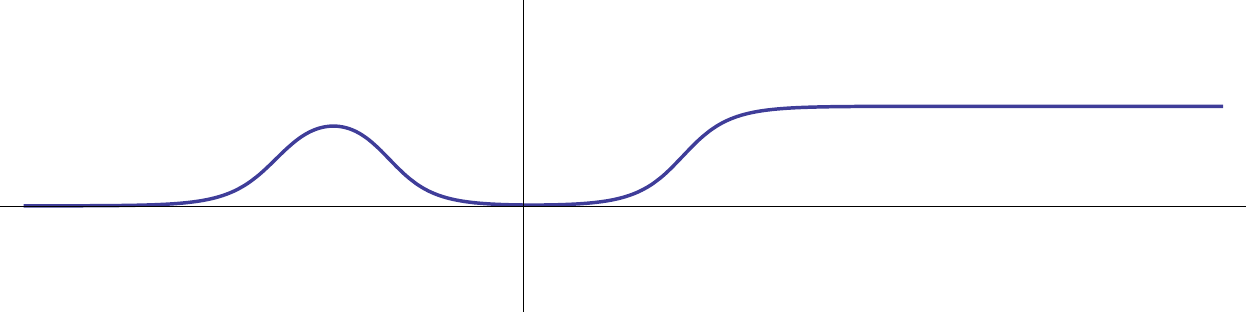}\\
         \includegraphics[clip, height=1.2cm]{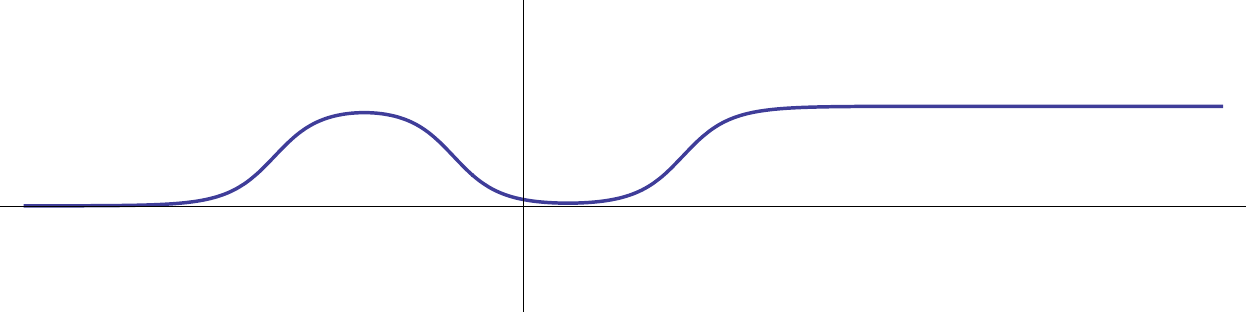}\\
         \includegraphics[clip, height=1.2cm]{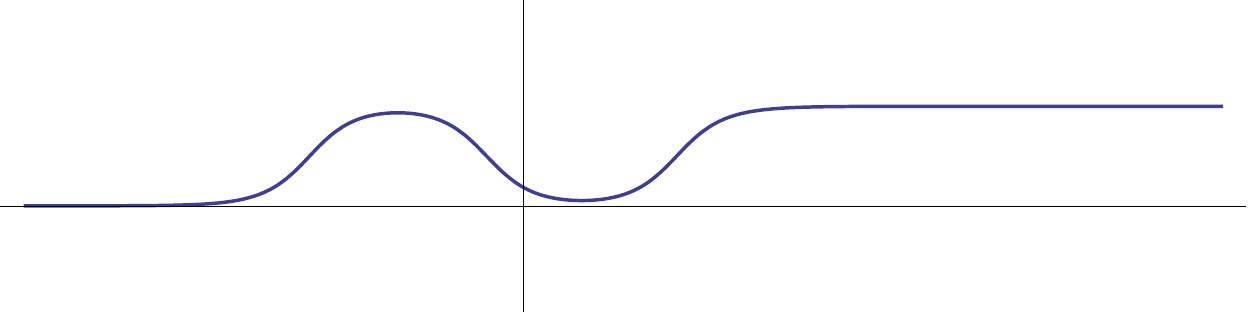}\\
         \includegraphics[clip, height=1.2cm]{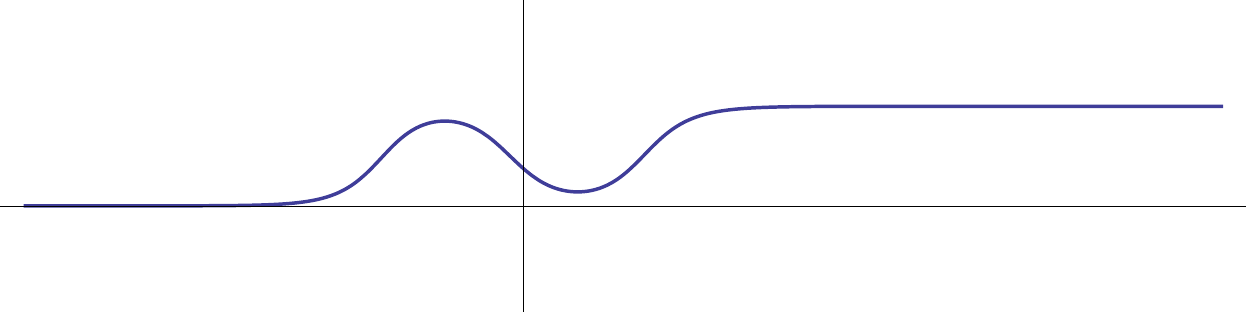}\\
         \includegraphics[clip, height=1.2cm]{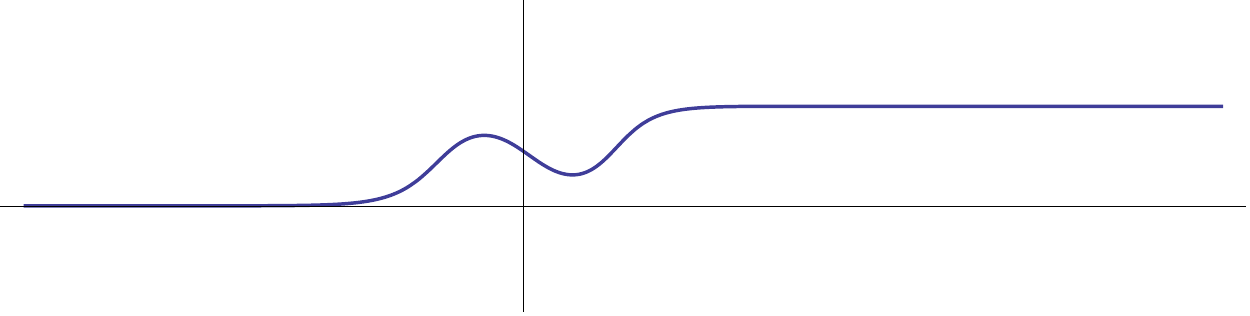}\\
         \includegraphics[clip, height=1.2cm]{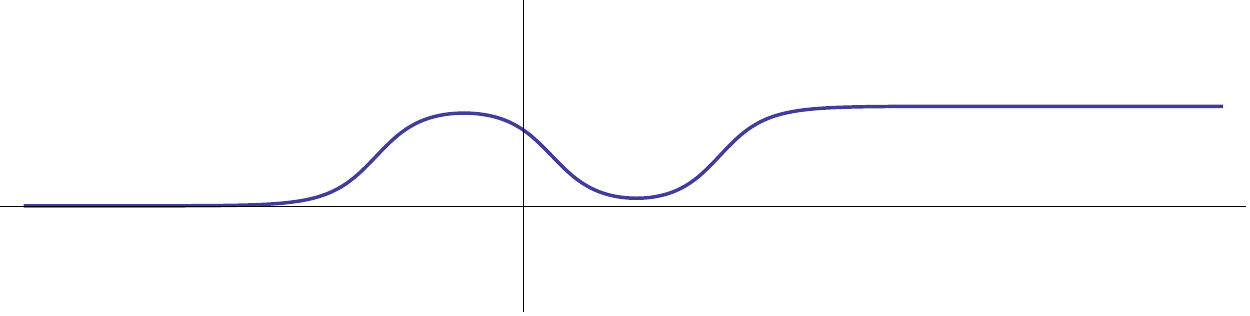}\\
        \includegraphics[clip, height=1.2cm]{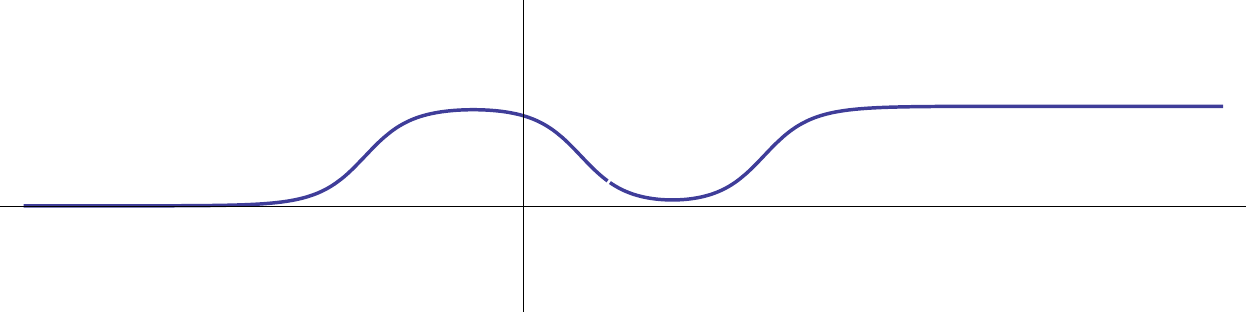}\\
        \includegraphics[clip, height=1.2cm]{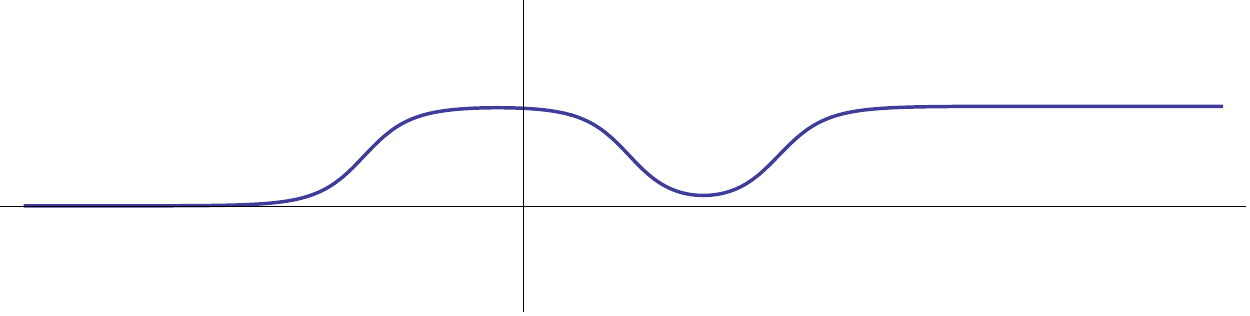}\\
        \includegraphics[clip, height=1.2cm]{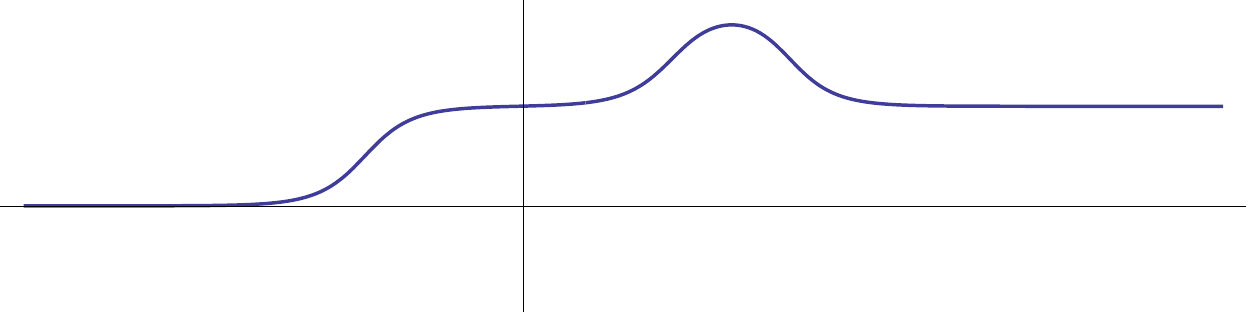}\\
        \includegraphics[clip, height=1.2cm]{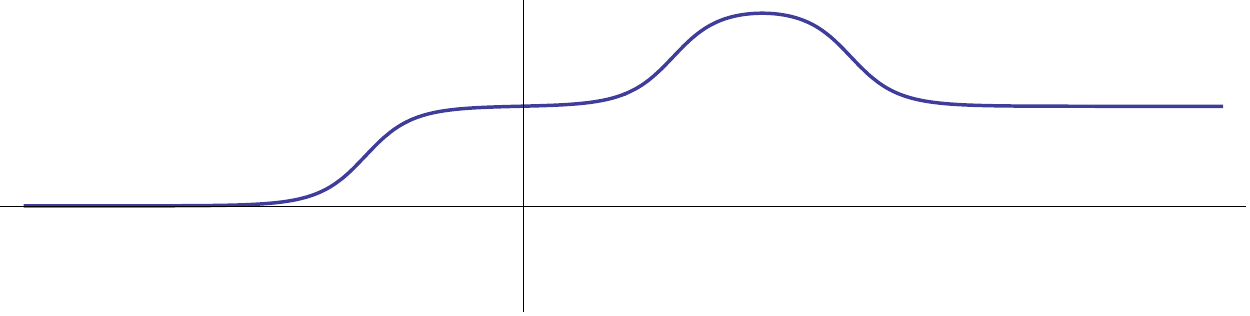}\\
        \includegraphics[clip, height=1.2cm]{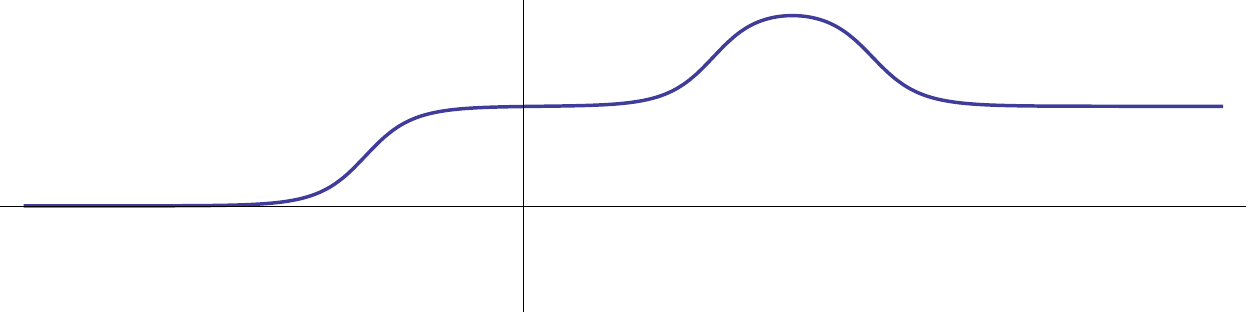}\\
        \includegraphics[clip, height=1.2cm]{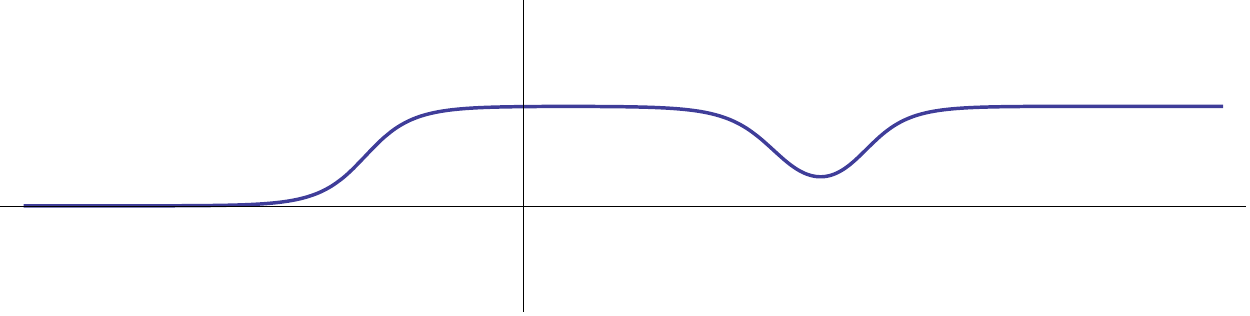}\\
        \includegraphics[clip, height=1.2cm]{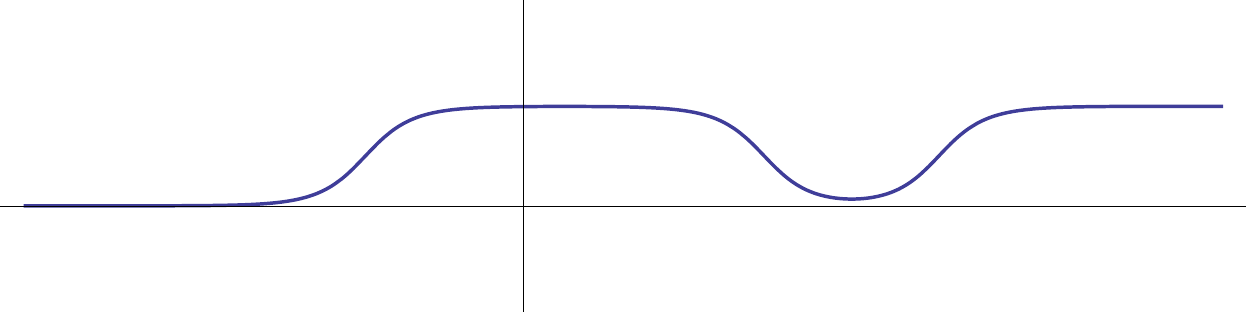}
         \end{center}
      \end{minipage}
     \end{tabular} 
\caption{\capfont The scattering of the magnetic meson and the isolated monopole.
We set $gv=1$,
$m=1/3$ and $\omega=1/10$, $b=1/10$, $k=0$, $x^1\in[-2.5,2.5]$, $x^3\in[-50,70]$.
The snapshots are taken for $x^0 \in[-200,250]$ with $\delta t = 30$ interval.}
\label{fig:meson_mono}
\end{center}
\end{figure}

In order to illustrate the usefulness of our dictionary,  
let us also give a solution of three body scattering. 
Three body system is in general very complicated. 
However, due to the  
power of integrability, we can easily describe the three body  
collision of a magnetic meson and a monopole: 
\beq
\tilde \Theta &=& 2 \tan ^{-1}e^{\theta_2} +
2 \tan ^{-1}\left[\frac{a_1+a_3}{a_1-a_3} 
\tan \left\{\tan ^{-1}\frac{\left(a_1+a_2\right) 
\sinh \left(\theta _1-\theta _2\right)/2}{\left(a_1-a_2\right) 
\cosh \left(\theta _1+\theta _2\right)/2}\right.\right.\non
&&\left.\left.-\tan ^{-1}\frac{\left(a_2+a_3\right) 
\sinh \left(\theta _2-\theta _3\right)/2}{\left(a_2-a_3\right) 
\cosh \left(\theta _2+\theta _3\right)/2}\right\}\right],\\
\theta_n &=& \frac{a_n^2+1}{2 a_n} \left(m x^3+\frac{a_n^2-1}{a_n^2+1} 
m x^0\right)
\quad (n=1,2,3),\\
a_1 &=& \sqrt{\frac{1-b}{1+b}} \left(\sqrt{1-\omega^2}-i \omega\right),\\
a_2 &=& \sqrt{\frac{1-k}{1+k}},\\
a_3 &=& \sqrt{\frac{1-b}{1+b}} \left(\sqrt{1-\omega^2}+i \omega\right).
\eeq
Here $b$ is the velocity of the magnetic meson, 
$k$ is that of the isolated monopole, 
and $\omega$ is the frequency of the magnetic meson. 
The solution is shown in Fig.~\ref{fig:meson_mono}. 
Of course, this is an example and one can easily add any 
number of monopoles and anti-monopoles by using well 
established methods, such as the B\"acklund transformation.

Let us discuss a number of points about our results and 
future directions of our research in the following. 

(I) 
We have studied the specific parameter 
region $m \ll gv$. We have made this choice in order to utilize 
the rigid-body approximation, which leads to 
a nice mapping of the $1+3$ dimensional problem into the 
integrable sine-Gordon system in $1+1$ dimensions. 
However, we can study only the low energy dynamics 
 with our approximation.
In order to go beyond the approximation, 
we need either to include higher order corrections or to 
solve the full equations of motion. 
Although solving $1+3$ dimensional second order 
differential equations is in general not 
an easy task, 
it is worth doing for the following reasons.
Firstly, it is a direct check 
of the validity of our approximation. 
Secondly, the monopoles in the parameter region $m \gg gv$ are 
close to spherical, and 
their dynamics can be very different from those of the 
slender monopoles studied in this work. 
Numerical works may be needed, since there is little 
chance to solve the full equations of motion in the whole 
range of parameter space by analytic methods. 
However, so far, no numerical solution even for the static 
single monopole in the non-Abelian superconductor has been 
constructed (only approximate analytical solution is 
known \cite{Cipriani:2012pa}). 
We hope to do the numerical works and to report the results 
elsewhere.

(II) 
We have studied the scattering of the slender 
monopole and anti-monopole but have not studied the 
dynamics of two monopoles. 
This is because the monopoles are put in the non-Abelian 
superconductor. 
Due to the flux conservation, two identical monopoles 
are not allowed to be next to each other 
on a vortex-string. 
Either an anti-monopole should be sandwiched between them 
or the monopoles 
should sit on different vortex strings. 
The former type of configuration is precisely those studied 
in this work. 
The latter configuration gives almost 
decoupled monopoles and are uninteresting. 
On the other hand, the gauge theory with higher rank gauge group 
such as $U(N)$ with $N\ge 3$ provides a new possibility: 
there are several species of monopoles which can be put 
next to each other on a vortex string. 
We leave an interesting dynamics of such monopoles for future investigations.

(III) 
In this work, we have studied the monopole and vortex string 
in the $U(2)$ gauge theory, where only Abelian vortices and Abelian 
monopoles are possible. 
Non-Abelian vortex is defined as a vortex with non-Abelian 
orientational moduli, and non-Abelian monopoles are those that 
can interact with such non-Abelian vortices. 
These non-Abelian vortices and non-Abelian 
monopoles can arise, if higher rank 
gauge theories such as $U(3)$ are considered.
In general, dynamics of Abelian solitons and non-Abelian 
solitons are quite different. 
For instance, non-Abelian domain walls have been studied 
in comparison to Abelian domain walls in Ref.\cite{Eto:2008dm}. 
Moreover, the non-Abelian vortex and non-Abelian monopole are 
believed to be very relevant to the question of confinemnt. 
The dynamics of non-Abelian monopoles in the 
non-Abelian superconductor has not been 
examined, but it is worthwhile to study it in detail. 
We also leave this problem as a future work.

(IV) The solutions which we considered in this work are all 
noncompact because they are accompanied by the infinitely 
long vortex string. 
It is also interesting to study the dynamics of monopoles 
on a curved vortex string. 
Especially, one may consider a vortex ring (which is called 
a vorton) \cite{Radu:2008pp}. 
It is interesting to pursue the similarity 
between bound states of topological solitons 
and bound states (mesons) of elementary constituents (quarks).

\clearpage

\clearpage


\section*{Acknowledgements} 
M.\ A.\ and F.\ B.\ is supported by the Research Program MSM6840770029 and 
by the project of International Cooperation ATLAS-CERN of 
the Ministry of Education, Youth and Sports of the Czech Republic.
This work is supported by Grant-in Aid for Scientific Research 
No.25400280 (M.\ A.\ ), No.23740226 and No.26800119 (M.\ E.\ ),  and
No. 25400241 (N.\ S.\ )  from the Ministry of Education, 
Culture, Sports, Science and Technology  (MEXT) of Japan.

\end{document}